# Language Models for Automated Classification of Brain MRI Reports and Growth Chart Generation

Maryam Daniali, PhD[1], Shivaram Karandikar[2,3,4], Dabriel Zimmerman[2,3,4], J. Eric Schmitt MD, PhD[4,6], Matthew J. Buczek, MS[2,3,4], Benjamin Jung, PhD[2,3,4], Laura Mercedes[2,3,4], Jakob Seidlitz, PhD[2,3,4], Vanessa Troiani[7], Lena Dorfschmidt, PhD[2,3,4], Eren Kafadar[2,3,4], Remo Williams, MS[2,3,4], Susan Sotardi, MD, MSE[5,6], Arastoo Vosough MD, PhD[5,6], Scott Haag, PhD[1,8], Jenna M. Schabdach, PhD[2,3], Aaron Alexander-Bloch, MD, PhD[2,3,4]

**Author Affiliations**
1 Department of Biomedical and Health Informatics, Children's Hospital of Philadelphia (CHOP), PA
2 Lifespan Brain Institute (LiBI) of the Children's Hospital of Philadelphia (CHOP) and Penn Medicine, Philadelphia, PA
3 Department of Child and Adolescent Psychiatry and Behavioral Science
Children's Hospital of Philadelphia, Philadelphia, PA;
4 Department of Psychiatry, University of Pennsylvania, Philadelphia, PA

5 Department of Radiology, Children's Hospital of Philadelphia
6 Department of Radiology, University of Pennsylvania
7 Department of Developmental Medicine, Geisinger, Lewisburg, PA
8 Department of Computer Science, Drexel University.

Address correspondence either to M.D. (email: danialim@chop.edu) or to A.A.B., Richards Medical Research Laboratories, 3700 Hamilton Walk, Philadelphia, PA 19104 (email: aaron.alexander-bloch@pennmedicine.upenn.edu).

**Acknowledgements**: Undergraduate research volunteers Harry Hearn, Sepp Kohler, Leila Abdel-Qader, Madison Dengel, Angela Um, Julia Katowitz, Naomi Shifman, and Alexa DeJean contributed to the manual annotation of radiology reports. Viveknarayanan Padmanabhan of the CHOP Translational Informatics Group (TRiG) served as an honest broker for radiology scan requests, which were managed by Johnny Mclaughlin in CHOP Radiology.

**Disclosures**: A.A.B holds equity in Centile Bioscience. J.S. is a director of and holds equity in Centile Bioscience.

# Abstract

## Background

Clinically acquired brain magnetic resonance images (MRIs) and their associated radiology reports are a valuable but underutilized resource in multiple contexts, including translational brain imaging research, due to the challenges of manual analysis and data heterogeneity. Automated machine learning methods are needed to unlock the potential of these data for large-scale studies.

## Methods

We developed a series of fine-tuned language models (LMs) to classify brain MRI reports as normal or abnormal. We trained and evaluated BERT, BioBERT, ClinicalBERT, and RadBERT on a dataset of 44,661 reports from the Children's Hospital of Philadelphia (CHOP). We also explored the reasoning capabilities of a leading LM, Gemini 1.5-Pro, for normal report detection. Finally, we applied fully automated image de-identification, curation, processing, and statistical modeling to generate brain growth charts from LM-classified normal scans and compared them to charts derived from human annotations.

## Results

Fine-tuned LMs achieved high performance in classifying radiology reports (Accuracy >99.5%, F1-Score >97%). There was an impact of class imbalance due to training data being skewed toward abnormal reports, which was mitigated by unbalanced training with minority class weighting. Performance remained robust on out-of-distribution data when evaluated on data from a different year and a different institution with a non-overlapping age range, with only a small decrease in precision and F1-score. Full text provided significantly more information than only the summary "Impression" section of reports, resulting in a higher performance. Compared to BERT models, Gemini 1.5-Pro showed promising categorization capabilities, particularly when using clinical inference with a structured sub-question approach. Brain growth charts generated from LM-classified scans were essentially identical with those derived from human-annotated classification (r = 0.99, p < 2.2e-16).

## Conclusions

Our fine-tuned LMs offer a scalable and efficient method for analyzing radiology reports, enabling the automated classification of abnormal and normal brain MRIs in large clinical datasets. One application of these models is automated generation of brain growth charts, which can be used to benchmark quantitative brain image features against typical brain development, overcoming limitations in scalability associated with manual analysis. Further research is needed to address the remaining challenges of data heterogeneity and temporal generalizability and to optimize LM reasoning capabilities.

# Introduction

Clinically acquired radiology scans of the brain represent a rich, yet largely untapped resource for biomedical research and clinical neuroscience. The wealth of information contained within the scans is augmented by their associated reports describing brain pathology, written by expert radiologists. However, the widespread adoption of radiology data in research is significantly affected by limitations in data availability, accessibility, and the inherent challenges associated with their technical heterogeneity and variable quality [1]. Current limitations include strict data privacy regulations, challenges in data sharing across institutions, and the overwhelming amount of data that requires analysis.

Radiology reports, containing crucial information derived from imaging studies like MRI scans, play a vital role in clinical decision-making. Clinicians rely on these reports to identify abnormalities and patterns of specific conditions, determine appropriate courses of action, monitor disease evolution, and assess treatment effectiveness [2–4]. We have previously demonstrated that manual (i.e., human-annotated) classification of normal brain MRIs from these reports can be used to generate a cohort of "clinical controls" (i.e., patients with limited imaging pathology) that are comparable to those based on research-quality MRI data[5]. However, manual analysis of radiology reports is time-consuming, prone to human error, and lacks scalability, limiting the potential for large-scale research initiatives.

Language models (LMs), particularly Transformer-based architectures such as Bidirectional Encoder Representations from Transformers (BERT) [6], have revolutionized Natural Language Processing (NLP). Contextualized embeddings generated by these models represent a significant advancement over their non-contextualized predecessors (e.g., Word2Vec [7]), offering superior performance across a range of NLP tasks. This improvement stems from the ability of LMs to generate context-dependent representations of words and sentences. In the resource-constrained domain of clinical NLP, where annotated data is scarce, the transfer learning paradigm facilitated by pre-training on massive corpora offers a compelling solution. Traditional methods typically necessitate tens if not hundreds of thousands of annotated documents for effective classifier training. In contrast, fine-tuning pre-trained Transformer models significantly reduces this data requirement, reducing the inherent data scarcity problem in clinical applications. Domain-specific adaptations, such as BioBERT[8], ClinicalBERT[9], and RadBERT [10], demonstrate the effectiveness of this approach by further enhancing performance within their respective domains by pre-training on biomedical, clinical, and radiological datasets (see Supplementary Appendix Section S7). LMs offer a promising solution to automate and enhance the analysis of unstructured text data such as radiology reports. LMs excel at NLP tasks, making them suitable for various tasks related to radiology reports, including but not limited to categorizing reports based on predefined criteria to facilitate large-scale analysis and comparisons. Furthermore, LMs can potentially identify subtle patterns and correlations within the text data that may be missed by manual analysis.

In this study, we investigate the potential of LMs to automate the classification of normal and abnormal brain MRI reports. Additionally, we combine report classification with automated image processing and statistical modeling to create brain growth charts from normal MRI scans. The study leverages a set of LMs, fine-tuned on human-annotated data, and their ensemble system, to replicate the process of categorizing radiology reports and creating growth charts previously performed by human annotators. This approach offers a potential solution to address the scalability and reproducibility issues associated with manual analysis, to unlock the data resource of clinically acquired brain MRIs.

# Methods

This study was determined to be exempt from review by Children's Hospital of Philadelphia (CHOP) Institutional Review Board because it consisted of secondary analyses of preexisting clinical data (Supplementary Appendix, see Sections S3 and S4).

## Data

The initial dataset of brain MRI reports was sourced from the CHOP electronic health record (EHR) and comprised results from 318,502 brain MRI scan sessions conducted between 1992 and 2023 (Supplementary Appendix, see Section S3).[11] Each report included detailed findings and overall impressions, signed by a licensed, board-certified neuroradiologist. A random sample of reports was selected for human annotation. Abnormal reports were described in terms of tissue abnormalities, their radiographic properties, and anatomical locations, captured either through a filtering process based on relevant keywords, referred to as a coarse text search (Supplementary Appendix, see Section S5) or human annotation. Reports with limited imaging pathology, referred to as "normal" reports, were explicitly noted as the absence of brain tissue abnormalities and were reviewed by multiple annotators for verification. Because the ultimate goal was to find brain MRIs for quantitative analysis that require high quality images, reports that referenced low quality scans or significant imaging artifacts were also considered abnormal.

To date, 44,661 reports have been examined (average age 11.74 years, std 7.44 years; 54.37% male), of which 3,261 were labeled as normal reports, positive class, and 40,800 were labeled as abnormal reports, negative class (See Fig. 1 and Supplementary Fig. S1). The reports from negative and positive classes were further divided into training (80%), validation (10%), and test (10%) subsets for our experimental purposes, using 5 random seeds, resulting in a total of 17,909 distinct reports for testing across the seeds (see Supplementary S2). A total of 1,261 reports were identified as normal by either the human annotators or the LM system, and scans for these normal reports were obtained through an honest broker at the Radiology Department of CHOP.

We also used a Geisinger Medical Center dataset, consisting of 101 de-identified MRI radiology reports that were available as part of a preprocessing pipeline of a published study [12], to evaluate the generalizability of our method on data from an institution with different patient demographics (Supplementary Appendix, see Section S3).

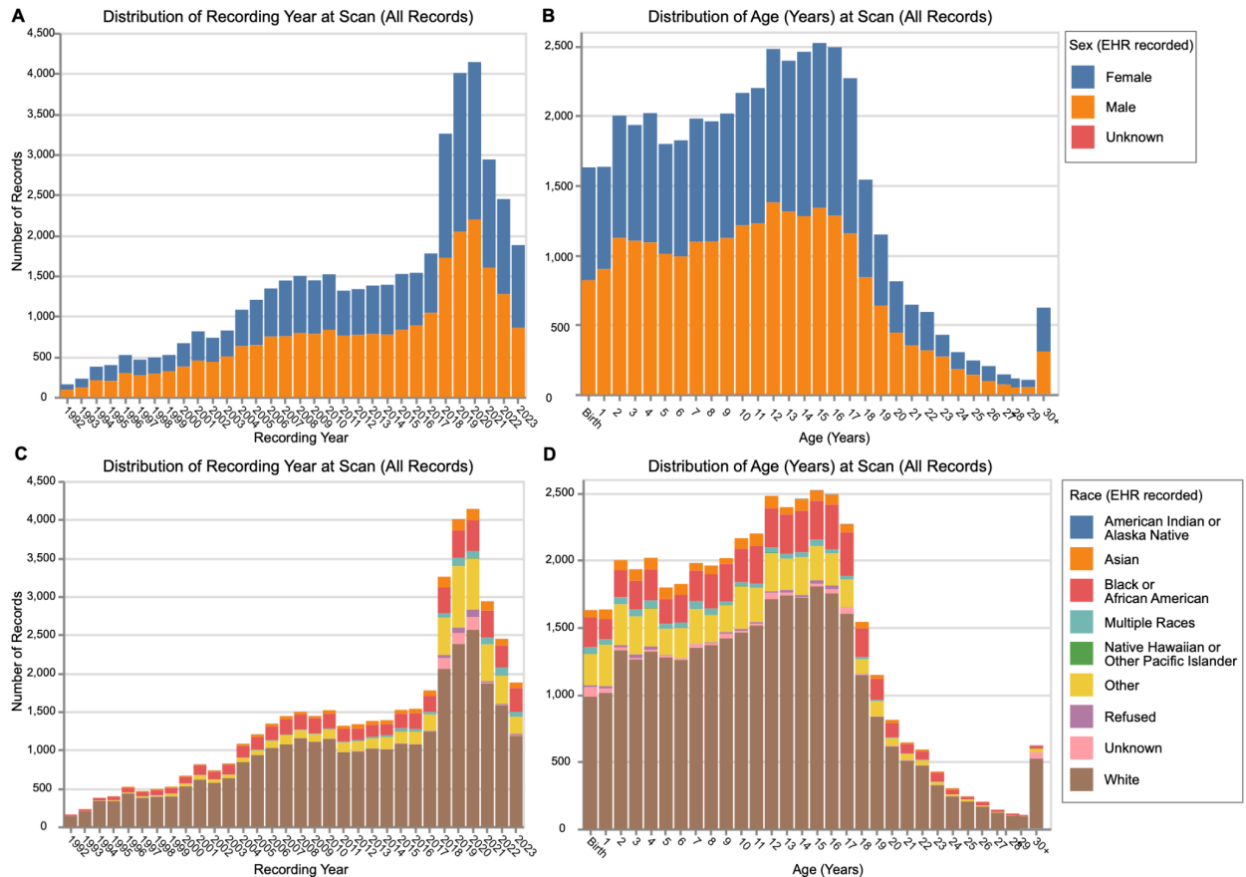

**Figure 1. Demographic and Temporal Distribution of the CHOP Study Cohort (All Graded Reports).** **(A)** Distribution of the number of records by recording year for the entire cohort, stratified by sex as recorded in electronic health records (EHR)s. **(B)** Distribution of patient age at time of scan, stratified by sex as recorded in EHRs. **(C)** Distribution of the number of records by recording year for the entire cohort, stratified by race as recorded in EHRs. **(D)** Distribution of patient age at time of scan, stratified by race as recorded in EHRs. Note that some sex and race categories with smaller proportions may not be clearly visible in the stacked bars, but are represented in the underlying data.

## Clinical MRI, Processing and Analysis

The delivery received from the radiology department consisted of 970 scan sessions from 24 scanners across the CHOP network. The 23% loss relative to the 1,261 sessions requested was attributed to missing data in the picture archiving and communication system (PACS) server. Sessions underwent de-identification with *LOCUTUS*[13], conversion from DICOM to NIfTI using *HeuDiConv*[14], curation into Brain Imaging Data Structure (BIDS)[15] format using *CuBIDS*[16], and whole-brain segmentation and cortical parcellation using SynthSeg+[17]. Details on MRI processing have been described elsewhere[5], and additional information is provided in Supplementary Appendix (see Section S9).

Brain growth charts for global and regional imaging phenotypes were generated using two approaches. The first approach only included volumes from T1-weighted magnetization-prepared rapid acquisition gradient echo (T1-MPRAGE) sequences. The second approach calculated the median volumes across all sequences in each scan session (Supplementary Appendix, see Section S9)[17]. Subsets were created for scan sessions marked as normal by human annotators or LMs. Due to high agreement between human annotators and LMs, the two resulting datasets shared approximately 92% of scan sessions. Growth charts were generated for both the Human and LM datasets using generalized additive models for location, scale, and shape (GAMLSS)[18]. Additional information on growth chart generation included in Supplementary Appendix (see Section S9).

## Experiments Details

To specialize BERT-based models (BERT, BioBERT, ClinicalBERT, and RadBERT) for categorizing normal radiology reports, we fine-tuned them on our data using standard procedures. Further implementation details are provided in Supplementary Appendix (see Sections S6 and S8). We also used a state-of-the-art large language model (LLM) with a large context window, Gemini 1.5-Pro [19], to evaluate its classification and reasoning capabilities relevant to normal radiology report detection.

To investigate the impact of data characteristics on the performance of language models in analyzing radiology reports, we conducted six experiments (Exp) focused on: **Exp 1 & 2)** the effect of balanced versus imbalanced training data; **Exp 3)** the generalizability of models to out-of-distribution (OOD) data, considering report year, institution, and patient age; **Exp 4)** the relative contributions of impression sections and full narrative reports; **Exp 5)** the exploration of reasoning capabilities within LLMs for improved normal radiology report detection; **Exp 6)** the feasibility of generating accurate brain growth charts using the fine-tuned LMs. For each validation and test set, we assessed model performance using accuracy, specificity, sensitivity, precision, and F1-score, reporting both individual results for each random seed and the average across replicates. Each experiment was replicated five times with different random seeds to mitigate the influence of stochastic initialization. Mean and standard deviations are reported (Supplementary Appendix, see Section S10).

# Results

**Exp 1: Balanced Training:** To assess the effect of balanced training, we trained four BERT-based language models (BERT, BioBERT, ClinicalBERT, and RadBERT) on datasets with equal representation of normal and abnormal radiology reports (Fig. 2A). On balanced held-out test sets, all models exhibited high performance across all metrics (specificity, sensitivity, precision, and F1-score), with no significant differences between models. However, when evaluated on an imbalanced test set consisting solely of abnormal reports (simulating a real-world scenario with a higher proportion of abnormal cases), a substantial decrease in precision and F1-score was observed (Fig. 2B; Supplementary Fig. S3 and Table S1). This suggests that while balanced training yields excellent performance on balanced data, it may not generalize well to the class imbalance prevalent in clinical practice.

**Exp 2: Unbalanced Training with Minority Class Weighting:** To address the class imbalance inherent in real-world radiology data, we trained the four BERT-based models on a subset of the original, unbalanced dataset, employing a weighted loss function that assigned a tenfold higher penalty to misclassifications of normal reports (Fig. 2C). This strategy aimed to mitigate the potential bias towards the majority class (abnormal reports) observed in Exp 1. Remarkably, all four models demonstrated robust performance across all metrics on the unbalanced held-out test set, even with the significantly higher proportion of abnormal cases (>92%).This finding contrasts sharply with the results from Exp 1, where balanced training led to a substantial performance drop on an imbalanced test set. Notably, no significant differences in performance were observed between BERT-based models (Supplementary Fig. S4 and Table S1). The successful application of minority class weighting in this experiment highlights its effectiveness in mitigating class imbalance and improving the generalization performance of the models on real-world data distributions. We found the tenfold higher penalty empirically; further investigation is needed to determine the optimal weighting strategy for different levels of class imbalance and datasets.

**Exp 3: Out-of-Distribution Testing:** To assess temporal generalizability, models trained using the unbalanced training protocol (Exp 2) on data from before 2023 were evaluated on both held out pre-2023 data and 2023 data. This allowed us to quantify the performance degradation on unseen data representing a distinct temporal distribution. Performance on pre-2023 data served as a baseline (Supplementary Table S1). The models maintained near-perfect accuracy, sensitivity, and specificity on the out-of-distribution (OOD) 2023 data (Fig. 2D). However, we observed a decrease in precision (by 20.20% on average) and F1-score (by 12.69% on average) on the 2023 data compared to the pre-2023 data. This suggests some sensitivity to temporal shifts in data characteristics. Importantly, the unbalanced training approach (Exp 2) yielded a higher precision (21.39% increase on average) and F1-score (13.09% increase on average) on the 2023 data compared to the balanced training approach from Exp 1(Fig. 2E), indicating its improved robustness to temporal OOD generalization. We also performed out-of-distribution testing of the model trained in Exp 2 using the Geisinger dataset [12] (adult data from a different institution). All models maintained near-perfect accuracy and specificity (Fig. 2F). However, RadBERT demonstrated superior performance compared to other models, suggesting the effectiveness of its pre-training data containing adult radiology reports.

**Exp 4: Impression vs Full Report:** By convention, the Impression section of radiology reports contains a concise summary of the overall conclusions of the signing radiologist. While this section may contain sufficient information to classify reports, it is also possible that the rest of the report provides additional information that is important for report classification. To examine the relative contribution of impression and full report sections to model performance, we trained our four BERT-based models on datasets consisting of two input variations: impression-only and full report (impression plus narrative). Following the training protocol established in Exps 1-3, we observed consistent performance degradation across all four models when trained on impression-only data compared to full-report data. Specifically, all models exhibited notably lower sensitivity (by 8.76% on average), precision (by 3.23% on average), and F1-score (by 6.17% on average) on unseen, unbalanced test sets when trained on the impression section alone (Supplementary Tables S1 and S2). These performance differences were not significant between models. This indicates that while the impression section contains valuable information, the full narrative report is crucial for achieving optimal diagnostic accuracy in our task.

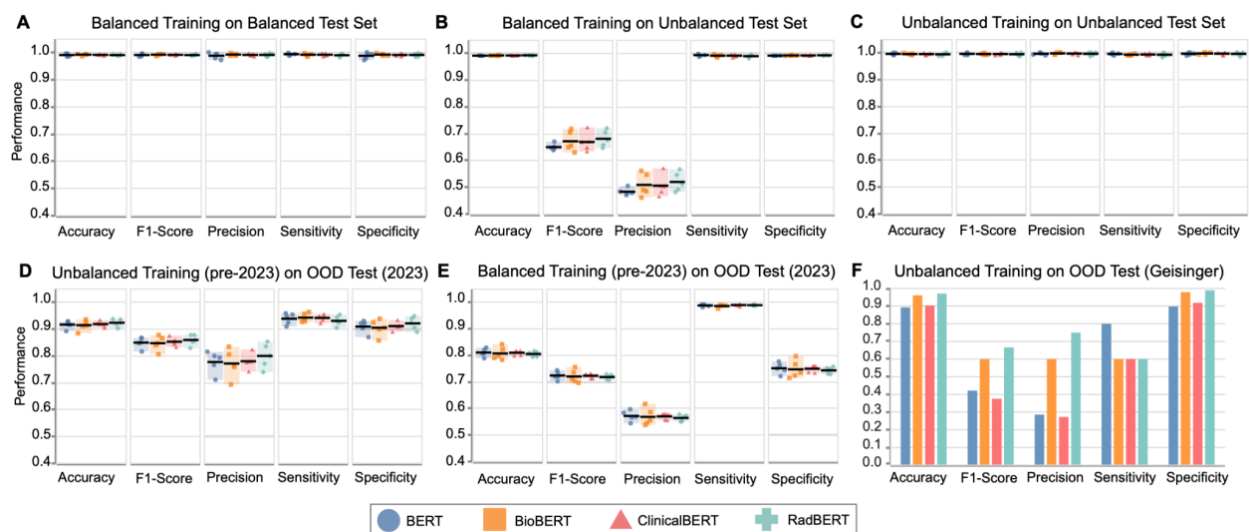

**Figure 2. Performance of Fine-tuned BERT-based Language Models on Normal Radiology Report Categorization.** (**A**) Performance on a balanced test set after training on balanced data; Exp 1. (**B**) Performance on an unbalanced test set after training on balanced data, suggesting lower generalizability to unbalanced test scenarios; Exp 1. (**C**) Performance on an unbalanced test set after training on unbalanced data; Exp 2. (**D**) Performance on an out-of-distribution (OOD) 2023 test set after unbalanced training on pre-2023 data; Exp 3. (**E**) Performance on the OOD 2023 test set after balanced training on pre-2023 data; Exp 3. (**F**) Performance on an OOD Geisinger dataset using the model trained on CHOP data in Exp 2; Exp 3. Dots represent performance over 5 random seeds, with solid black lines indicating the average. Bar plots show mean performance across seeds.

**Exp 5: Reasoning Capabilities of LLMs in Normal Report Detection:** To investigate the performance limitations of our fine-tuned BERT-based LMs and identify failure modes, we evaluated the capabilities of a leading large language model (LLM), Gemini 1.5-Pro, in classifying normal brain radiology reports. The evaluation set comprised 41 reports from our test set that were unanimously misclassified by all four BERT-based models. We employed two prompting strategies: a single direct inquiry (Q1) and a five-step, clinically-informed, sub-question approach (Q1-Q5; Fig. 3A; Supplementary Appendix, Section S10, Tables S3 and S4). Each report was assessed using both strategies. Manual re-review of the original annotations by two independent annotators– with discrepancies resolved by consensus between them– also established a revised gold standard, revealing that the BERT-based models exhibited higher accuracy (26/41 reports, 63.41% accuracy) and inter-rater reliability than the initial human annotators (15/41, 35.58% accuracy), highlighting the inherent complexity of this task. Gemini 1.5-Pro's performance was then assessed against this revised gold standard (Fig. 3B; Supplementary Appendix, Section S10, Table S5).

The direct inquiry achieved 60.97% accuracy (25/41). Error analysis indicated challenges related to imaging artifacts (e.g., orthodontic hardware) and the presence of previously noted, but clinically insignificant, pathology on repeat scans. The stepwise approach yielded a higher accuracy of 75.60% (31/41), suggesting that decomposing the task into smaller, more manageable sub-questions enhanced the LLM's ability to reason through the complexities of interpreting radiology reports and more accurately identify normal findings or imaging artifacts.

Of note, two cases demonstrated correct classification with faulty reasoning. Representative examples of the stepwise analysis for two reports are provided in Table 1 (Supplementary Table S6 contains all records).

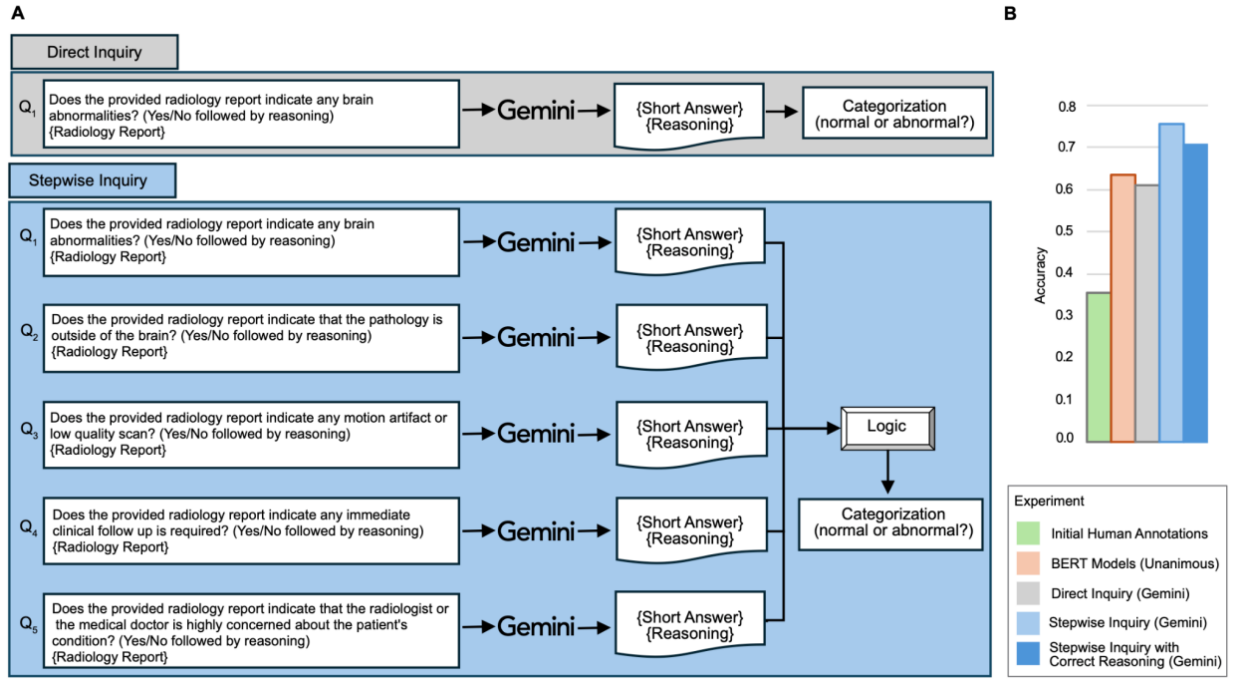

**Figure 3. Normal Radiology Report Classification Using Direct and Stepwise Inquiry**. **(A)** Flowchart illustrating direct (single question, Q1) and stepwise (five clinically informed questions, Q1-Q5) with Gemini 1.5-Pro to classify normal brain radiology reports. The stepwise inquiry categorized a report as "normal" only if the initial query about brain pathology or the follow up about non-brain pathology indicated no brain abnormalities (Q1: "no" OR Q2: "yes") and the remaining queries (Q3-Q5) are consistently negative. **(B)** Accuracy of different methods compared to the revised gold standards edge cases: initial human annotations, unanimous BERT models predictions, direct inquiry with Gemini, stepwise inquiry with Gemini, and stepwise inquiry with Gemini with correct reasoning.

**Table 1. Stepwise Reasoning Improves Gemini 1.5-Pro's Accuracy in Identifying Normal Brain Radiology Reports.** Illustrative examples show stepwise inquiry (Q1-Q5, Agg.) corrects initial misclassifications by direct inquiry (Q1) when compared to revised gold standard (Ref.) on whether a provided report is abnormal (i.e., contains significant brain pathology or is a low-quality scan). Correct categorizations are marked in bold. Report texts are updated and shortened due to space constraints (see Supplementary Table S6 for full records in their original format).

| ID | Report | Ref. | Answer Q1 Direct Inquiry | Answer Q2 | Answer Q3 | Answer Q4 | Answer Q5 | Agg. |
|---|---|---|---|---|---|---|---|---|
| 5 | Degraded study due to orthodontic hardware. Unremarkable brain MRI prior to and following intravenous contrast. No deep venous sinus thrombosis or flow-limiting intracranial venous stenosis. BRAIN MRI, WITHOUT AND WITH CONTRAST, INTRACRANIAL MR VENOGRAPHY CLINICAL INDICATION: Intractable migraines COMPARISON: None FINDINGS: Orthodontic hardware results in significant signal dropout and geometric distortion, degrading the examination and rendering the diffusion weighted images non-diagnostic. The ventricular system is normal in size, shape, and configuration. There is no intracranial mass lesion, mass effect, or midline shift. No abnormal fluid collections are identified. There are no areas of abnormal parenchymal signal intensity on these sequences. There are no areas of abnormal enhancement following contrast administration. | **Yes** | No. The report states "Unremarkable brain MRI prior to and following intravenous contrast". Although the study is degraded by orthodontic hardware, then visible portions of the brain showed No abnormalities. | No. The report specifically states "Unremarkable brain MRI prior to and following intravenous contrast" and focuses solely on findings within the brain such as the ventricular system, intracranial venous structures, and sinuses. | **Yes** The report explicitly states "Degraded study due to orthodontic hardware" and "Orthodontic hardware results in significant signal dropout and geometric distortion, degrading the examination and rendering the diffusion-weighted images Non-diagnostic." | No. The report explicitly states "Unremarkable brain MRI prior to and following intravenous contrast" and While the study is degraded due to orthodontic hardware, No acute findings requiring immediate clinical follow-up are reported. | No. The report explicitly states "Unremarkable brain MRI" and finds No signs of serious conditions like thrombosis or stenosis. While the orthodontic hardware degraded the study, the visible findings were Not concerning to the radiologist. | **Yes** |
| 38 | No definite abnormality on this screening brain MRI. SCREENING BRAIN MRI, WITHOUT CONTRAST: CLINICAL INDICATION: Headache and vomiting. History of microcephaly and failure to thrive with dysmorphic features and scoliosis, seizures, developmental delay. COMPARISON: Previous Brain MRI FINDINGS: This examination is not of the typical quality of a full diagnostic MRI and, as such, pathology may be missed. The ventricular system is normal in size, shape, and configuration. There are no intracranial mass lesions, and there is no mass effect or midline shift. No abnormal fluid collections are identified. No definite areas of abnormal parenchymal signal. No definite abnormality is seen in the visualized portions of the orbits. The middle ear cavities and mastoid air cells are clear. The paranasal sinuses are clear. | **Yes** | No. Reasoning: The report explicitly states "No definite abnormality on this screening brain MRI". | No. Reasoning: The report explicitly states it is a "screening brain MRI" and focuses solely on findings within the brain.There is No mention of imaging or findings related to any areas outside of the brain. | **Yes**. The report explicitly states, "This examination is Not of the typical quality of a full diagnostic MRI and, as such, pathology may be missed." This indicates the images were likely Not optimal, possibly due to motion artifact, the "fast screening protocol", or other factors impacting image quality. | No. Reasoning: "No definite abnormality" - This statement explicitly states that No significant issues were found on the scan. While the patient has a complex medical history and presented with headache and vomiting, the report doesn't link these symptoms to any specific finding on the MRI. Important Note: While the MRI itself doesn't necessitate immediate action, the patient's symptoms and history warrant clinical follow-up with their physician. | No. Reasoning: "No definite abnormality" is stated multiple times throughout the report, indicating a lack of strong findings suggesting a serious acute issue.. Important Note: While the report doesn't indicate immediate, high-level concern, it doesn't rule out potential underlying issues. | **Yes** |

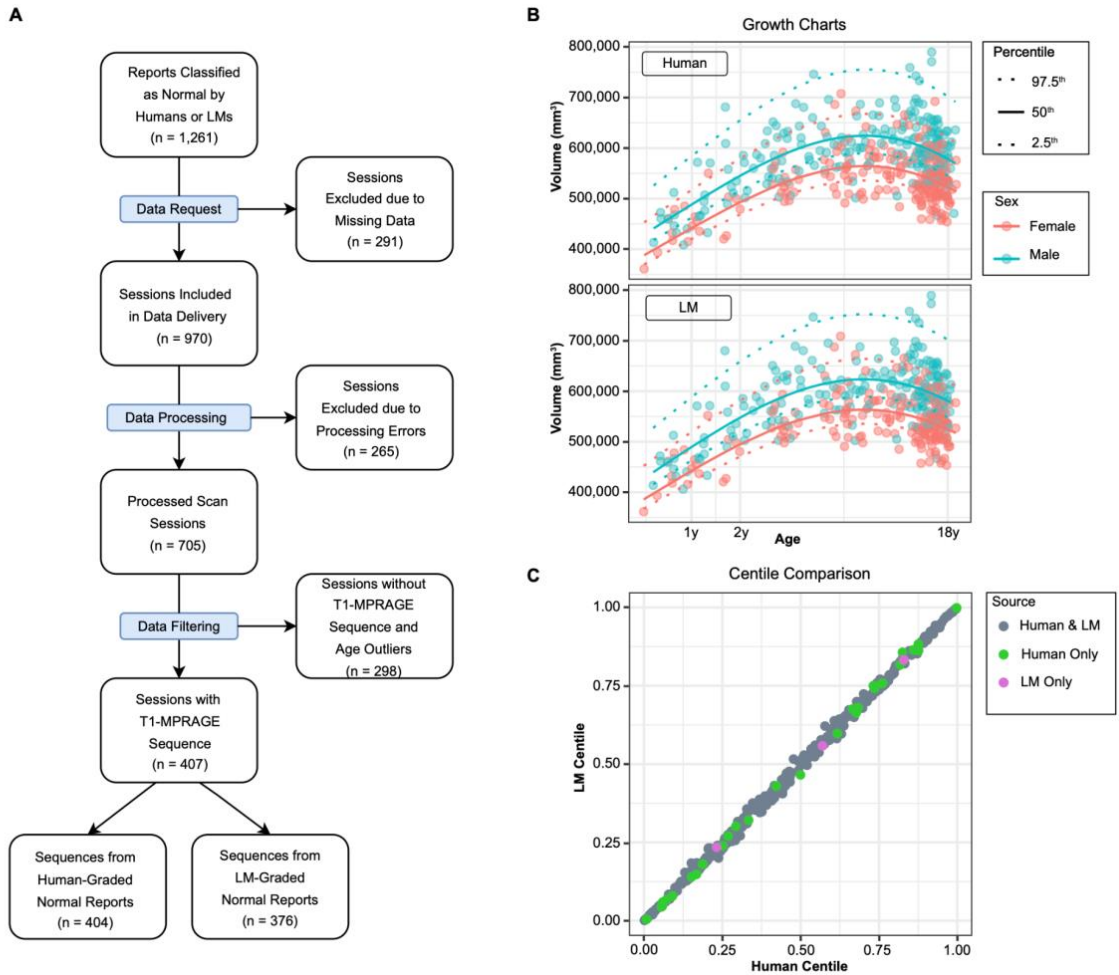

**Figure 4. Generation of Brain Growth Charts from Reports Classified by Language Models versus Human Annotators.** (**A**) The flowchart details the data curation and processing pipeline used to generate growth charts for normal reports graded by humans and language models (LMs). Data processing includes DICOM-to-NIFTI conversion, BIDS data curation, and SynthSeg+ image processing with quality control (QC) thresholding. (**B**) Growth chart models fit with generalized additive models for location, shape, and scale (GAMLSS) for cortical gray matter volume derived from T1-weighted MPRAGE sequences based on normal reports classified by human (top panel) and LMs (bottom panel). (**C**) Scatterplot shows comparison of derived centile scores of the same patient MRIs benchmarked against either human (x-axis) or LM (y-axis) growth chart models.

**Exp 6: Brain Growth Charts: Language Models vs Human Annotations:** Processing and quality control of the data received from the radiology department (Fig. 4A) yielded 705 scan sessions from 22 scanners, a 27% reduction from the initial delivery. Details are provided in Supplementary Appendix (see Section S9). Of the 705 sessions available, 407 sessions included T1-MPRAGE sequences, comprising our primary dataset for the creation of growth charts (n = 407 subjects, 207 male, ages between 135 and 7100 days, 19 scanners). 404 sessions in this dataset were classified as normal by human annotators, while 376 sessions were classified as normal by LMs.

Across all imaging phenotypes, the growth charts produced by radiology reports classified by LMs were essentially identical to those produced by radiology reports classified by human annotators (Figs. 4B and 4C; Supplementary Figs. S6-S11). Centile scores were generated for the union of the LM and Human datasets, which strongly converged between the growth charts (Cortical Gray Matter Volume, r=0.99, p<2.2e-16; Subcortical Gray Matter Volume, r=0.99, p<2.2e-16; White Matter Volume, r=0.99, p<2.2e-16). The strength of the correlations indicated that the LM-classified growth charts effectively captured the distribution of the total dataset, despite a lower sample size compared to Human-classified growth charts. Results using T1-MPRAGE sequences converged with results using the median volumes across all sequences in each scanning session (see Supplementary Figs. S6-S11).

# Discussion

This study demonstrates the potential of a novel workflow involving fine-tuned language models (LMs) to automate retrospective analysis of clinically acquired brain MRIs and their corresponding radiology reports. Our models showed high out-of-distribution (OOD) performance, generalization across different temporal periods and across hospital systems. Resulting brain growth charts were indistinguishable from those based on human annotators. Our automated approach has now been deployed to process 331,598 reports, vastly scaling the analytical capacity for a crucial task previously reliant on time-consuming and resource-intensive manual annotation.

One important application of our models is in the generation of brain growth charts, reference models that can be used to benchmark brain quantitative features of brain MRIs of individual patients [5,20]. The need for manual annotation of radiology reports to create a large corpus of "control" MRI scans creates significant bottlenecks, delaying large-scale research endeavors aimed at understanding brain development in health and disease. Our LM-based approach offers a compelling alternative, accelerating the creation of resources such as brain growth charts derived from readily available clinical data.

There are several relevant considerations for future research in this area. Further research is needed to evaluate the efficacy of a fine-tuned large scale language model (LLM) for the task of normal radiology report categorization. Our chosen ensemble of fine-tuned BERT-based models, optimized for this specific task, delivered satisfactory results for our project. This strategy prioritizes practical applicability and resource efficiency while significantly lowering computational cost compared to fine-tuning state of the arts LLMs, aligning with efforts to reduce the carbon footprint associated with computationally intensive deep learning approaches. Future directions should include exploring the integration of advanced LLMs, and investigations into optimizing the weighting strategy for imbalanced datasets beyond the tenfold higher penalty empirically identified in this study. The impact of the observed performance degradation on out-of-distribution radiology reports, and the factors contributing to misclassification by the LLM (such as misinterpretation of technical terminology) require further investigation and will contribute to more refined and robust models in future iterations. The presence of spelling and grammatical errors in the radiology reports represents a potential limitation. While LMs have demonstrated robustness to this type of noisy data, mainly through their tokenizers, the consistent presence of such errors, potentially introduced during transcription, could have influenced models' performance. Future studies could benefit from incorporating a pre-processing step to correct such errors before training or evaluation. However, we believe that these errors may represent a real-world constraint for clinicians leveraging LLMs in practice. One inherent limitation of our study is the inability to share the underlying clinical data due to stringent privacy regulations. This restriction might impact the immediate reproducibility of our results, but the detailed and transparent methodology will enable others to replicate our approach with comparable datasets. Collectively, our work represents a critical step forward to unlock the wealth of information contained in clinically acquired brain MRI scans and the accompanying unstructured radiology reports.

# Supplementary

## S1.  Author Contribution

M.D. designed and implemented the machine learning study, performed the data analysis, and wrote the initial draft of the manuscript. S.K. conducted the brain imaging experiments, detailed the experimental methods in the manuscript, and contributed to the interpretation of results. A.A.B. was involved in planning the study, contributed to the interpretation of results, and supervised the work. All authors reviewed and approved the final version of the manuscript.

## S2.  Funding

This work is supported by NIMH R01MH133843, R01MH134896, and the CHOP Research Institute.

## S3.  Data

The scan and report request were submitted to the CHOP Department of Radiology by an honest broker, a designated individual authorized to access clinical data for the purpose of de-identification, using biomedical informatics resources hosted at the hospital. The radiology reports used in this study contained some spelling and grammatical errors, commonly encountered in clinical practice due to the transcription software. These errors were present in the reports as provided to the models. See Table 1 and Table S6 for examples of the reports used with associated spelling errors.

For the Geisinger dataset, we used all available radiology reports from unique patients (with an average age of 59.82 years) labeled either "normal" or "abnormal". This was prior to data reduction based on image quality and other parameters for the study, and thus, additional patient reports were available for analysis here compared to the final published records (n=69) in the original study.

## S4.  Data Availability

The electronic radiology reports and MRI scans of CHOP and Geisinger cannot be made available publicly due to patient privacy restrictions.

## S5.  Annotation Process Details

Following our data description, "normal" reports—the minority class—were considered the positive class and the "abnormal" reports were considered the negative class. A coarse text search was used for initial labeling. Radiology reports with "brain" in the procedure description and any of the following keywords in the report's findings section were flagged as "abnormal", the negative class: *chemotherapy*, *resect*, *craniotomy*, *craniectomy*, *surgical cavity*, *post surgery*. Keyword search was case-insensitive and permitted combinations of the listed terms. Because the ultimate goal was to find normal brain MRIs for quantitative analyses that require high quality images, reports that referenced low quality scans or significant imaging artifacts were also considered abnormal.

During the annotation process, in addition to labeling reports as negative (graded as 0) or positive (graded as 2), annotators had the option to assign a grade of 1 to indicate uncertainty or ambiguity in the findings. In our analysis, we calculated the average grade for reports that underwent multiple reviews. Reports with an average grade of >1.5 were classified as positive, while those with an average grade of <0.5 were considered as negative.

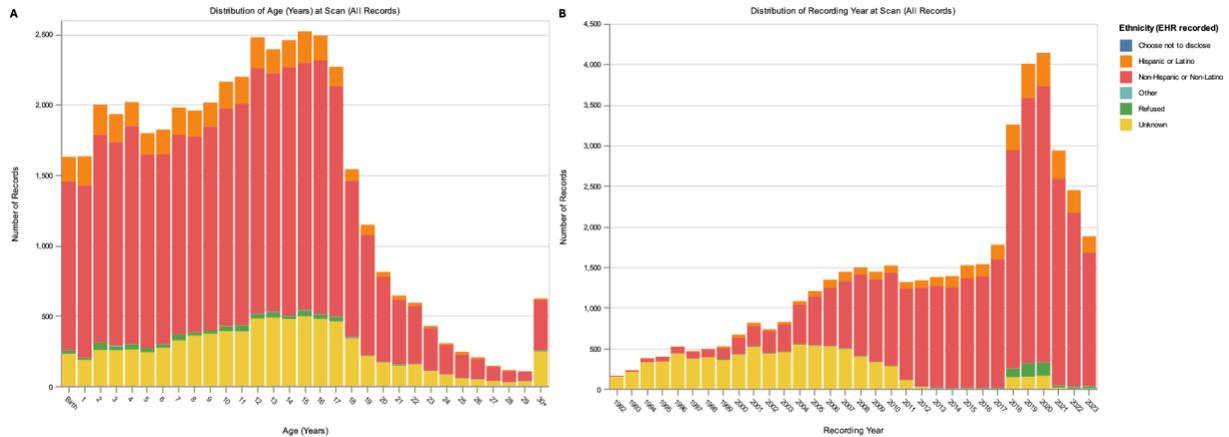

**Figure S1. Demographic and Temporal Distribution of the CHOP Study Cohort (All Graded Reports). (A)** Distribution of patient age at time of scan, for the entire cohort stratified by ethnicity as recorded in EHRs. **(B)** Distribution of the number of records by recording year for the entire cohort, stratified by ethnicity as recorded in EHRs. Note that some ethnicity categories with smaller proportions may not be clearly visible in the stacked bars, but are represented in the underlying data.

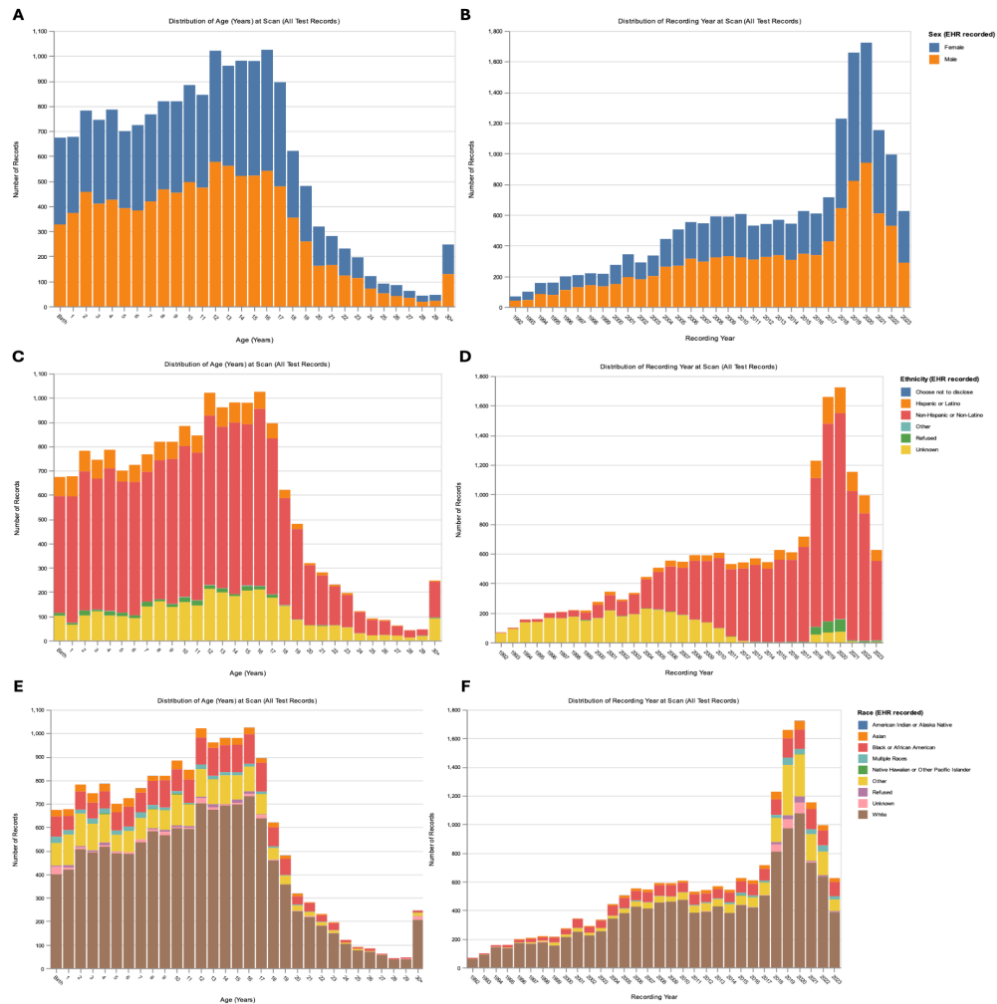

**Figure S2. Demographic and Temporal Distribution of The CHOP Test Cohort (Union Across All Random Seeds).** (**A**) Distribution of patient age at time of scan for the test cohort, stratified by sex as recorded in EHRs. (**B**) Distribution of the number of records by recording year for the test cohort, stratified by sex as recorded in EHRs. (**C**) Distribution of patient age at time of scan for the test cohort, stratified by ethnicity as recorded in EHRs. (**D**) Distribution of the number of records by recording year for the test cohort, stratified by ethnicity as recorded in EHRs. (**E**) Distribution of patient age at time of scan for the test cohort, stratified by race as recorded in EHRs. (**F**) Distribution of the number of records by recording year for the test cohort, stratified by race as recorded in EHRs. Note that some categories with smaller proportions may not be clearly visible in the stacked bars, but are represented in the underlying data.

## S6. Code availability

Upon acceptance of this manuscript, the source code developed for this project, without the corresponding data, will be made available on GitHub.

## S7. Detailed background of language models

BERT is a foundational transformer-based language model pre-trained on a massive corpus of text and code from sources such as Wikipedia. Such a broad pre-training mechanism allows BERT to learn robust representations of words and their contextual relationships, allowing for its application across a wide variety of NLP tasks, including text classification, question answering, and named entity recognition. Its general-purpose nature makes it adaptable to numerous domains after fine-tuning.

BioBERT builds upon BERT's architecture but is specifically pre-trained on a large dataset of biomedical literature, including PubMed abstracts and full-text articles from PubMed Central[8]. Such focused pre-training allowed BioBERT to achieve superior performance compared to the general BERT model on tasks such as biomedical named entity recognition, making it useful when dealing with specialized terminology common in biological and medical data.
Similar to BioBERT, ClinicalBERT is a domain-adapted version of BERT pre-trained on a corpus of clinical notes [9]. This specialization results in improved performance on NLP tasks in healthcare settings, such as clinical text classification and information extraction from electronic health records (EHRs).

RadBERT is a family of transformer-based language models specifically pre-trained on a large corpus of radiology reports [10,21]. RadBERT leverages the strengths of transfer learning by initiating pre-training from existing language models and then fine-tuning them on millions of radiology reports from the U.S. Department of Veterans Affairs (VA) health care system nation-wide. RadBERT's performance was evaluated across three diverse radiology NLP tasks including abnormal sentence classification, report coding, and report summarization. The results demonstrate that RadBERT outperforms several established biomedical and general-purpose language models, highlighting the effectiveness of domain-specific pre-training for radiology applications.

Gemini is a family of multimodal models developed by Google that demonstrate impressive capabilities across various modalities, including image, audio, video, and text understanding [22]. Gemini family models achieve state-of-the-art performance on a wide range of benchmarks and demonstrate proficiency in complex reasoning and cross-modal understanding. The pre-training data for Gemini models contained a petabyte-scale collection of web documents, books, code, images, audio, and video, leveraging multimodal and multilingual data sources. Post-training also incorporated supervised fine-tuning, training using human feedback, and reinforcement learning from human feedback to further enhance the model's capabilities, allowing Gemini to achieve superior performance across a range of complex reasoning tasks. However, Gemini models have hundreds of billions of parameters and fine-tuning them on specific domains is very compute-intensive.

## S8. Details of fine-tuning the BERT-based models

For the fine-tuning, each training on the full train set of around 43 k reports for our longest experiment, unbalanced training with minority class weighting, takes an average of 97.81 minutes for RadBERT, 54.91 minutes for ClinialBERT, 97.63 minutes for BioBERT, and 94.63 minutes for BERT using [Hugging Face libraries](#) on a single NVIDIA L4 GPU.

## S9.   Details of Clinical MRI Processing and Analysis

**Missing Data:** Although the initial scan request consisted of 1,261 reports, the delivery from the radiology department included 970 scan sessions. The 23% loss in scan sessions was attributed to missing data in the PACS server. Scan sessions that were not delivered may have come from an external site, or were older scans that were not migrated to digital storage.

**Quality Control**: The output obtained from SynthSeg+ included automated quality control scores for eight categories: general white matter, general grey matter, general cerebrospinal fluid, cerebellum, brainstem, thalamus, putamen/pallidum, and hippocampus/amygdala[17]. Sequences with a QC score less than 0.65 for any given category were excluded from further analysis[17].

**Median:** An alternative approach to generating brain growth charts utilized the median volume of imaging phenotypes[17]. Following quality control, the median volume was computed across all sequences in a scan session. This method was robust to outliers arising from sequence-specific artifacts and processing errors. Additionally, the resulting dataset included subjects without a T1-MPRAGE sequence, who were previously excluded from analysis.

**GAMLSS**: Growth Charts were generated for scans selected by Humans and the LM system using generalized additive models for location, scale, and shape (GAMLSS)[18]. Each imaging phenotype was fitted to a customized version of the generalized gamma distribution ([GGalt](GGalt)) with age and sex as fixed effects and scanner ID included as a random effect[20]. To account for the limited age range of the dataset and prevent overfitting, model selection was restricted to first and second-order fractional polynomials. Human and LM growth chart models generated using GAMLSS included centile scores and normalized volumes across all phenotypes. Normative trajectories were visualized for both subsets of scans, displaying the median, 2.5th, and 97.5th percentile curves. Additionally, each model was run on the opposite subset of scans in order to determine similarity between centile scores.

## S10.   Details of the experiments

**Exp 1) Balanced Training:** This experiment assessed the impact of balanced class representation on model performance. Training, validation, and test sets were balanced to contain equal proportions of normal and abnormal radiology reports. Four BERT-based models (BERT, BioBERT, ClinicalBERT, and RadBERT) were trained on this balanced data and evaluated on: a) a balanced unseen validation and test sets; and b) an imbalanced unseen test set comprising only abnormal reports (to assess the model's performance on the majority class when trained on balanced data). Performance was measured using specificity, sensitivity, precision, and F1-score.

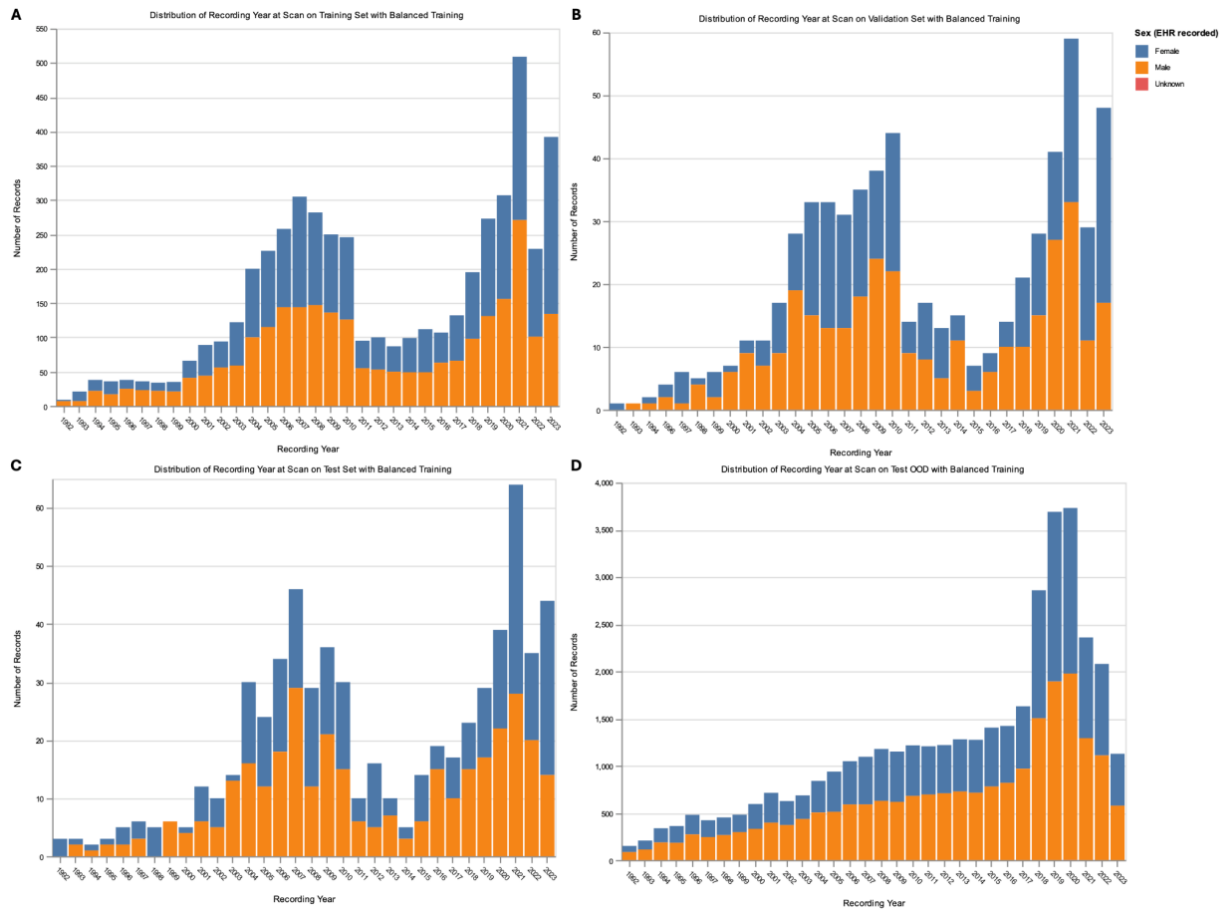

**Figure S3. Demographic and Temporal Distribution of CHOP Data Sets Used in the Balanced Training Paradigm.** (**A**) Distribution of recording year at scan, stratified by sex, for the training set in one random seed used for the balanced training paradigm. (**B**) Distribution of recording year at scan, stratified by sex, for the validation set. (**C**) Distribution of recording year at scan, stratified by sex, for the test set. (**D**) Distribution of recording year at scan, stratified by sex, for the out-of-distribution (OOD) test set, demonstrating a temporal distribution distinct from the balanced training, validation, and test sets, which potentially led to a lower performance compared to balanced validation and test sets.

**Exp 2) Unbalanced Training with Minority Class Weighting:** To simulate real-world scenarios with class imbalance, this experiment utilized the original, unbalanced distribution of normal and abnormal reports in the training, validation, and test sets; see Fig. S4. To mitigate potential bias towards the majority class (abnormal reports), we employed a weighted loss function with a tenfold higher penalty for misclassifying normal reports. The same four BERT-based models from Exp 1 were trained using this weighted loss function and evaluated on unseen, unbalanced validation and test sets. Similar performance metrics were used for this experiment.

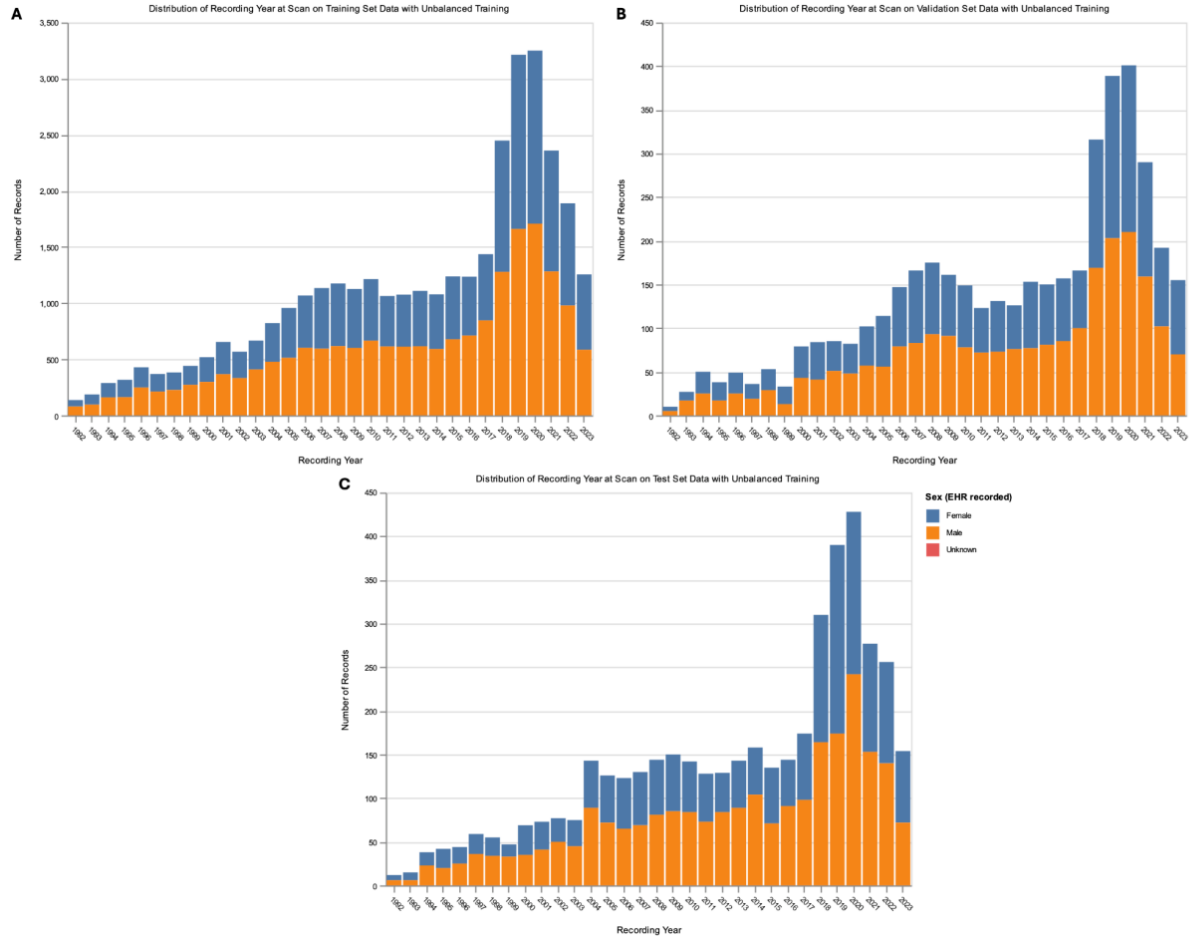

**Figure S4. Demographic and Temporal Distribution of CHOP Data Sets Used in the Unbalanced Training Paradigm**. **(A)** Distribution of recording year at scan, stratified by sex, for the training set in one random seed used for the unbalanced training paradigm. **(B)** Distribution of recording year at scan, stratified by sex, for the validation set. **(C)** Distribution of recording year at scan, stratified by sex, for the test set, demonstrating a similar distribution to the training and validation sets.

**Exp 3) Out-of-Distribution Testing: Report Year:** To evaluate temporal generalizability, we trained models (using the unbalanced training protocol from Exp 2) on data excluding reports from 2023. Model performance was then assessed separately on pre-2023 data and 2023 data. This allowed us to assess the degree of performance degradation on unseen data representing a different temporal distribution.

**Exp 3.2) Out-of-Distribution Testing: Institution & Age:** We further assessed generalizability by evaluating the performance of models (trained as in Exp 2) on reports from Geisinger Health System for adult patients—a population and institution not represented in the training data. We compared performance on the Geisinger data with performance on the original (training) dataset to assess robustness to institutional and demographic variability.

**Table S1:** Models' performance on validation, test, and test out-of-distribution (OOD) full report text (impression and narrative), using either all years' records or records excluding year 2023. Train/Val distribution indicates if the training and validation and test data were balanced over the two classes. If balanced, Test OOD reflects the remaining records which contain unbalanced sets of records. The results demonstrate low precision on test out-of-distribution (OOD) as a drawback of balanced-data training.

| Train/Val Experiment | Train/Val Distribution | Evaluation Set | Metric | BERT | BioBERT | ClinicalBERT | RadBERT |
|---|---|---|---|---|---|---|---|
| All Years | Balanced | Validation | Accuracy | 99.04 +/- 0.41 | 99.04 +/- 0.37 | 98.89 +/- 0.34 | 99.27 +/- 0.29 |
| All Years | Balanced | Test | Accuracy | 99.01 +/- 0.52 | 99.14 +/- 0.33 | 99.08 +/- 0.24 | 98.98 +/- 0.21 |
| All Years | Balanced | Test OOD | Accuracy | 99.11 +/- 0.03 | 99.19 +/- 0.14 | 99.18 +/- 0.12 | 99.22 +/- 0.11 |
| All Years | Balanced | Validation | F1 | 99.07 +/- 0.39 | 99.06 +/- 0.38 | 98.91 +/- 0.34 | 99.28 +/- 0.3 |
| All Years | Balanced | Test | F1 | 99.02 +/- 0.52 | 99.15 +/- 0.33 | 99.09 +/- 0.24 | 98.99 +/- 0.22 |
| All Years | Balanced | Test OOD | F1 | 64.97 +/- 1.22 | 67.13 +/- 3.79 | 66.87 +/- 3.42 | 68.07 +/- 3.11 |
| All Years | Balanced | Validation | Precision | 98.83 +/- 0.7 | 99.19 +/- 0.53 | 99.0 +/- 0.56 | 99.2 +/- 0.32 |
| All Years | Balanced | Test | Precision | 98.75 +/- 1.13 | 99.18 +/- 0.41 | 99.05 +/- 0.48 | 99.05 +/- 0.56 |
| All Years | Balanced | Test OOD | Precision | 48.28 +/- 1.24 | 50.84 +/- 4.04 | 50.54 +/- 3.76 | 51.95 +/- 3.39 |
| All Years | Balanced | Validation | Sensitivity | 99.31 +/- 0.36 | 98.94 +/- 0.38 | 98.81 +/- 0.75 | 99.37 +/- 0.56 |
| All Years | Balanced | Test | Sensitivity | 99.31 +/- 0.41 | 99.12 +/- 0.46 | 99.12 +/- 0.56 | 98.93 +/- 0.53 |
| All Years | Balanced | Test OOD | Sensitivity | 99.31 +/- 0.41 | 99.12 +/- 0.46 | 99.12 +/- 0.56 | 98.93 +/- 0.53 |
| All Years | Balanced | Validation | Specificity | 98.76 +/- 0.83 | 99.16 +/- 0.59 | 98.96 +/- 0.63 | 99.15 +/- 0.38 |
| All Years | Balanced | Test | Specificity | 98.72 +/- 1.2 | 99.17 +/- 0.42 | 99.04 +/- 0.49 | 99.04 +/- 0.59 |
| All Years | Balanced | Test OOD | Specificity | 99.11 +/- 0.03 | 99.19 +/- 0.14 | 99.18 +/- 0.12 | 99.23 +/- 0.11 |
| All Years | Unbalanced | Validation | Accuracy | 99.55 +/- 0.09 | 99.49 +/- 0.05 | 99.53 +/- 0.13 | 99.58 +/- 0.08 |
| All Years | Unbalanced | Test | Accuracy | 99.5 +/- 0.07 | 99.49 +/- 0.15 | 99.53 +/- 0.11 | 99.59 +/- 0.09 |
| All Years | Unbalanced | Validation | F1 | 96.8 +/- 0.69 | 96.31 +/- 0.41 | 96.62 +/- 0.93 | 96.94 +/- 0.6 |
| All Years | Unbalanced | Test | F1 | 96.39 +/- 0.51 | 96.27 +/- 1.09 | 96.58 +/- 0.88 | 97.06 +/- 0.58 |
| All Years | Unbalanced | Validation | Precision | 97.33 +/- 0.92 | 97.67 +/- 0.73 | 97.75 +/- 0.85 | 98.03 +/- 1.01 |
| All Years | Unbalanced | Test | Precision | 97.17 +/- 0.72 | 97.46 +/- 0.91 | 97.36 +/- 0.45 | 97.88 +/- 0.96 |
| All Years | Unbalanced | Validation | Sensitivity | 96.3 +/- 1.75 | 95.0 +/- 1.2 | 95.53 +/- 1.81 | 95.91 +/- 1.88 |
| All Years | Unbalanced | Test | Sensitivity | 95.64 +/- 1.26 | 95.13 +/- 1.8 | 95.84 +/- 2.09 | 96.26 +/- 0.79 |
| All Years | Unbalanced | Validation | Specificity | 99.8 +/- 0.08 | 99.83 +/- 0.06 | 99.83 +/- 0.07 | 99.85 +/- 0.08 |
| All Years | Unbalanced | Test | Specificity | 99.79 +/- 0.07 | 99.81 +/- 0.07 | 99.8 +/- 0.05 | 99.84 +/- 0.08 |
| Excluding 2023 | Balanced | Validation | Accuracy | 99.52 +/- 0.42 | 99.45 +/- 0.29 | 99.45 +/- 0.47 | 99.38 +/- 0.61 |
| Excluding 2023 | Balanced | Test | Accuracy | 99.71 +/- 0.21 | 99.38 +/- 0.38 | 99.6 +/- 0.24 | 99.67 +/- 0.24 |
| Excluding 2023 | Balanced | Test OOD | Accuracy | 80.97 +/- 1.53 | 80.62 +/- 2.42 | 80.87 +/- 0.98 | 80.41 +/- 0.99 |
| Excluding 2023 | Balanced | Validation | F1 | 99.54 +/- 0.41 | 99.47 +/- 0.28 | 99.47 +/- 0.45 | 99.4 +/- 0.6 |
| Excluding 2023 | Balanced | Test | F1 | 99.71 +/- 0.2 | 99.38 +/- 0.39 | 99.61 +/- 0.22 | 99.68 +/- 0.22 |
| Excluding 2023 | Balanced | Test OOD | F1 | 72.29 +/- 1.58 | 71.93 +/- 2.49 | 72.2 +/- 0.97 | 71.71 +/- 1.03 |
| Excluding 2023 | Balanced | Validation | Precision | 99.58 +/- 0.43 | 99.72 +/- 0.37 | 99.65 +/- 0.47 | 99.58 +/- 0.59 |
| Excluding 2023 | Balanced | Test | Precision | 99.86 +/- 0.18 | 99.57 +/- 0.15 | 99.57 +/- 0.28 | 99.64 +/- 0.23 |
| Excluding 2023 | Balanced | Test OOD | Precision | 57.05 +/- 1.89 | 56.71 +/- 3.1 | 56.87 +/- 1.21 | 56.28 +/- 1.18 |
| Excluding 2023 | Balanced | Validation | Sensitivity | 99.5 +/- 0.41 | 99.22 +/- 0.29 | 99.29 +/- 0.5 | 99.22 +/- 0.69 |
| Excluding 2023 | Balanced | Test | Sensitivity | 99.57 +/- 0.3 | 99.2 +/- 0.8 | 99.64 +/- 0.24 | 99.72 +/- 0.29 |
| Excluding 2023 | Balanced | Test OOD | Sensitivity | 98.71 +/- 0.44 | 98.51 +/- 0.55 | 98.86 +/- 0.38 | 98.81 +/- 0.11 |
| Excluding 2023 | Balanced | Validation | Specificity | 99.54 +/- 0.5 | 99.7 +/- 0.42 | 99.62 +/- 0.54 | 99.54 +/- 0.69 |
| Excluding 2023 | Balanced | Test | Specificity | 99.85 +/- 0.21 | 99.56 +/- 0.16 | 99.55 +/- 0.32 | 99.62 +/- 0.28 |
| Excluding 2023 | Balanced | Test OOD | Specificity | 75.02 +/- 2.08 | 74.62 +/- 3.39 | 74.84 +/- 1.4 | 74.24 +/- 1.31 |
| Excluding 2023 | Unbalanced | Validation | Accuracy | 99.83 +/- 0.03 | 99.77 +/- 0.09 | 99.81 +/- 0.05 | 99.82 +/- 0.06 |
| Excluding 2023 | Unbalanced | Test | Accuracy | 99.73 +/- 0.12 | 99.71 +/- 0.06 | 99.72 +/- 0.09 | 99.76 +/- 0.07 |
| Excluding 2023 | Unbalanced | Test OOD | Accuracy | 91.56 +/- 1.48 | 91.35 +/- 1.97 | 91.77 +/- 1.06 | 92.26 +/- 1.42 |
| Excluding 2023 | Unbalanced | Validation | F1 | 98.65 +/- 0.22 | 98.22 +/- 0.68 | 98.54 +/- 0.39 | 98.58 +/- 0.47 |
| Excluding 2023 | Unbalanced | Test | F1 | 97.85 +/- 0.89 | 97.64 +/- 0.43 | 97.75 +/- 0.72 | 98.05 +/- 0.62 |
| Excluding 2023 | Unbalanced | Test OOD | F1 | 84.86 +/- 2.05 | 84.64 +/- 2.75 | 85.19 +/- 1.61 | 85.84 +/- 2.03 |
| Excluding 2023 | Unbalanced | Validation | Precision | 99.04 +/- 0.55 | 99.18 +/- 0.3 | 99.12 +/- 0.66 | 98.98 +/- 0.55 |
| Excluding 2023 | Unbalanced | Test | Precision | 97.97 +/- 1.08 | 98.48 +/- 0.63 | 98.24 +/- 0.4 | 98.61 +/- 0.9 |
| Excluding 2023 | Unbalanced | Test OOD | Precision | 77.63 +/- 3.87 | 77.06 +/- 5.02 | 77.91 +/- 2.76 | 79.9 +/- 4.39 |
| Excluding 2023 | Unbalanced | Validation | Sensitivity | 98.27 +/- 0.78 | 97.28 +/- 1.41 | 97.97 +/- 0.58 | 98.18 +/- 0.58 |
| Excluding 2023 | Unbalanced | Test | Sensitivity | 97.74 +/- 0.81 | 96.81 +/- 0.83 | 97.28 +/- 1.14 | 97.5 +/- 0.6 |
| Excluding 2023 | Unbalanced | Test OOD | Sensitivity | 93.76 +/- 2.1 | 94.16 +/- 1.67 | 94.06 +/- 1.37 | 92.97 +/- 1.91 |
| Excluding 2023 | Unbalanced | Validation | Specificity | 99.93 +/- 0.04 | 99.94 +/- 0.02 | 99.94 +/- 0.05 | 99.93 +/- 0.04 |
| Excluding 2023 | Unbalanced | Test | Specificity | 99.86 +/- 0.08 | 99.9 +/- 0.05 | 99.88 +/- 0.03 | 99.91 +/- 0.06 |
| Excluding 2023 | Unbalanced | Test OOD | Specificity | 90.82 +/- 2.41 | 90.41 +/- 3.09 | 91.0 +/- 1.56 | 92.02 +/- 2.44 |

**Exp 4) Impression vs Full Report:** This experiment investigated the contribution of impression and narrative (full report) sections to model performance. We trained the four BERT-based models on two different input variations: 1) the impression section only; 2) both the impression and narrative sections (full report). Performance was compared across the two input conditions on unseen, unbalanced test sets. This allowed us to quantitatively assess the relative importance of each section for accurate classification.

**Table S2:** Models' performance on validation, test, and test out-of-distribution (OOD) impression text only, using either all years' records or records excluding year 2023. Train/Val distribution indicates if the training and validation and test data were balanced over the two classes. If balanced, Test OOD reflects the remaining records which contain unbalanced sets of records. The results demonstrate lower precision compared to Table S2 which contains impression and narrative text.

| Train/Val Experiment | Train/Val Distribution | Evaluation Set | Metric | BERT | BioBERT | ClinicalBERT | RadBERT |
|---|---|---|---|---|---|---|---|
| All Years | Balanced | Validation | Accuracy | 98.21 +/- 0.5 | 98.18 +/- 0.51 | 98.08 +/- 0.32 | 98.27 +/- 0.29 |
| All Years | Balanced | Test | Accuracy | 98.21 +/- 0.36 | 98.37 +/- 0.41 | 98.02 +/- 0.75 | 98.41 +/- 0.41 |
| All Years | Balanced | Test OOD | Accuracy | 98.01 +/- 0.22 | 98.17 +/- 0.16 | 98.22 +/- 0.18 | 98.1 +/- 0.16 |
| All Years | Balanced | Validation | F1 | 98.24 +/- 0.47 | 98.21 +/- 0.48 | 98.11 +/- 0.31 | 98.31 +/- 0.26 |
| All Years | Balanced | Test | F1 | 98.23 +/- 0.38 | 98.38 +/- 0.42 | 98.02 +/- 0.78 | 98.42 +/- 0.4 |
| All Years | Balanced | Test OOD | F1 | 49.44 +/- 3.09 | 51.54 +/- 2.38 | 52.04 +/- 2.61 | 50.55 +/- 1.99 |
| All Years | Balanced | Validation | Precision | 99.06 +/- 0.62 | 99.06 +/- 0.83 | 99.05 +/- 0.41 | 98.8 +/- 0.25 |
| All Years | Balanced | Test | Precision | 98.35 +/- 0.75 | 98.67 +/- 0.85 | 98.59 +/- 0.84 | 98.73 +/- 0.75 |
| All Years | Balanced | Test OOD | Precision | 33.09 +/- 2.57 | 34.98 +/- 2.13 | 35.54 +/- 2.41 | 34.08 +/- 1.83 |
| All Years | Balanced | Validation | Sensitivity | 97.45 +/- 0.62 | 97.38 +/- 0.15 | 97.19 +/- 0.33 | 97.82 +/- 0.4 |
| All Years | Balanced | Test | Sensitivity | 98.11 +/- 0.49 | 98.11 +/- 0.97 | 97.47 +/- 1.29 | 98.11 +/- 1.0 |
| All Years | Balanced | Test OOD | Sensitivity | 98.11 +/- 0.49 | 98.11 +/- 0.97 | 97.47 +/- 1.29 | 98.11 +/- 1.0 |
| All Years | Balanced | Validation | Specificity | 99.01 +/- 0.71 | 99.01 +/- 0.93 | 99.01 +/- 0.47 | 98.76 +/- 0.27 |
| All Years | Balanced | Test | Specificity | 98.33 +/- 0.77 | 98.66 +/- 0.9 | 98.59 +/- 0.89 | 98.72 +/- 0.82 |
| All Years | Balanced | Test OOD | Specificity | 98.01 +/- 0.23 | 98.17 +/- 0.17 | 98.22 +/- 0.2 | 98.1 +/- 0.17 |
| All Years | Unbalanced | Validation | Accuracy | 98.51 +/- 0.27 | 98.4 +/- 0.14 | 98.22 +/- 0.26 | 98.58 +/- 0.16 |
| All Years | Unbalanced | Test | Accuracy | 98.57 +/- 0.25 | 98.39 +/- 0.13 | 98.3 +/- 0.2 | 98.64 +/- 0.07 |
| All Years | Unbalanced | Validation | F1 | 90.48 +/- 1.94 | 89.63 +/- 0.97 | 88.13 +/- 1.29 | 90.86 +/- 1.15 |
| All Years | Unbalanced | Test | F1 | 91.17 +/- 1.41 | 89.88 +/- 0.66 | 89.03 +/- 1.2 | 91.52 +/- 0.39 |
| All Years | Unbalanced | Validation | Precision | 93.1 +/- 2.02 | 93.73 +/- 2.02 | 95.89 +/- 0.83 | 94.28 +/- 1.68 |
| All Years | Unbalanced | Test | Precision | 93.54 +/- 1.6 | 93.85 +/- 1.43 | 95.28 +/- 0.95 | 94.25 +/- 1.5 |
| All Years | Unbalanced | Validation | Sensitivity | 88.04 +/- 2.48 | 85.93 +/- 1.75 | 81.58 +/- 2.4 | 87.72 +/- 2.14 |
| All Years | Unbalanced | Test | Sensitivity | 88.99 +/- 2.89 | 86.25 +/- 1.43 | 83.58 +/- 2.2 | 88.99 +/- 1.85 |
| All Years | Unbalanced | Validation | Specificity | 99.43 +/- 0.18 | 99.49 +/- 0.19 | 99.69 +/- 0.06 | 99.53 +/- 0.15 |
| All Years | Unbalanced | Test | Specificity | 99.44 +/- 0.17 | 99.48 +/- 0.15 | 99.62 +/- 0.09 | 99.5 +/- 0.17 |
| Excluding 2023 | Balanced | Validation | Accuracy | 98.17 +/- 0.83 | 98.35 +/- 0.75 | 98.17 +/- 0.47 | 98.24 +/- 0.53 |
| Excluding 2023 | Balanced | Test | Accuracy | 97.77 +/- 0.68 | 98.32 +/- 0.64 | 98.1 +/- 0.74 | 98.24 +/- 0.38 |
| Excluding 2023 | Balanced | Test OOD | Accuracy | 86.13 +/- 1.32 | 86.54 +/- 1.65 | 86.79 +/- 1.34 | 86.16 +/- 1.29 |
| Excluding 2023 | Balanced | Validation | F1 | 98.22 +/- 0.81 | 98.4 +/- 0.72 | 98.21 +/- 0.45 | 98.29 +/- 0.51 |
| Excluding 2023 | Balanced | Test | F1 | 97.8 +/- 0.65 | 98.34 +/- 0.58 | 98.11 +/- 0.74 | 98.27 +/- 0.39 |
| Excluding 2023 | Balanced | Test OOD | F1 | 78.12 +/- 1.5 | 78.72 +/- 1.97 | 78.86 +/- 1.57 | 78.13 +/- 1.52 |
| Excluding 2023 | Balanced | Validation | Precision | 98.5 +/- 0.87 | 98.65 +/- 0.73 | 98.78 +/- 0.7 | 98.5 +/- 0.65 |
| Excluding 2023 | Balanced | Test | Precision | 98.11 +/- 1.24 | 98.77 +/- 0.45 | 98.69 +/- 0.82 | 98.41 +/- 0.85 |
| Excluding 2023 | Balanced | Test OOD | Precision | 64.91 +/- 2.21 | 65.56 +/- 2.83 | 66.16 +/- 2.31 | 64.96 +/- 2.12 |
| Excluding 2023 | Balanced | Validation | Sensitivity | 97.94 +/- 1.1 | 98.16 +/- 1.43 | 97.66 +/- 0.58 | 98.08 +/- 0.53 |
| Excluding 2023 | Balanced | Test | Sensitivity | 97.5 +/- 1.13 | 97.93 +/- 0.99 | 97.55 +/- 1.35 | 98.13 +/- 1.11 |
| Excluding 2023 | Balanced | Test OOD | Sensitivity | 98.17 +/- 1.03 | 98.61 +/- 0.71 | 97.67 +/- 0.83 | 98.07 +/- 0.85 |
| Excluding 2023 | Balanced | Validation | Specificity | 98.41 +/- 0.97 | 98.56 +/- 0.82 | 98.71 +/- 0.79 | 98.41 +/- 0.72 |
| Excluding 2023 | Balanced | Test | Specificity | 98.08 +/- 1.37 | 98.73 +/- 0.52 | 98.66 +/- 0.89 | 98.37 +/- 0.91 |
| Excluding 2023 | Balanced | Test OOD | Specificity | 82.08 +/- 2.0 | 82.48 +/- 2.41 | 83.13 +/- 1.99 | 82.15 +/- 1.87 |
| Excluding 2023 | Unbalanced | Validation | Accuracy | 98.82 +/- 0.15 | 98.84 +/- 0.13 | 98.75 +/- 0.2 | 98.95 +/- 0.18 |
| Excluding 2023 | Unbalanced | Test | Accuracy | 98.8 +/- 0.15 | 98.8 +/- 0.2 | 98.77 +/- 0.13 | 98.94 +/- 0.07 |
| Excluding 2023 | Unbalanced | Test OOD | Accuracy | 88.98 +/- 0.64 | 91.36 +/- 1.21 | 88.92 +/- 0.63 | 90.08 +/- 1.23 |
| Excluding 2023 | Unbalanced | Validation | F1 | 91.89 +/- 1.08 | 92.22 +/- 0.96 | 91.36 +/- 1.26 | 92.9 +/- 1.22 |
| Excluding 2023 | Unbalanced | Test | F1 | 91.75 +/- 1.22 | 91.9 +/- 1.48 | 91.48 +/- 1.23 | 92.83 +/- 0.73 |
| Excluding 2023 | Unbalanced | Test OOD | F1 | 76.66 +/- 1.79 | 83.03 +/- 3.14 | 76.1 +/- 1.47 | 79.71 +/- 3.41 |
| Excluding 2023 | Unbalanced | Validation | Precision | 95.55 +/- 0.97 | 93.8 +/- 3.25 | 95.68 +/- 1.49 | 94.9 +/- 0.62 |
| Excluding 2023 | Unbalanced | Test | Precision | 95.86 +/- 1.38 | 94.15 +/- 3.13 | 96.39 +/- 0.85 | 95.62 +/- 1.33 |
| Excluding 2023 | Unbalanced | Test OOD | Precision | 82.06 +/- 0.81 | 81.94 +/- 2.7 | 83.26 +/- 1.42 | 81.82 +/- 1.08 |
| Excluding 2023 | Unbalanced | Validation | Sensitivity | 88.54 +/- 2.22 | 90.89 +/- 3.83 | 87.42 +/- 1.11 | 91.0 +/- 2.32 |
| Excluding 2023 | Unbalanced | Test | Sensitivity | 88.01 +/- 2.5 | 89.91 +/- 3.28 | 87.07 +/- 2.19 | 90.23 +/- 1.72 |
| Excluding 2023 | Unbalanced | Test OOD | Sensitivity | 71.98 +/- 3.16 | 84.6 +/- 7.51 | 70.1 +/- 1.91 | 77.97 +/- 6.74 |
| Excluding 2023 | Unbalanced | Validation | Specificity | 99.66 +/- 0.06 | 99.5 +/- 0.29 | 99.68 +/- 0.12 | 99.6 +/- 0.06 |
| Excluding 2023 | Unbalanced | Test | Specificity | 99.68 +/- 0.13 | 99.53 +/- 0.28 | 99.73 +/- 0.07 | 99.66 +/- 0.11 |
| Excluding 2023 | Unbalanced | Test OOD | Specificity | 94.7 +/- 0.4 | 93.64 +/- 1.66 | 95.25 +/- 0.49 | 94.15 +/- 0.82 |

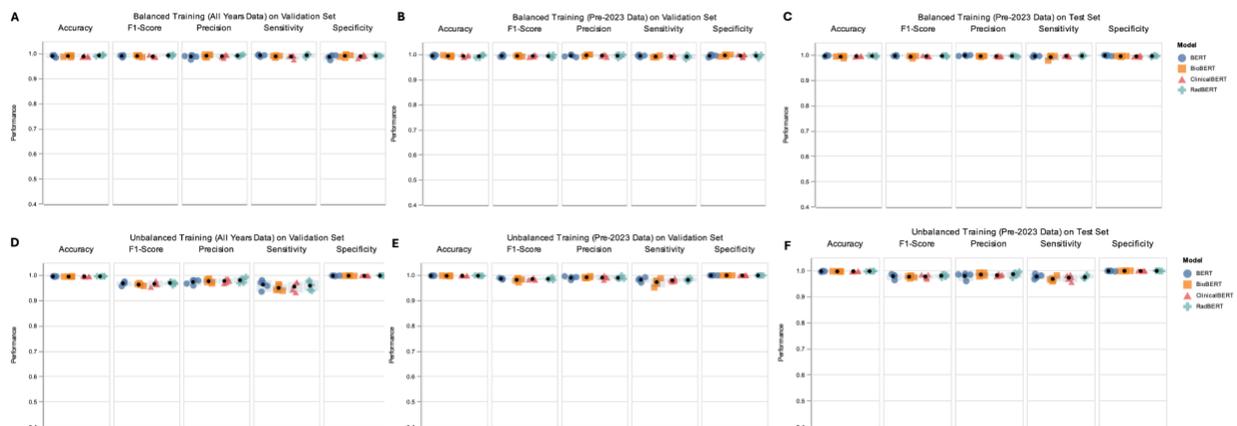

**Figure S5. Performance of Fine-tuned BERT-based Language Models on Normal Radiology Report Categorization (Remaining Experiments).** (**A**) Performance on a balanced validation set after training on balanced data. (**B**) Performance on a balanced pre-2023 validation set after training on balanced pre-2023 data. (**C**) Performance on a balanced pre-2023 test set after training on balanced pre-2023 data. (**D**) Performance on unbalanced validation set after training on unbalanced data. (**E**) Performance on pre-2023 validation set after unbalanced training on pre-2023 data. (**F**) Performance on pre-2023 test set after unbalanced training on pre-2023 data. Colored dots represent performance over 5 random seeds, with black dots indicating

the average.

**Exp 5) Reasoning Capabilities of LLMs in Normal Report Detection:** To further investigate the performance of our fine-tuned BERT-based LMs and find edge cases that resulted in their failure, we explored the performance of very large LLMs such as Gemini 1.5-Pro identifying normal brain radiology reports and their ability in providing reasoning relevant to this detection on failure cases. We developed a set of prompts (e.g., clinical inference) to examine the models' ability to identify subtle inconsistencies or indications on the presence or absence of abnormalities through multiple steps. We asked two independent reviewers to re-do the annotation on the 41 reports that were misclassified by all 4 models which revealed a 36.58% overlap with previous labels (Cohen's Kappa: -0.0370), illustrating the difficulty of the classification task for edge cases. Characteristic features of edge cases included the presence of orthodontic hardware causing imaging artifacts (N=3/41), other concerns about scan image quality (N=6/41), significant pathology present outside of the brain (N=20/41), and follow up scans where the report references previously observed pathology that might have subsequently normalized (N=25/41).

We then used the new labels as the revised gold standard and quantitatively analyzed the models' responses, focusing on the accuracy and consistency of their reasoning. For the direct prompting strategy, we used Q1 and for the indirect prompting (stepwise), we used Q1, Q2, Q3, Q4, and Q5 specified in Table S3.

**Table S3. Questions used for direct and indirect (stepwise) reasoning using Gemini 1.5-Pro's in categorizing normal brain radiology reports.**

| ID | Indication of brain abnormality | Question |
|---|---|---|
| Q1 | Yes | Does the provided radiology report indicate any brain abnormalities? (Yes/No followed by reasoning) |
| Q2 | No (Inverse Effect) | Does the provided radiology report indicate that the pathology is outside of the brain? (Yes/No followed by reasoning) |
| Q3 | Yes | Does the provided radiology report indicate any motion artifact or low quality scan? (Yes/No followed by reasoning) |
| Q4 | Yes | Does the provided radiology report indicate any immediate clinical follow up is required? (Yes/No followed by reasoning) |
| Q5 | Yes | Does the provided radiology report indicate that the radiologist or the medical doctor is highly concerned about the patient's condition? (Yes/No followed by reasoning) |

For the stepwise approach, a report was classified as "normal" only if the initial query about brain pathology or the follow up about non-brain pathology indicated no brain abnormalities (Q1: "no" OR Q2: "yes") and the remaining sub-questions returned a consistent response (Q3-Q5: "no"); see Table S4. The direct inquiry (Q1) achieved an accuracy of 60.97% (25/41) in identifying normal reports against the revised gold standard. Error analysis on the direct inquiry identified two major sources of misclassification: (a) the presence of imaging artifacts, particularly those caused by orthodontic appliances, which resulted in the model incorrectly classifying reports as abnormal; and (b) the persistence of previously documented, but clinically insignificant, pathology in follow-up scans.

**Table S4: The logic used in the stepwise classification. '-' represents an input condition that is irrelevant to the output, i.e., "don't care".**

| Q1 | Q2 | Q3 | Q4 | Q5 | Q1/Q2 Condition | Q3-Q5 Condition | Classified as "Normal"? |
|---|---|---|---|---|---|---|---|
| No | No | No | No | No | True | True | Yes |
| No | Yes | No | No | No | True | True | Yes |
| Yes | Yes | No | No | No | True | True | Yes |
| Yes | No | - | - | - | False | - | No |
| - | - | Yes | - | - | - | False | No |
| - | - | - | Yes | - | - | False | No |
| - | - | - | - | Yes | - | False | No |

The sub-question approach, reframing the task with smaller and more manageable sub-questions, yielded a higher accuracy of 75.60% (see Table S5). The improved performance with the sub-question approach might be attributed to increased focus on specific aspects of the report following the clinical inference. Interestingly, while achieving higher overall accuracy, two instances were identified where the sub-question approach yielded a correct classification but with logically inconsistent or clinically implausible reasoning, highlighting the need for further refinement in prompting techniques and model interpretability.

**Table S5. Performance results using the revised gold stand for BERT-based LMs, Gemini 1.5-Pro using direct inquiry (Q1) and sub-question approach (Q1-Q5) in categorizing normal brain radiology reports.**

| Metric | BERT-based LMs (Unanimous Voting) | Gemini (Direct Inquiry) | Gemini (Sub-question Approach) |
|---|---|---|---|
| True Positives (TP) | 3 | 6 | 7 |
| False Positives (FP) | 10 | 14 | 9 |
| True Negatives (TN) | 23 | 19 | 24 |
| False Negatives (FN) | 5 | 2 | 1 |
| Accuracy | 0.634 | 0.609 | 0.756 |
| Specificity | 0.696 | 0.950 | 0.727 |
| Sensitivity | 0.375 | 0.750 | 0.875 |
| Precision | 0.230 | 0.300 | 0.437 |
| F1 Score | 0.285 | 0.428 | 0.583 |

The breakdown of results across the individual sub-questions over all 41 reports is presented in Table S6. Further investigation is needed to explore the generalizability of these findings to a larger and more diverse dataset and other LLM architectures.

**Exp 6) Brain Growth Charts: Language Models vs Human Annotations**

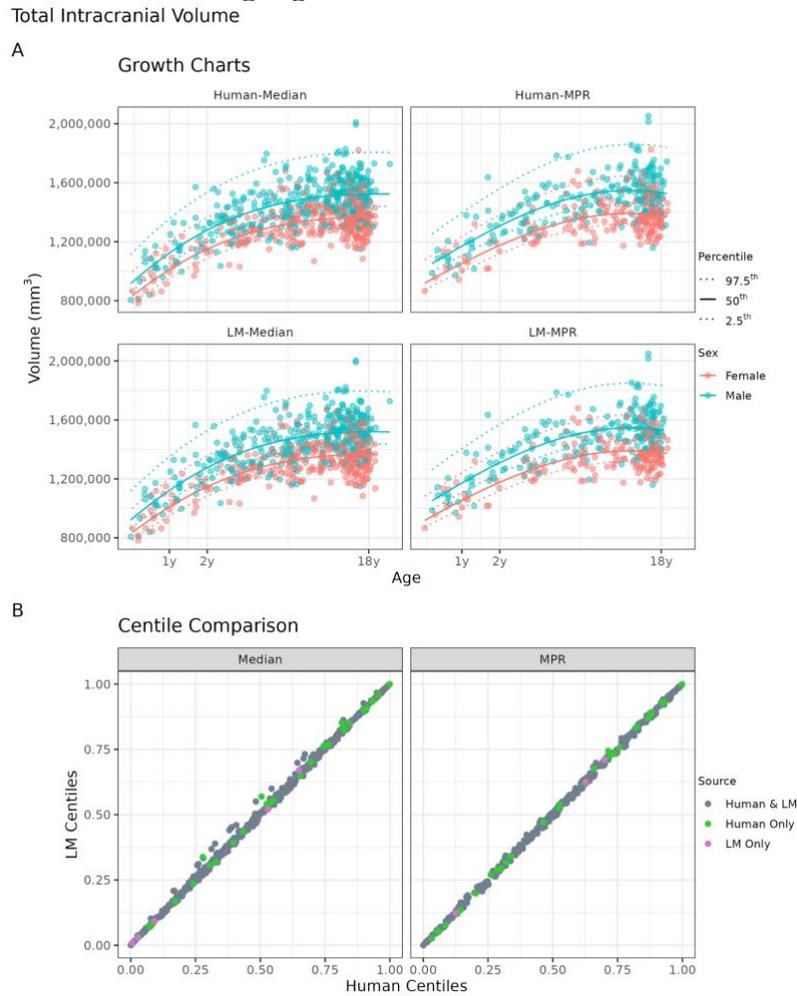

**Figure S6. Comparison of brain growth charts for total intracranial volume from reports classified by language models versus human annotators.** (**A**) Growth chart models fit with generalized additive models for location, shape, and scale (GAMLSS) for total intracranial volume derived from median across patient scan session (Median) and T1-weighted MPRAGE sequences (MPR) based on normal reports classified by human (top panel) and LMs (bottom panel). (**B**) Scatterplots show comparison of centile scores derived from median and T1-MPRAGE volumes benchmarked against human (x-axis) and LM (y-axis) growth chart models.

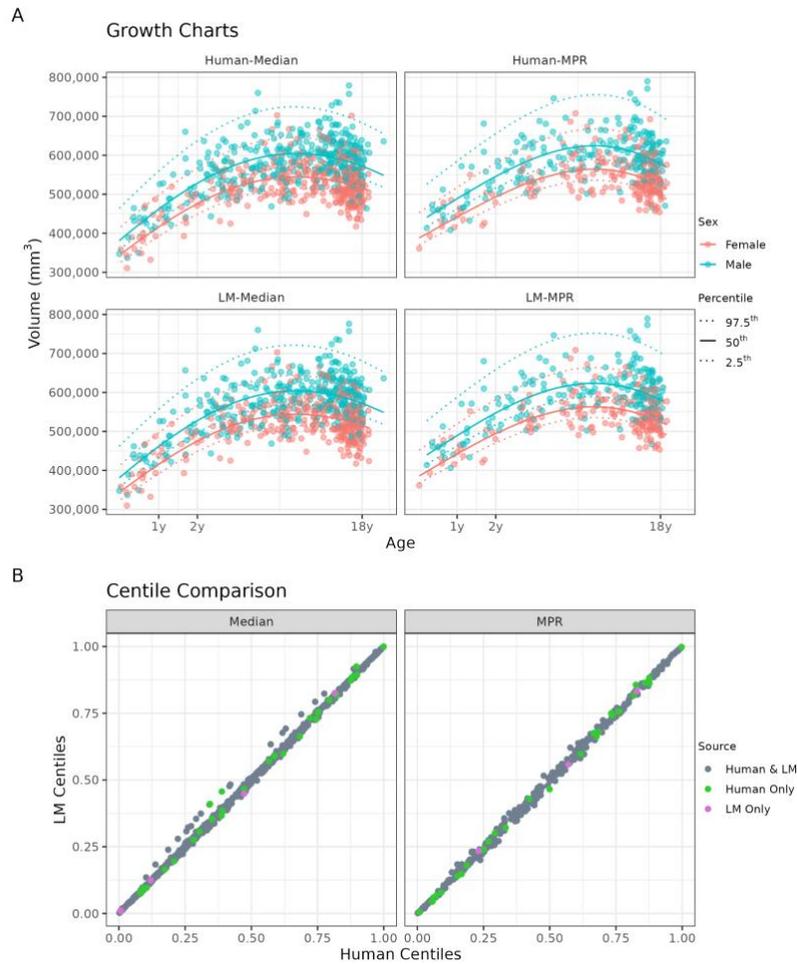

**Figure S7. Comparison of brain growth charts for cortical gray matter volume from reports classified by language models versus human annotators. (A)** Growth chart models fit with generalized additive models for location, shape, and scale (GAMLSS) for cortical gray matter volume derived from median across patient scan session (Median) and T1-weighted MPRAGE sequences (MPR) based on normal reports classified by human (top panel) and LMs (bottom panel). **(B)** Scatterplots show comparison of centile scores derived from median and T1-MPRAGE volumes benchmarked against human (x-axis) and LM (y-axis) growth chart models.

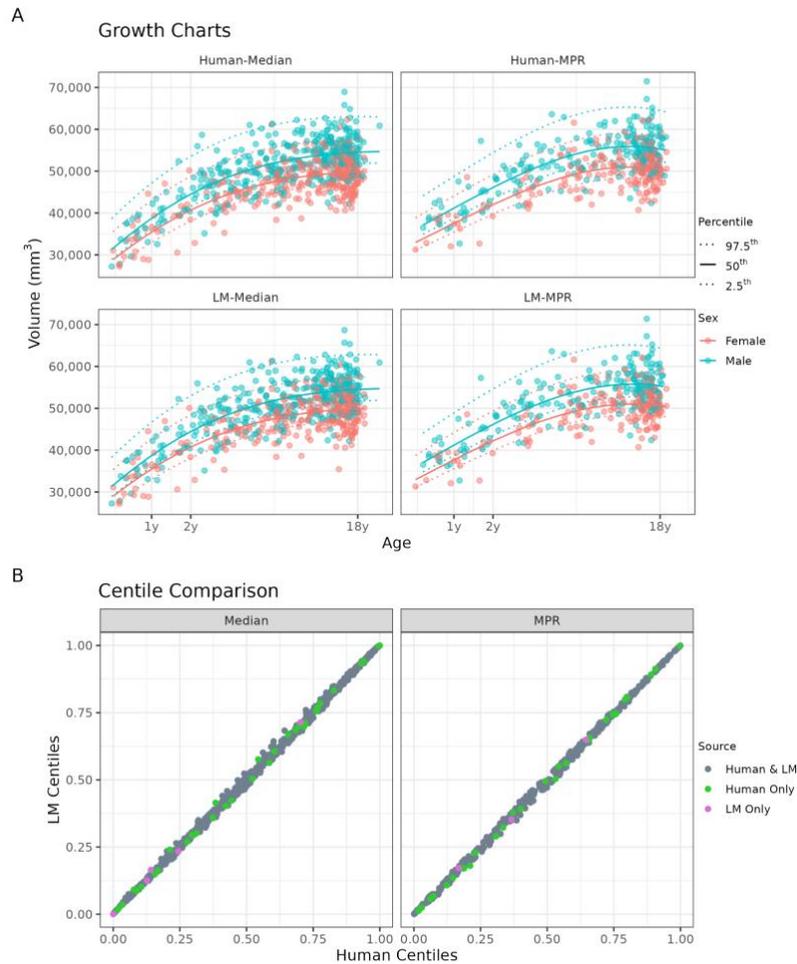

**Figure S8. Comparison of brain growth charts for subcortical gray matter volume from reports classified by language models versus human annotators. (A)** Growth chart models fit with generalized additive models for location, shape, and scale (GAMLSS) for subcortical gray matter volume derived from median across patient scan session (Median) and T1-weighted MPRAGE sequences (MPR) based on normal reports classified by human (top panel) and LMs (bottom panel). **(B)** Scatterplots show comparison of centile scores derived from median and T1-MPRAGE volumes benchmarked against human (x-axis) and LM (y-axis) growth chart models.

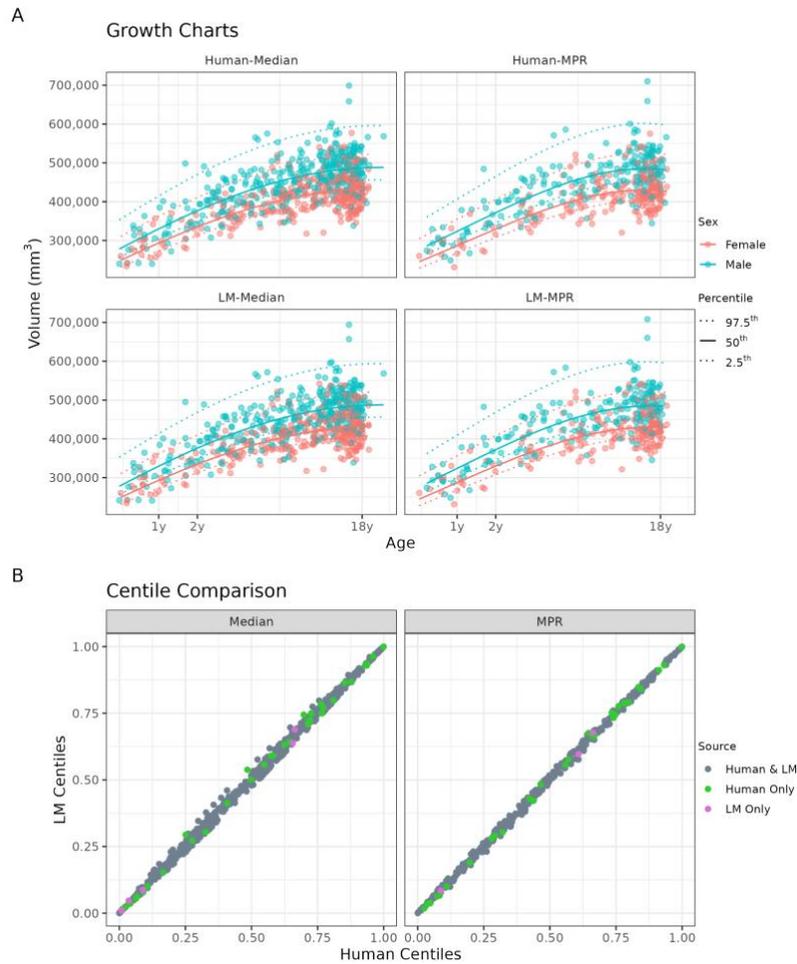

**Figure S9. Comparison of brain growth charts for white matter volume from reports classified by language models versus human annotators.** **(A)** Growth chart models fit with generalized additive models for location, shape, and scale (GAMLSS) for white matter volume derived from median across patient scan session (Median) and T1-weighted MPRAGE sequences (MPR) based on normal reports classified by human (top panel) and LMs (bottom panel). **(B)** Scatterplots show comparison of centile scores derived from median and T1-MPRAGE volumes benchmarked against human (x-axis) and LM (y-axis) growth chart models.

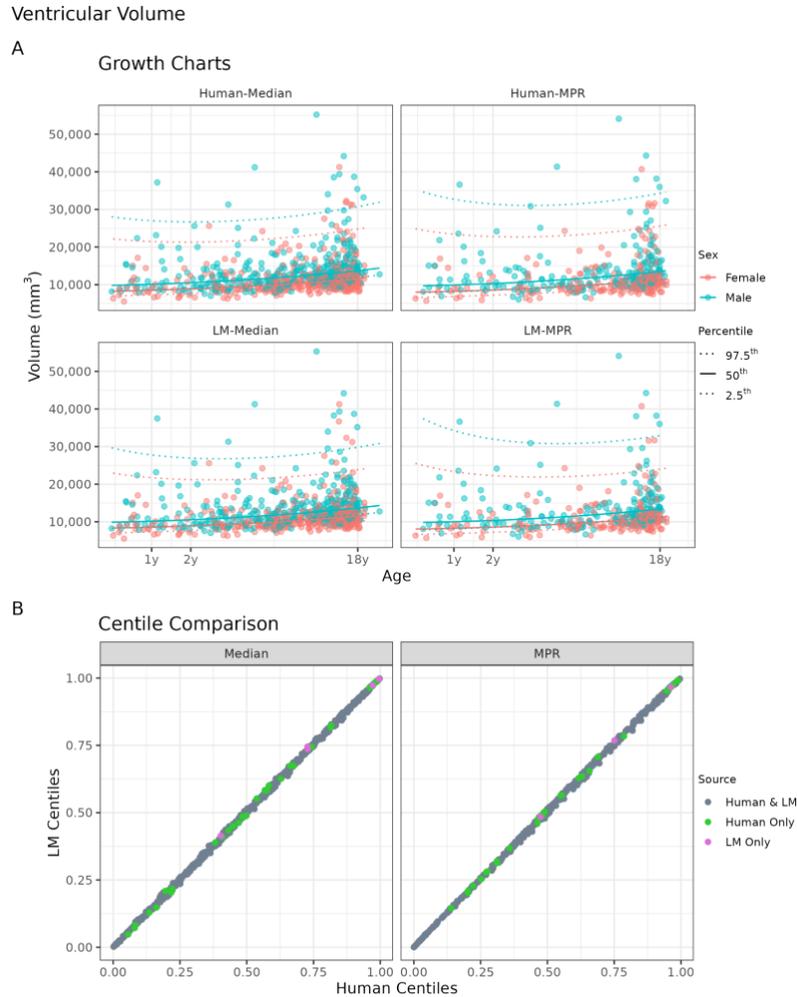

**Figure S10. Comparison of brain growth charts for ventricular volume from reports classified by language models versus human annotators.**
**(A)** Growth chart models fit with generalized additive models for location, shape, and scale (GAMLSS) for ventricular volume derived from median across patient scan session (Median) and T1-weighted MPRAGE sequences (MPR) based on normal reports classified by human (top panel) and LMs (bottom panel). **(B)** Scatterplots show comparison of centile scores derived from median and T1-MPRAGE volumes benchmarked against human (x-axis) and LM (y-axis) growth chart models.

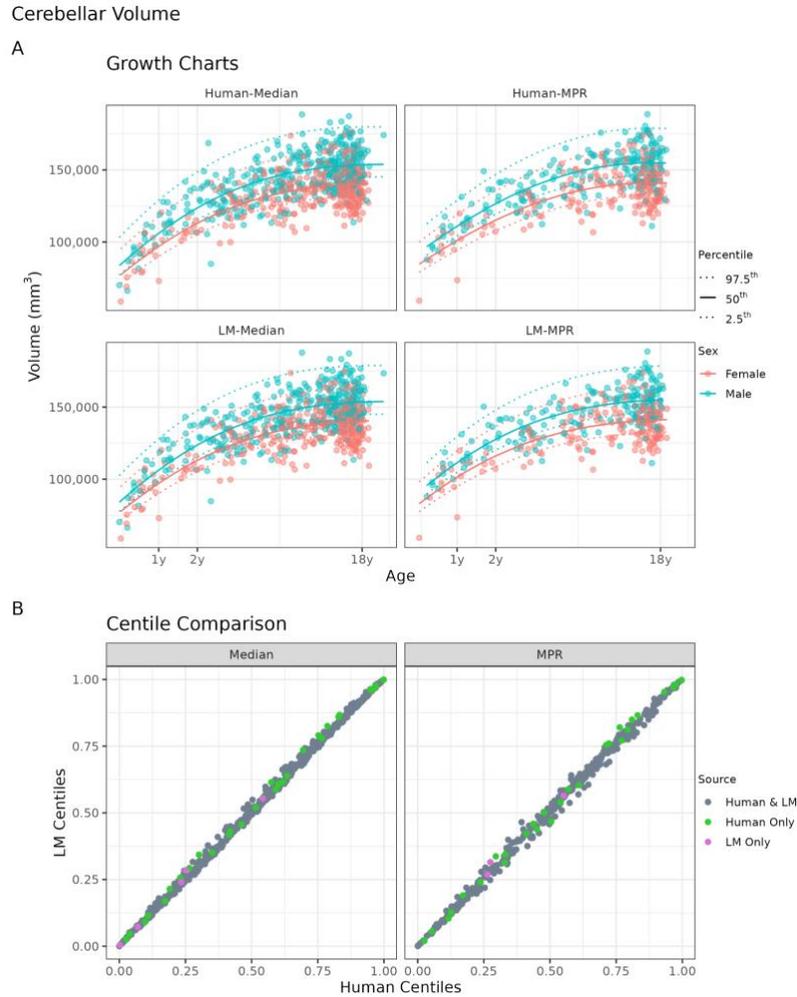

**Figure S11. Comparison of brain growth charts for cerebellar volume from reports classified by language models versus human annotators.**
**(A)** Growth chart models fit with generalized additive models for location, shape, and scale (GAMLSS) for cerebellar volume derived from median across patient scan session (Median) and T1-weighted MPRAGE sequences (MPR) based on normal reports classified by human (top panel) and LMs (bottom panel). **(B)** Scatterplots show comparison of centile scores derived from median and T1-MPRAGE volumes benchmarked against human (x-axis) and LM (y-axis) growth chart models.

**Table S6:** Gemini 1.5-Pro's responses to direct inquiry (Q1) and stepwise reasoning prompts (Q1-Q5) compared to manual annotations by human referee for 41 brain radiology reports discussed in Experiment 5. While direct inquiry (Answer Q1) yielded 60.97% accuracy, the sub-question approach (Aggregate Stepwise) resulted in a higher accuracy of 75.60%. Correct categorizations are marked in bold and incorrect reasoning are highlighted in blue (records 25 and 29). The reports are shown as they were presented to LMs and contain some grammatical and spelling errors likely due to the transcription software used in the clinic, however, patient health information (PHI) is scrubbed from the reports and replaced with dashed lines.

| # | Report | Abnormal? (human referee) | Answer Q1 Direct Inquiry | Answer Q2 | Answer Q3 | Answer Q4 | Answer Q5 | Aggregate Stepwise |
|---|---|---|---|---|---|---|---|---|
| 1 | Compared to previous examination, the imaging appearance has not changed. Stable nonenhancing area of signal abnormality involving the right internal capsule and medial globus pallidus. Nonspecific foci of signal abnormality in cerebral hemispheres, pons, and middle cerebellar peduncles. Stable plexiform neurofibroma in the right masseter muscle. Unremarkable MRI of orbits. Unremarkable intracranial MRA. END OF IMPRESSION: MRI BRAIN AND ORBITS, WITHOUT CONTRAST:MR ANGIOGRAPHY OF THE BRAIN, WITHOUT CONTRAST: CLINICAL INDICATION: Possible neurofibromatosis type 1 TECHNIQUE: BRAIN: 3D Sagittal T1 gradient echo, axial and coronal TSE T2,axial FLAIR, axial T1, axial spin echo T1, arterial spin labeled perfusion imaging, axial diffusion tensor imaging, post-contrast axial spin echo T1. ORBITS: Axial T1 gradient echo reformations, axial and coronal TSET2 with fat suppression, and post-contrast axial and coronal T1 With fat suppression. All imaging was performed on a 3.0 TeslasystemBrain MRA: MRA examination of the brain was performed utilizing a 3 Tesla system via 3D time-of-flight intracranial magnetic resonance angiography of the circle of Willis, with volume of interest 3D reconstructions using the maximum intensity projection method.MRA was deemed medically necessary based on clinicalhistory. COMPARISON: Brain MRI —-—— FINDINGS: BRAIN: Compared to previous examination, the imaging appearance has Notchanged. There is redemonstration of a T2 hyperintense region involving theright genu of the internal capsule along with the medial aspect ofthe globus pallidus. There is a rim of T1 shortening on theNoncontrast images due to magnetization transfer effect. There isNo contrast enhancement. Striated faint areas of T2 FLAIRhyperintensity are again demonstrated in pons. There is alsoredemonstration of poorly defined areas of T2 FLAIR hyperintensity in cerebellar peduncles. Again Noted are foci of T2 hyperintensity in the left anterior temporal subcortical white matter and scattered foci of T2 Hyperintensity in the subcortical frontal white matter bilaterally, which are Nonspecific. The ventricular system is Normal in size, shape, and configuration. and there is. No abnormal fluid collections are identified. There are No areas of abnormal parenchymal enhancement following contrast administration. There is asymmetric enlargement of the right masseter muscle which demonstrates heterogeneous T1 and T2 signal, without heterogeneous enhancement. Allowing for differences in positioning, there is No significant change compared to previous examination. . The previously seen opacification of left mastoid air cells is resolved. | Yes | **Yes.** Reasoning: Although the report Notes that the "imaging appearance has Not changed" since the previous examination, it goes on to describe several areas of signal abnormality: Right internal capsule and medial globus pallidus: "Stable Nonenhancing area of signal abnormality". While stable, this is still an abnormality. Cerebral hemispheres, pons, and middle cerebellar peduncles: "Nonspecific foci of signal abnormality". Again, even if Nonspecific, these are abnormalities. Right masseter muscle: "Stable plexiform neurofibroma". This is a type of tumor, indicating an abnormality. The report aims to highlight that these findings are unchanged, Not that they are Normal. | Yes. The report explicitly states "Stable plexiform neurofibroma in the right masseter muscle". The masseter muscle is a facial muscle responsible for chewing and is located outside the brain. | No. The report does Not mention any motion artifacts or issues that would suggest a low-quality scan. In fact, the report specifically states that "The major intracranial arterial structures demonstrate Normal flow-related enhancement", which indicates a clear and usable image was acquired. | No. The report explicitly states that the imaging appearance has Not changed compared to the previous examination and describes the findings as stable. There is No mention of any acute findings or the need for urgent follow-up. | No. The report repeatedly uses the word "stable" to describe the findings, indicating that the patient's condition has Not worsened since the previous examination. While the report Notes some abnormalities, the language does Not convey a sense of urgency or high concern. | No |

| # | Report | Abnormal? (human referee) | Answer Q1 Direct Inquiry | Answer Q2 | Answer Q3 | Answer Q4 | Answer Q5 | Aggregate Stepwise |
|---|---|---|---|---|---|---|---|---|
|  | Again visualized is mucosal sinus disease in thespheNoid sinus. There are scattered mildly enlarged cervical lymphNodes. The middle ear cavities and remaining paranasal sinuses are clear. ORBITS:Evaluation of the orbital contents and visual pathways reveals Nofocal mass, or signal abnormality.  The extraocular muscles, opticnerve-sheath complexes, and optic chiasm appear Normal in morphology and signal intensity.  The globes appear Normal. BRAIN MRA:The major intracranial arterial structures demonstrate Normal Flow-related enhancement. Some degree of tortuosity of intracranial arteries is demonstrated. There is No appreciable alteration in course or caliber.  No areas of aneurysmal outpouching are appreciated.  Hypoplasia of right PCOM is demonstrated. Left vertebral artery is larger than right. It should be Noted that MRA may be insensitive for small (<4 mm)aneurysms. |  |  |  |  |  |  |  |
| 2 | Orthodontic hardware results in significant signal dropout and geometric distortion, degrading the examination.Unremarkable brain MRI prior to and following intraveNous contrast.BRAIN MRI, WITHOUT AND WITH CONTRAST CLINICAL INDICATION: Severe headaches with right hand weakness and numbness.TECHNIQUE: Sagittal T1, axial and coronal TSE T2, axial and coronalFLAIR, axial T1, post-contrast T1 axial, coronal, and sagittal, and axial diffusion-weighted imaging were performed on a 1.5 Tesla system.COMPARISON: NoneFINDINGS:Orthodontic hardware results in significant signal dropout and geometric distortion, degrading the examination.The ventricular system is Normal in size, shape, and configuration., mass effect, or midline shift..  There are No areas of abnormal parenchymal signal intensity on these sequences.  Allowingfor degraded diffusion images due to the dental braces, there are No areas of restricted diffusion within the posterior portion of the brain.There are No areas of abnormal enhancement following contrastadministration.. The middle ear cavities, mastoid air cells, and the visualized portions of the paranasal sinuses are clear.IMPRESSION:Orthodontic hardware results in significant signal dropout and geometric distortion, degrading the examination.Unremarkable brain MRI prior to and following intraveNous contrast.END OF IMPRESSION:70553 MR ANGIOGRAPHY OF THE BRAIN, WITHOUT CONTRAST:TECHNIQUE: MRA examination of the brain was performed utilizing a 1.5Tesla system via 3D time-of-flight intracranial magnetic resonance angiography of the circle of Willis, with volume of interest reconstructions, using the maximum intensity projection method.COMPARISON: NoneFINDINGS:Orthodontic hardware results in significant signal dropout and geometric distortion, degrading the examination.The major intracranial arterial structures demonstrate Normal Flow-related enhancement.  There is No appreciable alteration in course or caliber.  No areas of aneurysmal outpouching areappreciated.It should be Noted that MRA may be insensitive for small (<4 mm)aneurysms.IMPRESSION:Unremarkable intracranial MRA.END OF IMPRESSION—--— | Yes | No.

Reasoning: Although the orthodontic hardware caused signal dropout and distortion, the report repeatedly states that No brain abnormalities were detected. The radiologist clearly states "Unremarkable brain MRI prior to and following intraveNous contrast" and "Unremarkable intracranial MRA." | No.

Reasoning: Although the report mentions the presence of orthodontic hardware and its impact on the imaging quality, the report focuses solely on the brain and its vasculature. There is No mention of any structures or findings outside the brain. | Yes.

Reasoning: The report explicitly states "Orthodontic hardware results in significant signal dropout and geometric distortion, degrading the examination" in multiple sections. Signal dropout and geometric distortion are classic indicators of artifacts in MRI, negatively impacting image quality and interpretation. | No.

Reasoning: Although the orthodontic hardware degraded the examination quality, the report clearly states "Unremarkable brain MRI prior to and following intraveNous contrast" and "Unremarkable intracranial MRA". This means that No concerning findings were found despite the limitations. | No.

Although the report repeatedly mentions that the orthodontic hardware degraded the examination, the overall impression of both the brain MRI and the MRA is unremarkable. This means that No significant abnormalities were found despite the technical limitations.

While the patient's symptoms of severe headaches, weakness, and numbness are concerning and prompted the need for imaging, the report itself doesn't convey a sense of heightened concern from the radiologist. | **Yes** |

| # | Report | Abnormal? (human referee) | Answer Q1 Direct Inquiry | Answer Q2 | Answer Q3 | Answer Q4 | Answer Q5 | Aggregate Stepwise |
|---|---|---|---|---|---|---|---|---|
| 3 | Interval slight decrease in size of the intraorbital right optic nerve as well as decreased enhancement, with residual bilateral optic nerve tortuosity.Otherwise stable manifestations of neurofibromatosis 1, as detailed.BRAIN AND ORBIT MRI, WITHOUT AND WITH CONTRAST:CLINICAL INDICATION: Neurofibromatosis type 1, follow up.TECHNIQUE: BRAIN: 3D Sagittal T1 gradient echo, axial and coronalTSE T2, axial FLAIR, axial T1, axial spin echo T1, axial diffusion weighted imaging, post-contrast axial spin echo T1. ORBITS: AxialT1 gradient echo reformations, axial and coronal TSE T2 with fatsuppression, and post-contrast axial and coronal T1 with fat suppression. All imaging was performed on a 3.0 Tesla system.COMPARISON: Brain and orbit MRI —— .FINDINGS:BRAIN:Once again, there are stable, scattered foci of T2 hyperintensity in the bilateral cerebellar peduncles and right greater than left pons and medulla in keeping with spongiform dysplasia of neurofibromatosis type 1. The lesions are more apparent than on the prior examinations, likely related to technique, however Nonew lesions have developed. The columns of the fornix and mammillary bodies are prominent, with resultant fullness in the region of the anterior recess of the third ventricle. There is Nomidline shift, mass effect, or acute infarction. The ventricular system is stable.. TheadeNoid tonsils are prominent.  There is fluid of the mastoid air cells and Nodular mucosal thickening of the paranasal sinuses.ORBITS:The enlargement and tortuosity of the bilateral optic nerves,predominantly in the intraorbital portions, remains stable. The previously seen enlargement of the posterior intraorbital right optic nerve is less prominent, and enhancement also has significantly decreased in the interval when compared to bothApril and —— examinations.  The orbits are otherwise unremarkable.IMPRESSION:Interval slight decrease in size of the infraorbital right opticnerve as well as decreased enhancement, with residual bilateraloptic nerve tortuosity.Otherwise stable manifestations of neurofibromatosis 1, asdetailed.END OF IMPRESSION: | Yes | **Yes**. Reasoning: The report describes several findings consistent with neurofibromatosis type 1, which is a genetic disorder that causes tumors to grow on nerves: Optic nerve abnormalities: While there's a decrease in size and enhancement of the right optic nerve, the report still Notes "residual bilateral optic nerve tortuosity." Tortuosity (twisting) of the optic nerves is an abnormality often seen in neurofibromatosis type 1. Spongiform dysplasia: The report identifies "stable, scattered foci of T2 hyperintensity" consistent with spongiform dysplasia, aNother common finding in neurofibromatosis type 1. Prominent columns of the fornix and mamillary bodies: While Not necessarily abnormal in isolation, this finding in the context of other signs suggests a structural difference that could be related to the underlying condition. Important Note: This response is for informational purposes only and should Not be considered medical advice. Please consult with a healthcare professional for | No. Reasoning: The report focuses on the brain and orbits (eye sockets). While it Notes fluid in the mastoid air cells and mucosal thickening in the paranasal sinuses, these findings are mentioned in passing and are Not the primary focus of the report. The significant pathology, the optic nerve changes, is located within the orbits, which are considered part of the skull and closely associated with the brain. | No. The report does Not mention any motion artifacts or low image quality issues. In fact, the radiologist specifically Notes that some differences observed in the current exam compared to prior exams are "likely related to technique," implying satisfactory image quality for interpretation. | No. Reasoning: The report indicates a slight decrease in size and enhancement of a previously Noted finding (enlarged right optic nerve), which is generally a positive sign. The overall impression describes the changes as "stable manifestations of neurofibromatosis 1." While regular monitoring is essential for NF1 patients, this report doesn't suggest an acute issue requiring immediate clinical follow-up. | No. The report Notes a "slight decrease in size" and "decreased enhancement" of a previously Noted abnormality, which suggests improvement rather than worsening. Additionally, the report repeatedly uses the word "stable" to describe the patient's condition, indicating No cause for immediate concern. | **Yes** |

| # | Report | Abnormal? (human referee) | Answer Q1 Direct Inquiry | Answer Q2 | Answer Q3 | Answer Q4 | Answer Q5 | Aggregate Stepwise |
|---|---|---|---|---|---|---|---|---|
| | | | diagNosis and treatment. | | | | | |
| 4 | The major metabolites are Normal for age.MR SPECTROSCOPYCLINICAL INDICATION: Normal brain objective research study.TECHNIQUE: 3D CSI technique was utilized through the level thebasal ganglia bilaterally with TR 1600ms and TE 30ms. Single voxelspectroscopy with voxel placed over the left frontal centrumsemiovale, medial parietal gray matter, and left basal gangliawere performed. MRS was performed on a X.X Tesla system.COMPARISON: MRS from —— IMPRESSION:The major metabolites are Normal for age.END OF IMPRESSION:76390 | Yes | No. Reasoning: The report explicitly states "The major metabolites are Normal for age" in both the Impression and the body of the report. This means there were No significant abnormalities detected in the brain tissue based on the spectroscopy results. | No Reasoning: The report specifically states that the spectroscopy was performed "through the level of the basal ganglia bilaterally," "over the left frontal centrum semiovale," "medial parietal gray matter," and "left basal ganglia." All of these anatomical locations are within the brain. The report does Not mention any structures outside of the brain. | No. Reasoning: The report does Not mention any motion artifact or low-quality scan. It describes the technique used for the MRI spectroscopy and states that the major metabolites are Normal for age. There is No indication of any technical difficulties or issues with the quality of the images obtained. | No. Reasoning: The report clearly states "The major metabolites are Normal for age." This indicates that there are No abnormal findings that would necessitate immediate clinical follow-up. | No Reasoning: The report explicitly states "The major metabolites are Normal for age." Additionally, the clinical indication is listed as a "Normal brain objective research study." These phrases indicate that there are No significant abnormalities detected and No cause for immediate concern based on this imaging study. | No |
| 5 | Degraded study due to orthodontic hardware.Unremarkable brain MRI prior to and following intraveNouscontrast.No deep veNous sinus thrombosis or flow-limiting intracranialveNous steNosis.BRAIN MRI, WITHOUT AND WITH CONTRAST, INTRACRANIAL MR VENOGRAPHYCLINICAL INDICATION: Intractable migrainesTECHNIQUE:  Sagittal T1, axial and coronal TSE T2, axial andcoronal FLAIR, axial T1, post-contrast T1 axial, coronal, andsagittal, and axial diffusion-weighted imaging were performed on a1.5 Tesla system.2D time-of-flight intracranial magnetic resonance veNography wasperformed in the coronal and axial planes, with volume of interestreconstructions.COMPARISON: NoneFINDINGS:Orthodontic hardware results in significant signal dropout andgeometric distortion, degrading the examination and rendering thediffusion weighted images Non-diagNostic.The ventricular system is Normal in size, shape, andconfiguration.   lesion, mass effect,or midline shift.  . onthese sequences.There are No areas of abNormal enhancement following contrastadministration.. Thevisualized portions of the middle ear cavities, mastoid air cells,and visualized paranasal sinuses are grossly clear.The major intracranial | Yes | No. The report states "Unremarkable brain MRI prior to and following intraveNous contrast". Although the study is degraded by orthodontic hardware, the visible portions of the brain showed No abnormalities. | No. The report specifically states "Unremarkable brain MRI prior to and following intraveNous contrast" and focuses solely on findings within the brain such as the ventricular system, intracranial veNous structures, and | Yes The report explicitly states "Degraded study due to orthodontic hardware" and "Orthodontic hardware results in significant signal dropout and geometric distortion, degrading the examination and rendering the diffusion-weighted images Non-diagNostic." This indicates that the orthodontic hardware caused artifacts that | No. The report explicitly states "Unremarkable brain MRI prior to and following intraveNous contrast" and "No deep veNous sinus thrombosis or flow-limiting intracranial veNous steNosis." While the study is degraded due to orthodontic hardware, No | No. The report explicitly states "Unremarkable brain MRI" and finds No signs of serious conditions like thrombosis or steNosis. While the orthodontic hardware degraded the study, the visible findings were Not concerning to the radiologist. | **Yes** |

| # | Report | Abnormal? (human referee) | Answer Q1 Direct Inquiry | Answer Q2 | Answer Q3 | Answer Q4 | Answer Q5 | Aggregate Stepwise |
|---|--------|---------------------------|--------------------------|-----------|-----------|-----------|-----------|--------------------|
|   | veNous structures demonstrate Normalflow-related enhancement. There is No MRV evidence for deepveNous sinus thrombosis or flow-limiting veNous steNosis.Incidental Note is made of a hypoplastic left transverse sinus.The internal jugular veins are Not visualized on the MRV study dueto brace artifact.IMPRESSION:Degraded study due to orthodontic hardware.Unremarkable brain MRI prior to and following intraveNouscontrast.No deep veNous sinus thrombosis or flow-limiting intracranialveNous steNosis.END OF IMPRESSION: —--— | | | sinuses. While it mentions orthodontic hardware causing imaging issues and the internal jugular veins Not being visualized, these are in the context of their impact on visualizing structures within the brain, Not as indications of pathology outside of it. | negatively impacted the quality of the MRI scan. | acute findings requiring immediate clinical follow-up are reported. The patient is being evaluated for intractable migraines, and this study rules out some concerning causes. The ordering physician will likely want to discuss the results with the patient and determine if further investigation (potentially with different imaging modalities less affected by the braces) is warranted. | | |
| 6 | 1. Low T2 signal soft tissue within the left maxillary sinus likelyreflects postoperative obliteration, however, clinical correlation isrecommended to exclude recurrence. No evidence of mass at previoussite of tumor.2. Mucosal reactive changes within the nasal cavity and adjacentsinuses as described.3. Normal brain on MRA.70553/70541tl496History: juvenile angiofibroma, status post resection, follow up.Findings: T1 sagittal, T2 axial, FLAIR axial, fat saturated doubleecho coronal, T1 coronal, contrast enhanced T1 weighted magnetizationtransfer axial and coronal images throught the brain and skull basewere performed and are compared to a prior dated ——.Post op changes and defect are identified in the left nasal cavity andnasopharynx. There are reactive mucosal changes about the surgicaldefect. Enhancing relatively low signal on T2 weighted imaging softtissue is seen within the left maxillary sinus. This may reflectsurgical obliteration of the left maxillary sinus vs tumor recurrence.It is thought more likely to be post surgical in nature as the sinusis Not expanded and there is No extension of soft tissue into thenasal cavity. However, clinical correlation is recommended. There isreactive mucosa which is somewhat coalescent within the rightmaxillary sinus. Mucosal thickening is seen about multiple ethmoidair cells. There is reactive mucosa within the left spheNoid sinus.The brain is Normal in signal intensity and morphology. The ventriclesare Normal in size. No extra-axial fluid collection is identified.The MRA is Normal without evidence for tumor blush within the nasalcavity or questionable maxillary sinus.IMPRESSION:1. Low T2 signal soft tissue within the left maxillary sinus likelyreflects postoperative obliteration, however, clinical correlation isrecommended to exclude recurrence. No evidence of mass at previoussite of tumor.2. Mucosal reactive changes within the nasal | No | **No**. Reasoning: The report explicitly states "The brain is Normal in signal intensity and morphology" and "Normal brain on MRA." While there are findings in the sinuses potentially related to the patient's history of juvenile angiofibroma, these findings are Not in the brain itself. | Yes. Reasoning: The report focuses on the left maxillary sinus, a structure located in the facial bones, Not within the brain. The report describes "Low T2 signal soft tissue within the left maxillary sinus" and raises the possibility of either postoperative changes or tumor recurrence in that area. While the report also mentions a "Normal brain on MRA," the primary | No. The report does Not mention any motion artifacts or low image quality. In fact, it states that the images were compared to a prior study, implying sufficient quality for comparison. | Yes. Reasoning: Although the report leans towards the soft tissue in the left maxillary sinus being post-surgical changes, it explicitly recommends clinical correlation to rule out tumor recurrence. This means a clinician should evaluate the findings and determine if further action is necessary. | No. The report explicitly states the findings within the left maxillary sinus are "likely" postoperative changes. While they recommend clinical correlation to rule out recurrence, the language suggests it's a precautionary measure rather than a serious concern. The report doesn't use any urgent or alarming language, and the overall tone is neutral. | Yes |

| # | Report | Abnormal? (human referee) | Answer Q1 Direct Inquiry | Answer Q2 | Answer Q3 | Answer Q4 | Answer Q5 | Aggregate Stepwise |
|---|---|---|---|---|---|---|---|---|
| | cavity and adjacentsinuses as described.3. Normal brain on MRA.70553/70541tl496END OF IMPRESSION: | | | pathology discussed is located outside the brain. | | | | |
| 7 | 1. No acute intracranial abnormality. No areas of abnormalenhancement.2. Borderline Chiari I malformation with distorted architectureof the inferior right cerebellum secondary to crowding.3. Asymmetric prominence of the lateral ventricles.4. AbNormally small bilateral optic nerves and optic chiasm.5. Other findings as described above.Requires Act 112 Notification: TrueI have personally reviewed the images and agree with theinterpretation.Act 112^TrueSECOND OPINION INTERPRETATION OF AN OUTSIDE MR: HISTORY: Developmental delays, irritability COMPARISON: MRI brain without contrast ——TECHNIQUE:MR images of the brain dated ——are submitted forour interpretation. Images submitted include: Axial DWI, sagittalT1, axial T2, axial T1 and T2 FLAIR, axial SWI, coronal T2, axialT1 FLAIR postcontrast, coronal T1 postcontrast, and 1.5 mm axialand coronal T1. Images were acquired on a 3 Tesla magnet.FINDINGS:The architecture of the inferior right cerebellar hemisphere ismildly distorted secondary to crowding from a borderline Chiari Imalformation with the right cerebellar tonsil lower than theleft, approximately 5 mm below the foramen magnum. There isassociated effacement of the CSF spaces at the foramen magnum.The lateral ventricles are slightly prominent, left greater thanright, stable from prior examination. There is mild leftperiventricular white matter volume loss and some areas. Theapparent narrowing of the posterior aspect of the body of thecorpus callosum Noted previously could just be the isthmus. Againseen is a cavum septum pellucidum et vergae, an anatomic variant.Incidentally Noted is a small hypothalamic adhesion. No clearassociated subependymal Nodular heterotopia is seen., and there is No masseffect or midline shift. No abnormal extra-axial fluidcollections are identified. A few small foci of T2/FLAIR hyperintensity are again seen withinthe subcortical and deep white matter, stable and Nonspecific.One close to the body of the right lateral ventricle raises thepossibility of prior white matter insult, although Not definitiveand the etiology remains Nonspecific.There are No areas of abnormal enhancement following contrastadministration.. Thebilateral optic nerves and optic chiasm are abnormally small. Themiddle ear cavities and mastoid air cells are clear. Theparanasal sinuses are clear. A rounded focus of T2 hyperintensityin the right aspect of the spheNoid bone is stable and likelyrelated to the development of the spheNoid sinuses. | Yes | **Yes**<br><br>The radiology report indicates several brain abnormalities:<br><br>Borderline Chiari I malformation: This indicates a structural defect where the cerebellum protrudes into the spinal canal. Asymmetric prominence of the lateral ventricles: This suggests an imbalance of cerebrospinal fluid within the brain. AbNormally small bilateral optic nerves and optic chiasm: This finding points to potential underdevelopment or atrophy of the optic nerves, which are crucial for vision. Distorted architecture of the inferior right cerebellum: This indicates structural abnormality in a part of the brain responsible for motor control and coordination. Mild left periventricular white matter volume loss: This finding could indicate damage or loss of white matter, potentially affecting communication within the brain.<br>While some findings are described as "mild" or "borderline," they still represent deviations from a Normal, healthy brain structure and could | No.<br><br>The report focuses solely on findings within the brain. While some findings, such as the small optic nerves and chiasm, could be influenced by factors outside the brain, the report itself describes these as internal structures and doesn't point to any pathology externally. | No.<br><br>The report doesn't mention any motion artifacts or low-quality issues. In fact, the "Technique" section explicitly states that the images were reviewed and found to be adequate for interpretation. The report describes findings in detail, suggesting a clear visualization of the brain structures. | Yes.<br><br>Reasoning:<br><br>Borderline Chiari I malformation: While labeled "borderline," this still represents a structural abnormality where the cerebellum protrudes into the spinal canal. This can cause a variety of neurological symptoms and requires monitoring by a medical professional. Asymmetrical ventricles and periventricular white matter loss: These findings suggest potential underlying neurological issues that require further investigation and follow-up. AbNormally small optic nerves and chiasm: This is a significant finding that could indicate a serious underlying condition affecting vision. It requires immediate evaluation by an ophthalmologist and/or neurologist. It is crucial to emphasize that this report | Yes.<br><br>Reasoning: While the report states "No acute intracranial abnormality," several findings point to significant concerns:<br><br>Borderline Chiari I malformation: This condition involves the cerebellum pushing into the spinal canal, potentially affecting brain function. Asymmetrical ventricles and periventricular white matter loss: These findings suggest potential issues with cerebrospinal fluid flow and brain tissue health. AbNormally small optic nerves and chiasm: This is a serious concern, indicating potential problems with vision and possibly pointing to underlying neurological issues.<br>Act 112 Notification (True): While the specific meaning of this Act isn't provided, the fact it's triggered suggests a need for further review and potential intervention due to the findings. The report also mentions "developmental delays" and "irritability" in the patient's history, further supporting the presence of a concerning medical condition.<br><br>In summary, though No immediate life-threatening situation ("acute") is indicated, the | **Yes** |

| # | Report | Abnormal? (human referee) | Answer Q1 Direct Inquiry | Answer Q2 | Answer Q3 | Answer Q4 | Answer Q5 | Aggregate Stepwise |
|---|---|---|---|---|---|---|---|---|
| | | | potentially contribute to the patient's developmental delays and irritability. | | | indicates potentially significant neurological findings. Prompt clinical follow-up is necessary to determine the underlying cause of these abnormalities and to plan appropriate management and treatment. | report highlights multiple abnormalities requiring further investigation and management. | |
| 8 | 1. Expected interval evolution of the previously describedinfarction in the right basal ganglion and body of the rightcaudate nucleus. No new infarction or hemorrhage.2. Complete recanalization of the veNous sinus thrombosisinvolving the right sigmoid and transverse sinuses. Requires Act 112 Notification: FalseI have personally reviewed the images and agree with theinterpretation.Act 112^FalseBRAIN MRI, WITHOUT AND WITH CONTRAST:CLINICAL INDICATION: History of veNous sinus thrombosis, onanticoagulation. TECHNIQUE: Sagittal 3D T1 gradient echo with axial and coronalreformations, axial and coronal TSE T2, axial and coronal FLAIR,axial SWI with coronal and sagittal reformations, axial spin echoT1, arterial spin labeled perfusion imaging, post-contrastsagittal 3D T1 gradient echo with axial and coronal reformations,post-contrast axial spin echo T1 with fat suppression, axialdiffusion imaging were performed on a 3.0 Tesla system.COMPARISON: Multiple prior studies, the most recent MRI brainobtained at ————.FINDINGS:The head circumference as on the patient's chart was 40 cm, whichcorresponds to the 34 th percentile for age and gender matchedcontrols.There is expected interval evolution of the previously describedinfarction in the right basal ganglion and body of the rightcaudate nucleus. Residual linear T1 and T2 hyperintensity is seenin the right putamen. There is No new area of infarction orhemorrhage.Mild asymmetric prominent size of the right lateral ventricle isNoted, likely from ex vacuo dilatation. Prominent extra-axial CSFspaces are seen in the bilateral anterior temporal pole/middlecranial fossa regions. ,and .There is No abnormal enhancement.Complete recanalization of the veNous sinus thrombosis involvingthe right sigmoid and transverse sinuses is Noted. There ishypoplastic right transverse and sigmoid sinuses.The remaining visualized major intracranial vessels appearpatent. No definite abnormality is seen in the visualizedportions of the orbits. Opacification of the bilateral mastoidair cells and middle ear cavities is Noted, left more than right.The paranasal sinuses show mild diffuse mucosal thickening. Thereis an oral airway. | Yes | **Yes**. Reasoning: While the report indicates complete recanalization of the veNous sinus thrombosis, it also describes "expected interval evolution of the previously described infarction in the right basal ganglion and body of the right caudate nucleus." This means there is evidence of a prior infarction (tissue death due to obstructed blood supply) in the brain, which is an abnormality. | No. The report focuses solely on the brain and surrounding structures. Here's why: Infarction: Described in the right basal ganglion and caudate nucleus - these are structures within the brain. VeNous Sinus Thrombosis: Involves the right sigmoid and transverse sinuses - these are blood vessels specifically within the skull and drain blood from the brain. Lateral Ventricle/Extra-axial CSF spaces: These are fluid-filled spaces within the skull and around the brain. Intracranial Vessels: Refers to blood vessels | No. The report explicitly states "I have personally reviewed the images and agree with the interpretation." This implies the reviewing radiologist found the image quality sufficient to make a diagNosis and did Not observe significant motion artifacts hindering their assessment. | No Reasoning: The report indicates positive developments. The veNous sinus thrombosis has completely recanalized, and there is No new infarction or hemorrhage. While there are expected changes from the previous infarction, these do Not necessitate immediate clinical follow-up. The report suggests stable findings and continued monitoring as opposed to urgent intervention. | No. Reasoning: Positive Developments: The report indicates "expected interval evolution" of the previous infarction, meaning it's healing as anticipated. Importantly, there's "No new infarction or hemorrhage." Additionally, the veNous sinus thrombosis has "complete recanalization," signifying a successful resolution of the clotting. Residual Findings: While some findings like hypoplastic sinuses and prominent ventricles are mentioned, they are Not described in an alarming way. The report uses terms like "mild" and "likely from," indicating these might be residual effects rather than acute concerns. Lack of Alarmist Language: The report lacks phrases like "critical," "urgent," "worsening," or recommendations for immediate action, which are typical indicators of high concern. | **Yes** |

| # | Report | Abnormal? (human referee) | Answer Q1 Direct Inquiry | Answer Q2 | Answer Q3 | Answer Q4 | Answer Q5 | Aggregate Stepwise |
|---|---|---|---|---|---|---|---|---|
| | | | | within the skull, supplying the brain. Opacification of Mastoid Air Cells/Middle Ear Cavities: While these structures are outside the brain itself, they are still within the skull and closely related to the ear and sinuses. The report consistently describes findings within the skull, indicating the pathology is Not outside the brain. | | | In summary, the report suggests a positive trajectory in the patient's condition, with the resolution of serious issues and No signs of new problems. While follow-up and monitoring are likely necessary, the language doesn't convey a sense of high concern from the radiologist. | |
| 9 | New patchy white matter signal alteration in the cerebellarhemispheres, Nonspecific but concerning for parenchymalinvolvement of Langerhans cell histiocytosis. Improved signalalteration in the right medial thalamus. Stable signal alteration in the periventricular and deep cerebral white matter.Unremarkable MRI of the pituitary.Requires Act 112 Notification: TrueAct 112^TrueAct 112^TrueMRI BRAIN AND PITUITARY, WITH AND WITHOUT CONTRAST:CLINICAL INDICATION: Multi risk-organ involvement L CDH, one yearoff therapyTECHNIQUE: BRAIN: 3D sagittal T1 gradient echo with axial reformations,axial TSE T2, axial FLAIR, axial diffusion weighted imaging,post-contrast 3D sagittal T1 gradient echo with axialreformations, post-contrast axial spin echo T1. SELLA: Coronal TSE T2 with fat suppression, post-contrastcoronal and sagittal T1 with fat suppression. All imaging was performed on a 3 Tesla system.COMPARISON: MRI brain and pituitary ——FINDINGS: BRAIN:There are patchy areas of T2/FLAIR hyperintense signal Now seenwithin the cerebellar white matter bilaterally.There are are stable patchy Nonenhancing foci of T2/FLAIRhyperintensity in the periventricular and deep cerebral whitematter, including tiny foci of cystic changes within the leftfrontal centrum semiovale. Previously seen ill-defined area ofT2/FLAIR hyperintense signal in the medial aspect of the rightthalamus is decreased in conspicuity compared to prior study. There are No areas of abnormal enhancement following gadoliniumadministration.The ventricular system is Normal in size, shape andconfiguration. . Noabnormal fluid collections are identified. The major intracranial vessels are patent. No definiteabnormality is seen in the orbits. There is trace mucosalthickening in the paranasal sinuses. The mastoid air cells andmiddle ear cavities are clear.PITUITARY:The pituitary gland is Normal in size, and | Yes | **Yes**. Reasoning: "Patchy white matter signal alteration in the cerebellar hemispheres" - This finding indicates changes in the white matter of the brain, which is abnormal. While Nonspecific, it is being considered as a potential sign of Langerhans cell histiocytosis (LCH). "Concerning for parenchymal involvement of Langerhans cell histiocytosis" - This statement explicitly highlights the radiologist's concern that the observed abnormalities could be due to LCH affecting the brain tissue (parenchyma). | No. Reasoning: The report focuses solely on structures within the brain, specifically: Cerebellar hemispheres: Located in the back of the brain, responsible for coordination and movement. Thalamus: Deep brain structure involved in sensory processing and relaying information. Periventricular and deep cerebral white matter: These terms refer to areas within the | No. The report doesn't mention any motion artifacts or low-quality imaging. Instead, it describes the findings in detail, indicating that the images were of sufficient quality for interpretation. The report focuses on the identified abnormalities and their changes compared to the previous study. | Yes. Reasoning: The report indicates "new patchy white matter signal alteration in the cerebellar hemispheres," which is concerning for parenchymal involvement of Langerhans cell histiocytosis (LCH). While the report Notes improvement in the right medial thalamus, the new finding in the cerebellum necessitates immediate clinical follow-up to evaluate the progression of LCH and determine | Yes. Reasoning: While the report Notes some improvements ("Improved signal alteration in the right medial thalamus" and "Stable signal alteration in the periventricular and deep cerebral white matter"), the presence of "new patchy white matter signal alteration in the cerebellar hemispheres" raises concern. This concern is explicitly stated as "Nonspecific but concerning for parenchymal involvement of Langerhans cell histiocytosis." Additionally, the need for "Act 112 Notification" strongly suggests a significant finding requiring further attention | **Yes** |

| # | Report | Abnormal? (human referee) | Answer Q1 Direct Inquiry | Answer Q2 | Answer Q3 | Answer Q4 | Answer Q5 | Aggregate Stepwise |
|---|---|---|---|---|---|---|---|---|
| | homogeneous in signaland enhancement pattern. The . Thepituitary stalk is unremarkable. No discrete lesion isappreciated. | | "Stable signal alteration in the periventricular and deep cerebral white matter" - This indicates ongoing abnormalities in the white matter from a previous scan, suggesting a chronic condition. While some findings show improvement (e.g., decreased signal alteration in the right thalamus), the presence of new areas of concern in the cerebellum and persistent white matter abnormalities warrant a "Yes" response. | cerebrum (the largest part of the brain) containing nerve fibers. Pituitary gland: Although located at the base of the brain, it is considered a separate endocrine gland. The report does Not mention any structures outside the brain. | | appropriate management. | and potentially indicating a serious condition. | |
| 10 | 1. No significant interval change. Postoperative and treatmentrelated changes of the left orbit. 2. Redemonstration of enhancement in the left orbit and leftoptic nerve sheath which is unchanged in appearance and may berelated to postoperative and treatment-related changes. Nodefinite evidence of masslike lesions. Continued follow-up isrecommended.Requires Act 112 Notification: FalseI have personally reviewed the images and agree with theinterpretation.Act 112^FalseORBITAL MRI, WITHOUT AND WITH CONTRAST:CLINICAL INDICATION: History of relapsed RMS, Now off therapy.TECHNIQUE:  Sagittal T1 gradient echo, axial and coronal TSE T2with fat suppression, axial DWI/ADC, and post-contrast axial andcoronal T1 with fat suppression. All imaging was performed on a1.5 Tesla system.COMPARISON: MRI of the orbits from —— (3 T-System)FINDINGS:Prior enucleation of the left globe with left ocular implant andprosthesis in stable position. There is ill-defined T1hypointense/T2 hyperintense signal with corresponding enhancementwithin the superficial aspect of the left orbital cavitysurrounding the anterior aspect of the prosthesis, mostproNounced superiorly. There is also ill-defined leftretro-orbital reticulation with associated enhancement. There isNo corresponding restricted diffusion demonstrated in theaforementioned regions Nor there is evidence of obvious masslesions. Mild asymmetric atrophy of the left extraocular muscles.There is asymmetric atrophy of the intraorbital andintracanalicular segments of the left optic nerve, whichdemonstrates mildly T2 hyperintense signal limits intraorbital,intracanalicular, prechiasmatic and lesser degree left aspect ofthe chiasm. Stable minimal enhancement along the left optic nervesheath.Right orbit appear unremarkable.No definite abnormality is seen in the visualized portion of thebrain parenchyma. No abnormal intracranial enhancement. Theventricular system is stable in caliber. Mucosal thickening ofseveral ethmoid air cells, left greater than right, and leftmaxillary sinus. No air-fluid levels within the paranasalsinuses. | Yes | No. The report explicitly states "No definite abnormality is seen in the visualized portion of the brain parenchyma. No abnormal intracranial enhancement." This indicates that within the area of the brain visible in the scan, No abnormalities were detected. | No. The report explicitly states "No definite abnormality is seen in the visualized portion of the brain parenchyma". This indicates that the observed pathology, which includes postoperative and treatment-related changes like enhancement and ill-defined signal intensities, is confined to the left orbit (eye socket) and optic nerve, structures outside the brain itself. | No. The report does Not mention any motion artifacts or low-quality imaging. It describes the findings clearly and states that the images were reviewed by a radiologist. The report even specifically mentions that there is "No corresponding restricted diffusion," which is an indicator sometimes used to assess for the presence of motion artifacts. | No. While the report does recommend continued follow-up, it explicitly states there are "No significant interval changes" and "No definite evidence of masslike lesions." The observed enhancement is considered likely related to previous treatment and surgery. The language suggests routine monitoring rather than an immediate need for intervention. | No. The report explicitly states "No definite evidence of masslike lesions." While there are postoperative and treatment-related changes, the radiologist Notes that the enhancement seen is "unchanged in appearance" compared to the previous scan, indicating stability rather than a cause for immediate concern. The recommendation for "continued follow-up" suggests monitoring the situation for any future changes, but does Not imply a high level of concern at this time. | No |

| # | Report | Abnormal? (human referee) | Answer Q1 Direct Inquiry | Answer Q2 | Answer Q3 | Answer Q4 | Answer Q5 | Aggregate Stepwise |
|---|---|---|---|---|---|---|---|---|
|  | The middle ear cavities and mastoids are clear. Left posterior plagiocephaly. |  |  |  |  |  |  |  |
| 11 | Stable size, morphology and enhancement pattern of left occipital cortical/subcortical lesion, most likely low-grade glioma. No new lesions. Continuous surveillance is requested. Requires Act 112 Notification: FalseAct 112^FalseBRAIN MRI, WITHOUT AND WITH CONTRAST:CLINICAL INDICATION: Low-grade occipital glioma, surveillanceTECHNIQUE: Sagittal 3D T1 gradient echo with axial reformations,axial and coronal TSE T2, axial and coronal FLAIR, axial spinecho T1, arterial spin labeled perfusion imaging, post-contrastsagittal 3D T1 gradient echo with axial and coronal reformations,post-contrast axial spin echo T1 with fat suppression, axialdiffusion imaging were performed on a 3.0 Tesla system.COMPARISON: Numerous prior examinations, including most recentavailable MRI dated ——FINDINGS:Stable size (7.5 mm) and morphology of the FLAIR hyperintenselesion in the cortical subcortical region of the left occipitallobe. The lesion is stable since most recent examination in ——,however slightly progressed in size by 1 mm in every dimensionsince examination in ——. The enhancement portion of the lesiondemonstrates stable measurements, taken at similar locations of9.0 x 6.9 mm .No new lesions have developed. Stable punctate foci of FLAIRhyperintensity in the cerebral white matter are unchanged,Nonspecific.The ventricular system is Normal in size, shape, andconfiguration.  There are No additional intracranial mass lesionsand .  No abnormal fluidcollections are identified.  There are No new areas of abnormalparenchymal signal intensity on these sequences.  There are No new areas of abnormal enhancement following contrastadministration.. Nodefinite abnormality is seen in the visualized portions of theorbits. T., apart from trivial mucosalthickening in the bottom of the maxillary sinuses. Mild nasalseptum deviation towards the right. | Yes | **Yes.**<br><br>Reasoning: The report indicates a "stable size, morphology and enhancement pattern of left occipital cortical/subcortical lesion, most likely low-grade glioma." A glioma is a type of tumor that originates in the glial cells of the brain. While characterized as "low-grade" and stable, it still represents an abnormality. | No.<br><br>The report specifically states "Stable size, morphology and enhancement pattern of left occipital cortical/subcortical lesion, most likely low-grade glioma."<br><br>The occipital lobe is a part of the brain, and both the cortex and subcortical regions are located within the brain. Therefore, the pathology described is located within the brain, Not outside of it. | No.<br><br>The report doesn't mention any motion artifacts or low-quality issues. Instead, it describes the lesion's features clearly, compares them to prior scans, and specifies the techniques used, indicating a scan of sufficient quality for interpretation. | No.<br><br>The report explicitly states that the findings are stable and No new lesions are identified. Continuous surveillance is requested, implying routine follow-up, Not immediate action. | No.<br><br>The report indicates "stable size, morphology, and enhancement pattern" of the existing lesion. While there is a slight progression in size since ——, the report emphasizes stability since the most recent examination in ——. The language suggests routine monitoring ("continuous surveillance is requested") rather than urgent concern. | **Yes** |
| 12 | No structural abnormality identified as the source of thepatient's seizure...Requires Act 112 Notification: FalseI have personally reviewed the images and agree with theinterpretation.Act 112^FalseMRI BRAIN, WITHOUT AND WITH CONTRAST:CLINICAL INDICATION: Right sided sharp waves in rolandic region.History of scalp tumor removed. Pathology report from ——:Peripheral nerve sheath tumor in keeping with Intraneuralperineurioma. TECHNIQUE: Sagittal 3D T1 gradient echo with axial reformations,axial and angled coronal TSE T2, axial and angled coronal FLAIR,arterial spin labeled perfusion imaging, axial susceptibilityweighted imaging, post-contrast sagittal 3D T1 gradient echo withaxial and coronal reformations, post-contrast axial spin echo T1with fat suppression, and axial diffusion imaging were performedon a 3.0 Tesla system.COMPARISON: NoneFINDINGS:  Scattered Nonspecific subcortical subcm white matter foci of T2prolongation are seen in the right frontal lobe and left parietallobe,  or restricted diffusion. These are of uncertain etiology but may reflect minimal gliosisfrom remote insult of some kind, of doubtful current clinicalsignificance. ,asymmetric being the right more prominent than the left.  Noabnormal fluid collection is identified. There is No midlineshift, mass effect or acute infarction. The hippocampi | Yes | No.<br><br>The report explicitly states "No structural abnormality identified as the source of the patient's seizure."<br><br>While the report does mention "scattered Nonspecific subcortical subcm white matter foci," these are deemed to be "of uncertain etiology" and "of doubtful current clinical significance."<br><br>The report also Notes a hypointense signal in | No.<br><br>Reason: Although the report mentions a "prior scalp lesion excision" and a "focus of T1/T2 hypointense signal" in the scalp, it explicitly states "No evidence of recurrent scalp lesion." The report focuses on analyzing the brain for abnormalities | No.<br><br>The report does Not mention any motion artifacts or low image quality issues. It describes the images as adequate for interpretation and provides detailed findings, suggesting the scan quality was sufficient. | No.<br><br>Reasoning:<br><br>The report states "No structural abnormality identified as the source of the patient's seizure." While the report Notes some miNor findings (Nonspecific white matter foci, scalp hypointensity), these are described as:<br><br>"of doubtful current | No.<br><br>The report explicitly states "No structural abnormality identified as the source of the patient's seizure." While the report does mention some miNor findings (Nonspecific white matter foci, T1/T2 hypointense signal), these are described as "of uncertain etiology," "of doubtful current clinical significance," and potentially related to a previously excised scalp lesion. The overall tone and specific language | No |

| # | Report | Abnormal? (human referee) | Answer Q1 Direct Inquiry | Answer Q2 | Answer Q3 | Answer Q4 | Answer Q5 | Aggregate Stepwise |
|---|---|---|---|---|---|---|---|---|
|  | aresymmetric and Normal in size, signal, and internal architecture.. The cerebellar tonsilsterminate above the level of the foramen magnum. . Nodefinite abnormality is seen in the visualized portions of theorbits. The middle ear cavities, mastoid air cells, and paranasalsinuses are clear.There is a focus of T1/T2 hypointense signal, without associatedwell defined lesion or Nodular enhancement, located in the scalpsoft tissue of the high left convexity at the vertex, which mayrepresent the region of reported prior scalp lesion excision. Noevidence of recurrent scalp lesion. .. |  | the scalp, but attributes this to the "region of reported prior scalp lesion excision" and states there is "No evidence of recurrent scalp lesion." Therefore, while the report Notes some miNor findings, it concludes there are No significant brain abnormalities. | ("No structural abnormality identified as the source of the patient's seizure") and doesn't find evidence of the pathology extending beyond the scalp. |  | clinical significance" "may represent the region of reported prior scalp lesion excision" "No evidence of recurrent scalp lesion." The report does Not recommend any follow-up or express concern about any of the findings. Important Note: This is an AI interpretation of a radiology report, Not a medical diagNosis. Clinical decisions should always be made by a qualified healthcare professional in consultation with the patient and with access to their full medical history. | used suggest the findings are Not alarming. |  |
| 13 | Findings consistent with bilateral subependymal focal heterotopicgray matter, the larger involving the superior aspect of theright lateral ventricle posterior body, as detailed.Subtle T2 prolongation with questionable blurring of thecortical/subcortical junction in the right superior frontalgyrus. This may reflect signal alteration related to slightlyprominent perivascular spaces, however, prior subcortical whitematter insult is also a possibility or even subtle focal corticaldysplasia is Not entirely excluded. Requires Act 112 Notification: TrueI have personally reviewed the images and agree with theinterpretation.Act 112^TrueMRI BRAIN, WITHOUT CONTRAST:CLINICAL INDICATION: Focal slowing present onelectroencephalogram (EEG) - Rule out structural lesion. TECHNIQUE: Sagittal 3D T1 gradient echo with axial reformations,axial and angled coronal TSE T2, axial and angled coronal FLAIR,arterial spin labeled perfusion imaging, axial susceptibilityweighted imaging, and axial diffusion imaging were performed on a3.0 Tesla system.COMPARISON:  NoneFINDINGS: There is a lobulated Nodular lesion following the signalcharacteristics of cortex involving the superior surface of theright lateral ventricle posterior body, near the tail of theright caudate (series 6 image 30 and series 5 image 73), mostconsistent with a heterotopic gray matter. ANother focal lesionwith similar signal characteristics is see along the | Yes | **Yes**. Reasoning: The report identifies multiple abnormalities: Bilateral subependymal focal heterotopic gray matter: This indicates the presence of gray matter (which should Normally be on the surface of the brain) in an abnormal location (subependymal region, near the ventricles). Subtle T2 prolongation in the right superior frontal gyrus: This suggests a possible abnormality in this area, potentially related | No. The report describes findings exclusively within the brain. Here's why: Subependymal Heterotopic Gray Matter: This refers to misplaced brain tissue found in the ventricles (fluid-filled spaces) within the brain. Lateral Ventricles: These are | No. The report does Not mention any motion artifacts. In fact, the radiologist states they "have personally reviewed the images," implying sufficient image quality for interpretation. The report also mentions subtle findings like "blurring of the cortical/subcortical junction" and "slightly prominent perivascular spaces," which would be difficult to assess on low-quality images. | No. The report describes findings consistent with benign entities such as bilateral subependymal heterotopic gray matter and mega cisterna magna. While there is a mention of subtle T2 prolongation in the right superior frontal gyrus, the radiologist provides possible explanations such as prominent perivascular spaces and does | No. Here's why: Language is suggestive, Not definitive: The report uses terms like "subtle," "questionable," "may reflect," "possibility," and "Not entirely excluded." This indicates a level of uncertainty and suggests further investigation may be needed, but Not immediate alarm. Common findings: Heterotopic gray matter, the primary finding, is often benign and asymptomatic. Prominent perivascular spaces are also | **Yes** |

| # | Report | Abnormal? (human referee) | Answer Q1 Direct Inquiry | Answer Q2 | Answer Q3 | Answer Q4 | Answer Q5 | Aggregate Stepwise |
|---|---|---|---|---|---|---|---|---|
| | superioraspect of the left lateral ventricle body (series 6 image 27 andseries 5 image 118)There is a subtle T2 prolongation with underlying questionableblurring of the cortical/subcortical junction located in theright superior frontal gyrus (series 7, image 11; series 5, image89; series 15, image 35). Noting slightly prominent perivascularspaces in this region, best seen on coronal T2 images, series 16image 51. There is also asymmetric sulcal pattern involvingbilateral frontoparietal regions. . Noparenchymal lesion or abnormal fluid collection is identified.There is No midline shift, mass effect or acute infarction. Thehippocampi are symmetric and Normal in size, signal, and internalarchitecture.. There is prominence ofthe retrocerebellar CSF spaces ,likely a mega cisterna magna. The cerebellar tonsils terminateabove the level of the foramen magnum. . Nodefinite abnormality is seen in the visualized portions of theorbits. The middle ear cavities, mastoid air cells, and paranasalsinuses are essentially clear. | | to perivascular spaces, prior insult, or even cortical dysplasia (abnormal brain development). Asymmetric sulcal pattern involving bilateral frontoparietal regions: This indicates an asymmetry in the folds (sulci) of the brain, which can be a subtle sign of developmental variations. Prominence of the retrocerebellar CSF spaces: This likely represents a mega cisterna magna, which is a variation in the size of a fluid-filled space and may Not be clinically significant. While some findings might be considered variations rather than definitive pathologies, the report clearly indicates deviations from a Normal brain structure. | structures within the brain. Cortical/Subcortical Junction: This refers to the boundary between the outer layer (cortex) and the underlying layer (subcortical) of the brain. Superior Frontal Gyrus: This is a specific region within the brain's frontal lobe. Perivascular Spaces: These are spaces surrounding blood vessels within the brain. Subcortical White Matter: This refers to the white matter located beneath the brain's cortex. Focal Cortical Dysplasia: This is a developmental abnormality affecting the organization of the brain's cortex. Parenchyma: This term, in the context of the brain, refers to the functional tissue of the brain, excluding supporting structures. Hippocampi: These are structures within the brain, part of the limbic system. | | Not suggest immediate clinical follow-up. The report emphasizes the absence of any acute findings like mass effect or infarction. | relatively common and usually Not a cause for concern. No acute issues: The report clearly states "No parenchymal lesion or abnormal fluid collection," "No midline shift, mass effect or acute infarction." These are all positive findings ruling out serious immediate problems. Act 112 Notification: While this part of the report might seem concerning, it's related to a legal requirement (likely mandatory reporting in a specific jurisdiction) and doesn't necessarily reflect the severity of the findings. In summary: The report indicates some miNor abnormalities that warrant observation and potentially further investigation. However, the language doesn't suggest the radiologist or referring doctor is highly concerned about an immediate, serious condition. | |

| # | Report | Abnormal? (human referee) | Answer Q1 Direct Inquiry | Answer Q2 | Answer Q3 | Answer Q4 | Answer Q5 | Aggregate Stepwise |
|---|---|---|---|---|---|---|---|---|
| | | | | Retrocerebellar CSF Spaces: These are spaces filled with cerebrospinal fluid located at the back of the brain, near the cerebellum. Cerebellar Tonsils: These are structures at the bottom of the cerebellum, within the brain. Foramen Magnum: This is the opening at the base of the skull where the brainstem connects to the spinal cord. All the described findings and anatomical locations are consistently within the brain itself. | | | | |
| 14 | 1. Curvilinear focus of enhancement of the right globe, unchanged in size compared to prior study. Differences inconspicuity may be technical. 2. Unremarkable MRI of the pineal gland and suprasellar region. Requires Act 112 Notification: FalseI have personally reviewed the images and agree with theinterpretation.Act 112^FalseAct 112^FalseMRI BRAIN, WITHOUT AND WITH CONTRAST:CLINICAL INDICATION: Surveillance for unilateral leftretiNoblastoma status post treatment.TECHNIQUE: BRAIN: Sagittal T1, axial and coronal TSE T2, axialFLAIR, axial T1, axial diffusion weighted imaging, axial SWI,post-contrast axial T1. COMPARISON: Prior outside facility MRI from —— isavailable for correlation. FINDINGS:  BRAIN: The ventricular system is Normal in size, shape, andconfiguration.  lesion, masseffect, or midline shift.  No abnormal fluid collections areidentified.  There are No areas of abnormal parenchymal signalintensity or enhancement on these sequences.  The pineal gland is unremarkable. The suprasellar region iswithin Normal limits.Patient is status post left eye enucleation and prosthesis. Theleft optic nerve appears smaller in size than right but stablewhen compared to the prior exams.There is a curvilinear focus of enhancement in the posterolateralright globe [series 17 images 18-20] which appears moreconspicuous however, Not increased in size compared to priorstudy. The differences may be technical.The extraocular muscles, optic nerve-sheath complexes, and opticchiasm appear Normal in | Yes | No. Reasoning: While the report Notes a "curvilinear focus of enhancement in the posterolateral right globe," it attributes this to possible technical differences compared to the prior study and states it's Not increased in size. The report explicitly states "There are No areas of abnormal parenchymal signal intensity or enhancement" in the brain itself. Therefore, based on the information provided, there are No indications of brain abnormalities. | Yes. The report explicitly states "There is a curvilinear focus of enhancement in the posterolateral right globe". The globe of the eye is outside of the brain. | Yes. The report states, "There is a curvilinear focus of enhancement in the posterolateral right globe [series 17 images 18-20] which appears more conspicuous however, Not increased in size compared to prior study. **The differences may be technical.**" This statement suggests that the radiologist suspects the differences in appearance of the enhancement between the current and prior study *might* be due to technical factors related to image acquisition, such as motion artifacts, rather | No. Reasoning: Stable Finding: The report states the curvilinear focus of enhancement in the right globe is "unchanged in size compared to the prior study." This indicates stability and Not an acute change requiring immediate action. Possible Technical Variation: The report ackNowledges that "Differences in conspicuity may be technical," | No. Reasoning: Stable Findings: The report explicitly states the curvilinear focus of enhancement is "unchanged in size compared to the prior study." This indicates stability, Not worsening, of the observed finding. Technical Variation: The report ackNowledges that "Differences in conspicuity may be technical," suggesting the apparent change in the focus's appearance could be due to differences in imaging technique rather than a true change in the | **Yes** |

| # | Report | Abnormal? (human referee) | Answer Q1 Direct Inquiry | Answer Q2 | Answer Q3 | Answer Q4 | Answer Q5 | Aggregate Stepwise |
|---|--------|---------------------------|--------------------------|-----------|-----------|-----------|-----------|--------------------|
|   | signal intensity. . Themiddle ear cavities, mastoid air cells, and paranasal sinuses areclear. | | | | than a true change in the underlying condition. | indicating the appearance difference might be due to imaging factors rather than a real change in the lesion. Routine Surveillance: The clinical indication is "surveillance," implying this is a follow-up for a kNown issue (retiNoblastoma) and Not an emergency presentation. While the report doesn't necessitate immediate clinical follow-up, continued routine monitoring and discussion with the patient's primary care physician or oncologist are essential. | condition. Unremarkable Findings: Other areas of concern, such as the pineal gland, suprasellar region, and left optic nerve, are described as "unremarkable" or "stable." Surveillance Context: The clinical indication is "surveillance," implying the MRI is a routine check-up for a kNown pre-existing condition (treated retiNoblastoma) rather than an acute concern. Important Note: While the report does Not indicate high concern, it is crucial to remember that this analysis is based solely on the provided text. A definitive assessment of the patient's condition requires the expertise of a qualified medical professional who can consider the full clinical context, including the patient's history and other relevant information. | |
| 15 | Expansile lytic, mixed cystic and solid mass centered in theposterior elements of T4, with fluid/fluid levels. Markedepidural extension, resulting in severe spinal canal narrowing atthe T3-T5 level. No abnormal cord signal or enhancement. While the findings on this examination are compatible with ananeurysmal bone cyst, the areas of more solid enhancement withextensive epidural enhancement are somewhat unusual. Given thesuggestion of atypical findings on pathology, a telangiectaticosteosarcoma could be considered. However, a follow-upexamination dated ——, obtained one year after thisexamination dated ——, demonstrates only mildly increasedT2 signal within the posterior T4 vertebral body after resection,and this length of stability is reassuring for less aggressiveetiologies. Continued followup imaging is recommended.Unremarkable contrast-enhanced MRI of the brain.At the time of this dictation, the original report is Notavailable. Correlation with the original report is recommended toavoid discrepancies. An addendum can be issued upon request ifthe original report becomes available. If the patient's statushas changed, please consider repeat imaging.Requires Act 112 Notification: TrueI have personally reviewed the images and agree | Yes | No. The report explicitly states "Unremarkable contrast-enhanced MRI of the brain." Both the initial reading and the second opinion agree on this point. The focus of the report is a spinal lesion, initially suspected to be an aneurysmal bone cyst, which was later confirmed by pathology. | No. Reasoning: Although the report mentions "extension into the left T3-T4 with severe neural foraminal narrowing," and "extension into the right T3-T4 and T4-T5 neural foramina," this refers to extension of the mass from the T4 vertebra into | No. The report doesn't mention any motion artifacts or low image quality. In fact, the report describes the findings in detail, indicating that the images were of sufficient quality for interpretation. The radiologist was able to visualize and characterize the lesion, the spinal cord, and surrounding structures. | No. While the report describes a serious condition (an expansile lytic mixed solid and cystic lesion causing spinal canal narrowing), it states that a follow-up examination a year later showed only mild progression. The language indicates this stability is reassuring, | Yes. While the report ultimately indicates a likely benign diagNosis (aneurysmal bone cyst), it highlights several concerning features: "Marked epidural extension, resulting in severe spinal canal narrowing at the T3-T5 level": This indicates the mass is significantly compressing the spinal cord, which could lead to neurological damage. "The areas of more solid | **Yes** |

| # | Report | Abnormal? (human referee) | Answer Q1 Direct Inquiry | Answer Q2 | Answer Q3 | Answer Q4 | Answer Q5 | Aggregate Stepwise |
|---|---|---|---|---|---|---|---|---|
| | with theinterpretation.Act 112^TrueAct 112^TrueSECOND OPINION INTERPRETATION OF AN OUTSIDE MR: HISTORY: History of lytic T3 mass with intraspinal extension.Underwent T2-T4 laminectomy for resection of mass on ——.Pathology demonstrated aneurysmal bone cyst.TECHNIQUE: MR images of the brain and spine dated —— aresubmitted for our interpretation. Images submitted include:BRAIN: Axial and sagittal T1, axial T2, axial DWI, coronal FLAIR,axial T1 post contrast images of the brain were obtained. SPINE:Sagittal and coronal T2, sagittal STIR, axial and sagittal T1 preand postcontrast images of the spine were obtained. COMPARISON: CT dated ——and subsequent MR dated——FINDINGS:BRAIN:No abnormal parenchymal signal abnormality or enhancement. Noacute infarct, intracranial hemorrhage or extra axial collection,midline shift, or significant mass effect. The ventricles and sulci are Normal in size, shape, andconfiguration. .  The major intracranial flow voids are preserved. Trace mucosalthickening in the bilateral spheNoid sinuses. Otherwise, thevisualized paranasal sinuses, mastoid air cells, middle earcavities are clear. Visualized orbits are unremarkable.SPINE:There is a full complement of vertebrae through the cervical,thoracic, and lumbar spine. There is reversal of the Normalcervical lordosis. The Normal thoracic kyphosis and lumbarlordosis are preserved. There is No evidence of vertebral bodyfracture or spondylolisthesis.An expansile lytic mixed solid and cystic lesion with fluid/fluidlevels is seen centered within the posterior elements of the T4vertebral body, involving the bilateral pedicles, transverseprocesses, lamina, and spiNous process. The lesion demonstratesenhancing septations with areas of enhancing soft tissue. Thereis extension into the left T3-T4 with severe neural foraminalnarrowing and to a lesser extent left T4-T5 neural foramina withmild neural foraminal narrowing. There is also extension into theright T3-T4 and T4-T5 neural foramina without significantnarrowing. There is also suggestion of T2 hyperintense signalabnormality within the posterior elements at T5. There is extensive enhancement within the dorsal epidural spacefrom T1 to T8, resulting in severe spinal canal narrowing, withdeformation, compression, and rightward rotation of the cord,from T3 to T5. No obvious cord signal abnormality or enhancement.Enhancement is also seen within the overlying paraspinal softtissues. The remainder of the spinal cord is Normal in size and signal. The conus medullaris is Normal in shape and terminates at theL1-L2 level.  The cauda equina and filum terminale areunremarkable.The paraspinal soft tissues are otherwise unremarkable. Thekidneys are orthotopic without hydronephrosis. | | | the neural foramina. The neural foramina are bony openings within the spinal column where nerve roots exit the spinal canal.<br><br>The report focuses on the pathology centered in the T4 vertebra and its impact on the surrounding structures within the spinal canal. There is No indication of the pathology extending outside of the spinal column itself. | | suggesting immediate clinical follow-up is Not required. However, continued follow-up imaging is recommended, indicating the need for ongoing monitoring. | enhancement with extensive epidural enhancement are somewhat unusual": This suggests the possibility of a more aggressive tumor, such as a telangiectatic osteosarcoma. "Deformation, compression, and rightward rotation of the cord, from T3 to T5": This reinforces the severity of the spinal cord compression. Although the follow-up report is reassuring, the initial findings and the need for continued monitoring demonstrate a high level of concern from the radiologists regarding the patient's condition. | |
| 16 | Interval evolution of previously seen acute infarctions intofocal encephalomalacia/gliosis. No new lesions.MR ANGIOGRAPHY OF THE BRAIN, WITHOUT CONTRAST:CLINICAL INDICATION: Acute stroke in January 21, suspectedinflammatory vasculopathy (focal cerebral arteriopathy of theinflammatory subtype), follow-upTECHNIQUE: MRA examination of the brain was performed utilizing a3.0 Tesla system via 3D time-of-flight intracranial magneticresonance angiography of the circle of Willis, with volume ofinterest 3D reconstructions using the maximum intensityprojection method.MRA was deemed medically necessary based on abnormal priorimaging.COMPARISON:  MRA from ———FINDINGS:  The direct comparison is slightly difficult due to difference intechniques. | Yes | **Yes**.<br><br>Reasoning: Although the report mentions "No new lesions," it clearly states "interval evolution of previously seen acute infarctions into focal encephalomalacia/gliosis." This indicates existing brain | No.<br><br>The report focuses solely on the brain and its vasculature. Here's why:<br><br>"Focal encephalomalacia/gliosis" describes | No. The report does Not mention motion artifacts or low quality.<br><br>Reasoning:<br><br>Direct Statements: The report doesn't contain phrases like "motion artifact," "poor quality," or "suboptimal study" that typically indicate these | No.<br><br>Reasoning: The report indicates "mild interval progression" of previously observed irregularities in the left internal carotid terminus, proximal MCA, and A1 | No.<br><br>Reasoning:<br><br>Stable Condition: The report explicitly states "No new lesions" and describes the observed changes as "Mild interval progression." This indicates that the patient's condition has | **Yes** |

| # | Report | Abnormal? (human referee) | Answer Q1 Direct Inquiry | Answer Q2 | Answer Q3 | Answer Q4 | Answer Q5 | Aggregate Stepwise |
|---|---|---|---|---|---|---|---|---|
| | Previously seen narrowing and irregularity involving left A1 segment, proximal MCA and most distal left ICA seemingly progressed, especially of the A1 segment and ICA terminus. The vascular wall appears slightly irregular, however without definite abnormal enhancement The rest of the major intracranial arterial structures demonstrate Normal flow-related enhancement. There is No appreciable alteration in course or caliber. No areas of aneurysmal outpouching are appreciated. It should be Noted that MRA may be insensitive for small (<4 mm) aneurysms. IMPRESSION: Mild interval progression of previously seen irregularity and narrowing of the left internal carotid terminus, proximal MCA and A1 segments. No new lesions. Requires Act 112 Notification: FalseAct 112^FalseAct 112^FalseBRAIN MRI, WITHOUT AND WITH CONTRAST:CLINICAL INDICATION: Acute stroke in ——, follow-upTECHNIQUE: Sagittal 3D T1 gradient echo with axial reformations, axial and coronal TSE T2, axial and coronal FLAIR, axial spinecho T1, arterial spin labeled perfusion imaging, post-contrastsagittal 3D T1 gradient echo with axial and coronal reformations, post-contrast axial spin echo T1 with fat suppression, axialdiffusion imaging were performed on a 3.0 Tesla system. In addition, high-resolution vessel wall imaging was performed without and with contrast.COMPARISON: MRI from ——FINDINGS:There has been interval evolution of previously seen acuteischemic foci in the left caudothalamic groove and posterior leftputamen into the gliosis/encephalomalacia. No new lesions have developed.The ventricular system is Normal in size, shape, and configuration. and there is . No abnormal fluid collections are identified. There are No additional areas of abnormal parenchymal signal intensity on these sequences. There are No areas of abnormal enhancement following contrastadministration.No definite abnormality is seen in the visualized portions of theorbits. T.. | | abnormalities in the form of gliosis (scar tissue) and encephalomalacia (softening of brain tissue) as a result of prior infarctions (strokes). | changes within the brain tissue itself. The entire report centers on imaging findings within the skull: "intracranial," "circle of Willis," "major intracranial arterial structures," "parenchymal signal intensity," "ventricles." The report explicitly states No abnormalities outside the central nervous system: "No definite abnormality is seen in the visualized portions of the orbits" While the report mentions "inflammatory vasculopathy," it focuses on the manifestation of this condition within the brain's blood vessels, Not on potential causes or effects outside the brain. | issues. Comparison Difficulty: While the report mentions a "slightly difficult" comparison due to different techniques between the current and prior scans, this refers to the imaging parameters, Not image quality itself. Detailed Findings: The radiologist provides a thorough description of the findings, including subtle changes in vessel appearance. This level of detail suggests the images were clear eNough for accurate interpretation. | segments. However, there are No new lesions, and the report explicitly states "No immediate clinical follow-up is required." The findings suggest a slow progression of a kNown condition, which would likely be managed in a routine follow-up appointment rather than requiring immediate attention. | Not significantly worsened since the previous scan. Expected Progression: The report uses the phrase "interval evolution," suggesting that the observed changes (infarctions becoming encephalomalacia/gliosis) are a natural part of the healing process after a stroke. Lack of Alarming Language: The report does Not use language typically associated with high concern, such as "critical," "urgent," or "immediate attention." Important Note: While the report doesn't indicate high concern from the radiologist, it's crucial to remember that this is only one piece of the patient's medical picture. The treating physician will consider the entire clinical context, including the patient's symptoms and medical history, to make treatment decisions. | |
| 17 | Stable supratentorial ventriculomegaly. Stable narrowing of themidportion of the aqueduct of Sylvius. Stable Normal sized fourthventricle.Stable dysplastic appearance of the corpus callosum.Dilated central canal from C4 through the conus medullaris withdilation greatest at the C6-C7 measuring up to 3 mm.No evidence for fibrofatty infiltration of the filum terminale.Normal position of the conus does Not exclude spinal cordtethering.Requires Act 112 Notification: False I have personally reviewed the images and agree with theinterpretation.Act 112^FalseAct 112^FalseAct 112^FalseSCREENING BRAIN MRI, WITHOUT CONTRAST; MRI CERVICAL, THORACIC,AND LUMBAR SPINE, WITHOUT CONTRAST:CLINICAL INDICATION: Follow-up ventriculomegaly andsyringohydromyelia.TECHNIQUE: BRAIN: Axial diffusion- | Yes | **Yes**. Reasoning: The report describes several abnormalities, including: Stable supratentorial ventriculomegaly: This means that the fluid-filled spaces (ventricles) in the brain are enlarged, which | Yes. The report indicates dilation of the central canal of the spinal cord, which is a finding outside of the brain. Specifically, it Notes: "Dilated central canal | Yes, the report indicates potential low quality for part of the scan. Reasoning: The report explicitly states, "This examination is Not of the typical quality of a full diagNostic MRI and, as such, pathology may be missed." This statement suggests that the image quality may Not be | No. The report largely describes stable findings, meaning the abnormalities seen are unchanged from previous scans. This indicates that the patient's condition is Not acutely worsening. | No. The report repeatedly uses the word "stable" to describe the findings, indicating that the patient's condition has Not significantly worsened. While the report Notes some abnormalities (ventriculomegaly, aqueduct steNosis, | **Yes** |

| # | Report | Abnormal? (human referee) | Answer Q1 Direct Inquiry | Answer Q2 | Answer Q3 | Answer Q4 | Answer Q5 | Aggregate Stepwise |
|---|---|---|---|---|---|---|---|---|
| | weighted, axial GRE, axialFLAIR, coronal T2, and sagittal T1-weighted images of the brainwithout intraveNous contrast were performed on a 3.0 Teslasystem per fast screening protocol. SPINE: Sagittal T1 FLAIR,axial T1 of the lumbar spine, and sagittal T2 weighted imageswere performed on a 3.0 Tesla system.COMPARISON: Brain MRI ——. Ultrasound spine ——.Multiple prior renal ultrasound is —— FINDINGS: BRAIN:This examination is Not of the typical quality of a fulldiagNostic MRI and, as such, pathology may be missed.No definite change in enlargement of the supratentorialventricular system with the third ventricle measuringapproximately 15 mm in transverse dimension, unchanged. The rightatrium measures 21 mm and the left atrium measures 16 mm, alsounchanged. No periventricular signal abnormality to indicateinterstitial flow of CSF. The fourth ventricle remains Normal insize. As before, there is apparent narrowing at the midportion ofthe aqueduct of Sylvius. There is unchanged mild dilation of thetemporal horns which again demonstrate a somewhat atypicalconfiguration. Mild prominence of the foramina of Magendie isunchanged, which again raises the possibility of inferior vermianhypoplasia.There is persistent mild prominence of the prepontine cistern,bilateral sylvian fissures, and bilateral extratemporal CSFspaces.Stable appearance of the dysplastic corpus callosum with lack ofNormal contour of the genu and splenium. and there is No midlineshift. Patchy foci of T2 prolongation within the supratentorialwhite matter are unchanged and Nonspecific. No definite abnormality is seen in the visualized portions of theorbits. T.Diffuse paranasal sinus mucosal disease, progressed from prior.SPINE:There is a full complement of vertebrae through the cervical,thoracic, and lumbar spine. The vertebrae and intervertebraldiscs are Normal in size and signal. There is a long segment dilated central canal extending from theC4 level through the conus medullaris. Dilation is greatest atthe C6-C7 level where it measures up to 3 mm in AP dimension.The conus medullaris is Normal in shape and terminates at theL1-L2 disc level. No definite fatty infiltration seen in thefilum terminale. The paraspinal soft tissues are unremarkable. The kidneys areorthotopic. Right central caliectasis and dilation of the renalpelvis is similar to renal ultrasound from ——, accountingfor differences in technique. | | can be a sign of hydrocephalus. Stable narrowing of the midportion of the aqueduct of Sylvius: This narrowing can obstruct the flow of cerebrospinal fluid (CSF), potentially contributing to the ventriculomegaly. Stable dysplastic appearance of the corpus callosum: The corpus callosum connects the two hemispheres of the brain. A dysplastic appearance suggests it did Not form typically. Dilated central canal from C4 through the conus medullaris: This dilation suggests there might be an issue with CSF flow in the spinal cord, potentially syringohydromyelia. While some findings are described as "stable," they still represent deviations from a Normal brain anatomy. | from C4 through the conus medullaris with dilation greatest at the C6-C7 measuring up to 3 mm." The central canal is a structure within the spinal cord, and the cervical spine (C4-C7) is part of the spinal column, Not the brain. | optimal and could potentially obscure subtle findings. | The report does Not mention any new findings that would require immediate clinical follow-up. However, it's crucial to understand that this answer is for informational purposes only and should Not substitute professional medical advice. The patient should continue to follow up with their physician as scheduled to discuss the findings and manage any underlying conditions. | corpus callosum dysplasia, syringohydromyelia), the language suggests these are chronic and Not acutely concerning. The report also explicitly states "Requires Act 112 Notification: False," further suggesting a lack of immediate concern. | |
| 18 | Interval increase in size of the residual, presumably calcifiedlesion along the posterior medial aspect of the right globe, Nowdemonstrating more enhancement with No definite diffusionrestriction. Again Noted is contrast enhancement of the rightoptic nerve head and along the posterior aspect of the globewithin the intraconal space, Not significantly changed from priorexamination. Stable reduced caliber and signal alteration of theright optic nerve.Stable appearance of the brain with unchanged focal areas ofsusceptibility effect. No evidence of intraparenchymal newlesions or abnormal enhancement. Requires Act 112 Notification: TrueAct 112^TrueAct 112^TrueBRAIN AND ORBIT MRI, WITHOUT AND WITH CONTRAST:CLINICAL INDICATION: Right retiNoblastoma. Follow up. TECHNIQUE: BRAIN: 3D Sagittal T1 gradient echo, axial and coronalTSE T2, axial FLAIR, axial T1, axial spin echo T1, arterial spinlabeled perfusion imaging, axial diffusion tensor imaging,post-contrast axial spin echo T1. ORBITS: Axial T1 gradient echoreformations, axial and coronal TSE T2 with fat suppression, | Yes | **Yes**. Reasoning: The report indicates an "interval increase in size of the residual, presumably calcified lesion along the posterior medial aspect of the right globe" and "increased enhancement". While it states there is "No evidence of intraparenchymal new lesions or abnormal enhancement", the increased size and | Yes. The report specifically mentions "patchy enhancement along the posterior aspect of the globe in the intraconal space adjacent the optic nerve head perhaps within the subarachNoid space of the | No. The report does Not mention any motion artifacts or difficulties visualizing structures due to image quality. The radiologist describes the findings clearly and makes comparisons to prior scans, suggesting confidence in the images. | Yes. Reasoning: The report indicates an "interval increase in size" of the residual lesion, which is demonstrating "more enhancement." While there's no diffusion restriction, the change in size and enhancement pattern warrants | Yes. Here's why: Interval increase in size: The report clearly states the residual lesion has increased in size, which is a concerning sign in a follow-up for retiNoblastoma. More enhancement: Increased enhancement often suggests greater vascularity or changes in the blood-brain barrier, which can be associated | **Yes** |

| # | Report | Abnormal? (human referee) | Answer Q1 Direct Inquiry | Answer Q2 | Answer Q3 | Answer Q4 | Answer Q5 | Aggregate Stepwise |
|---|---|---|---|---|---|---|---|---|
| | andpost-contrast axial and coronal T1 with fat suppression. Allimaging was performed on a 3.0 Tesla system.COMPARISON: Most recent Brain and Orbit MR from ——FINDINGS: ORBITS: Redemonstration of the irregular residual lesion in theinferomedial portion of the right globe associated withsusceptibility effect, likely reflecting calcification. Comparedto prior examination has been increase in size of the componentin the posterior aspect of the globe, along the optic disc,currently measuring 6 x 3.5 mm compared to 4 x 2 mm on priorexamination. There is still No associated diffusion restriction,however, there has increased enhancement. The previously seenpunctate focus of enhancement medially has also increased insize.There is stable patchy enhancement along the posterior aspect ofthe globe in the intraconal space adjacent the optic nerve headperhaps within the subarachNoid space of the optic nerve sheaths,Not significantly change from most recent prior but slightly moreproNounced than examination from ——. No definiteenhancement of the intraorbital optic nerve which remains smallin caliber and associated increased T2 signal. The prechiasmaticright optic nerve from is also smaller in caliber. Linearhypointensity along the posterior globe may reflect sequela ofprior retinal hemorrhage.The left orbit, left optic nerve and optic chiasm remainsunremarkable. BRAIN: Again seen in the previously Noted scattered regions ofsusceptibility, consistent with blood products in thesupratentorial infratentorial brain, unchanged. No newparenchymal lesions or abnormal fluid collections are identified. The ventricles remain Normal in size. The myelination isappropriate for the age. There is No abnormal enhancement oracute infarction. . The pineal gland remains similar in size and signal. There is Nopineal mass.The mastoid air cells and middle ear cavities are clear. Mildmucosal thickening paranasal sinuses. Stable midline cyst at theroot of the tongue, likely thyroglossal duct cyst. | | enhancement of the existing lesion suggests an active process and therefore an abnormality. | optic nerve sheaths." The intraconal space, optic nerve head, and optic nerve sheaths are all structures outside of the brain itself. | | immediate clinical follow-up to assess for possible tumor recurrence or other complications. | with tumor activity. Requires Act 112 Notification: While the exact nature of Act 112 isn't specified, the fact that it's triggered indicates a need for reporting and likely points to concerns about potential malignancy or serious findings. Important Note: I canNot provide medical advice. This analysis is based solely on the provided report and should Not be interpreted as a definitive diagNosis or progNosis. It is essential to consult with the patient's physician for a complete understanding and appropriate management. | |
| 19 | Stable pineal cyst.Requires Act 112 Notification: FalseI have personally reviewed the images and agree with theinterpretation.Act 112^FalseBRAIN MRI, WITHOUT AND WITH CONTRAST:CLINICAL INDICATION: Pineal cystTECHNIQUE: Sagittal T1 MPRAGE with axial reformations, axial andsagittal TSE T2, axial and coronal FLAIR, axial spin echo T1,arterial spin labeled perfusion imaging, post-contrast T1sagittal MPRAGE with axial and coronal reformations,post-contrast axial spin echo T1, and axial diffusion tensorimaging were performed on a 3.0 Tesla system.COMPARISON: MR brain ——FINDINGS:There is No significant interval change in the appearance of thebrain.A pineal cyst is unchanged in size and signal characteristics. This again slightly narrows the cerebral aqueduct which remainspatent. The ventricular system is Normal in size, shape, andconfiguration. and thereis . No abnormal fluidcollections are identified. There are No areas of abnormalparenchymal signal intensity on these sequences. There are No areas of abnormal enhancement following contrastadministration.. Nodefinite abnormality is seen in the visualized portions of theorbits. T.. | Yes | No.<br><br>The report states that there is "No significant interval change" in the brain's appearance, and explicitly Notes that the pineal cyst is "unchanged in size and signal characteristics." While the cyst slightly narrows the cerebral aqueduct, it remains patent. The report further emphasizes No abnormal fluid collections, signal intensities, or enhancement post-contrast. Overall, the findings indicate a stable condition without signs of new or | No.<br><br>Reasoning: The report focuses solely on the brain, specifically mentioning structures like the pineal gland, cerebral aqueduct, and ventricles. There's No indication of imaging or findings related to areas outside the brain. | No.<br><br>The report does Not mention any motion artifact or low-quality imaging. The report explicitly states that the radiologist personally reviewed the images and that the ventricular system is Normal in "size, shape, and configuration," which would be difficult to assess on a poor quality scan. | No.<br><br>The report explicitly states that the pineal cyst is "stable" and "unchanged in size and signal characteristics". Additionally, it Notes that the cerebral aqueduct, while slightly narrowed by the cyst, remains patent. These findings suggest No acute changes or complications requiring immediate clinical follow-up. | No.<br><br>Reasoning:<br><br>Stable Condition: The report explicitly states the pineal cyst is "unchanged in size and signal characteristics" compared to a previous scan. This indicates stability and Not a cause for immediate concern. Patent Aqueduct: Although slightly narrowed by the cyst, the cerebral aqueduct (a crucial pathway for cerebrospinal fluid) is described as "patent" meaning open and functioning. This is a positive sign. | No |

| # | Report | Abnormal? (human referee) | Answer Q1 Direct Inquiry | Answer Q2 | Answer Q3 | Answer Q4 | Answer Q5 | Aggregate Stepwise |
|---|---|---|---|---|---|---|---|---|
| | | | worsening abnormalities. | | | However, this response should Not be considered medical advice. A medical professional should always be consulted for interpretation of radiological reports and to determine the appropriate clinical management. | No Other AbNormalities: The report repeatedly emphasizes the absence of other concerning findings: "ventricular system is Normal," "No abnormal fluid collections," "No areas of abnormal signal intensity," "No areas of abnormal enhancement." Act 112 Notification: False: While the exact nature of this Notification is unclear without context, the fact that it's marked "False" further suggests the case doesn't raise serious alarm bells requiring immediate action. Important Note: This analysis is based solely on the radiology report. Clinical context (patient symptoms, medical history, etc.) is crucial for a complete assessment. A medical professional should always be consulted for interpretation and management. | |
| 20 | 1. Complete absence of the septum pellucidum with secondaryfeatures as detailed above, developmental in nature. Noadditional features of septo-optic dysplasia/optic nervehypoplasia syndrome.2. No additional intracranial abnormality.3. Paranasal sinus findings as detailed. A small air-fluid levelsuch as that within the left maxillary sinus may indicate acutesinusitis in the appropriate clinical setting.Act 112^FalseMRI BRAIN, WITHOUT CONTRAST:CLINICAL INDICATION: —— with worsening daily headachesand neck pain for months, waking from sleep.TECHNIQUE: Sagittal 3D T1 gradient echo with axial reformations,axial and coronal TSE T2, axial FLAIR, arterial spin labeledperfusion imaging, and axial diffusion imaging were performed ona 3.0 Tesla system.COMPARISON: NoneFINDINGS: There is mild motion on some sequences. DiagNostic information isstill obtained.There is complete absence of the septum pellucidum, with acharacteristic flat contour of the superior margins of thefrontal horns of both ventricles. Portions of the fornices appearfused (body, portions of the crus, and possibly the columns).Ventricular system otherwise Normal. No abnormal parenchymal signal intensity.Posterior fossa structures Normal. Cerebellar tonsils areslightly low in position, extending into the foramen magnum by upto approximately 3 mm. This is most in | Yes | **Yes**. Reasoning: While the report states "No additional intracranial abnormality," the complete absence of the septum pellucidum IS an abnormality, even though it may be developmental and asymptomatic. | Yes. The report indicates findings within the paranasal sinuses, specifically mentioning air-fluid levels in the left maxillary sinus and mucosal thickening in various sinuses. These structures are Not within the brain. | No. While the report mentions "mild motion on some sequences", it explicitly states that "DiagNostic information is still obtained". This indicates that although there was some movement during the scan, it did Not significantly compromise the quality or diagNostic utility of the images. | Yes. While the septum pellucidum finding is Noted to be developmental and the cerebellar tonsillar ectopia is considered clinically insignificant, the report indicates potential acute sinusitis (air-fluid level in the left maxillary sinus). The report explicitly states this finding may indicate acute sinusitis in the | No. While the report Notes some abnormalities (absence of septum pellucidum, cerebellar tonsillar ectopia, sinus mucosal thickening), it characterizes them as developmental and/or unlikely to be clinically significant. The report explicitly states the cerebellar tonsillar ectopia is "usually considered of No clinical significance" and describes the sinus findings in a way that suggests they are common and Not | **Yes** |

| # | Report | Abnormal? (human referee) | Answer Q1 Direct Inquiry | Answer Q2 | Answer Q3 | Answer Q4 | Answer Q5 | Aggregate Stepwise |
|---|---|---|---|---|---|---|---|---|
|  | keeping with cerebellartonsillar ectopia, which is usually considered of No clinicalsignificance. There is a slight retroversion of the dens tip.Allowing for mild motion on the sagittal T1-weighted sequence,the posterior pituitary T1 bright spot appears Normal. There is aslight convex contour of the superior pituitary gland remainswithin the range of anticipated for age.Olfactory bulbs are present bilaterally.No intracranial mass; .No abnormal fluid collections. Visualized major intracranial vessels patent. No definite abnormality in visualized orbits. The optic nervesare Normal in caliber. No abnormality of the optic chiasm ortracts.There are a couple of polypoid T2 hyperintensities in the rightmaxillary sinus, the more lateral/posterior moderate in size, andthe more anterior small. There is lobular mucosal thickeningpredominantly inferiorly in the left maxillary sinus with a smallair-fluid level. Scattered mild mucosal thickening in the ethmoidair cells, and circumferentially within the left spheNoid sinus,slightly lobular. Trace mucosal thickening in the frontalsinuses, predominantly on the left. Trace fluid scattered in themastoid air cells bilaterally. |  |  |  |  | appropriate clinical setting. Therefore, clinical correlation and potential follow-up are necessary to address the sinusitis possibility. | alarming in isolation. The lack of any mention of concern regarding the septum pellucidum finding further points to this Not being a highly concerning report. |  |
| 21 | Imaging findings of mild papilledema, borderline optic nervesheath fluid volume, and narrowing of the distal transversesinuses. Dural veNous sinuses are patent.These findings suggest the possibility of idiopathic intracranialhypertension ("pseudotumor cerebri"). Correlate with the currentclinical status of the patient. Lumbar puncture with openingpressure measurement can be considered if clinically indicated....Examination: MRI of the brain with and without contrastINDICATION: HeadacheCOMPARISON: Head CT ——TECHNIQUE: MR imaging of the brain was obtained on a 3 Tesla MRIsystem before and after intraveNous contrast administration.Imaging sequences include axial diffusion-weighted imaging, 3-DT1 MPRAGE with multiplanar reformations before and aftercontrast, axial T1 spin-echo postcontrast with fat saturation,axial and coronal T2, axial T2 FLAIR with fat saturation, ASLperfusion.FINDINGS:Paranasal sinuses, mastoid air cells, and middle ear caviesappear aerated. No gross abnormality of arterial flow voids.Images such as axial T2 series 5 image 55 demonstrate anteriorprotrusion of the optic nerve heads concerning for papilledema.Axial T2 series 5 image 54 demonstrates borderline prominence ofthe optic nerve sheath fluid volume on both sides. There is Notan empty sella appearance.On postcontrast T1 MPRAGE images the dural veNous sinuses areseen to be patent, but with narrowing of the distal transversesinuses. Basilar cisternsare patent. No extra-axial fluid collection or mass effect on thebrain., and thegyral/sulcal pattern appears within Normal limits. Themyelination pattern is Normal. There are a few tiny frontal subcortical white matter FLAIRhyperintensities that have a Nonspecific appearance and are ofdoubtful clinical significance.... | Yes | **Yes** Reasoning: The report indicates several abnormalities, including: Mild papilledema: Swelling of the optic nerve head, suggestive of increased intracranial pressure. Borderline optic nerve sheath fluid volume: While Not definitively abnormal, this can be associated with increased intracranial pressure. Narrowing of the distal transverse sinuses: This can restrict cerebrospinal fluid flow and contribute to intracranial hypertension. The report explicitly states these findings suggest the possibility of idiopathic intracranial hypertension (IIH), a condition characterized by increased pressure in the skull. While further investigation | No. The report describes findings that are all within the structures of the head (brain, optic nerves, sinuses). The report specifically states "No extra-axial fluid collection or mass effect on the brain," indicating the pathology is Not outside the skull. | No. The report does Not mention any motion artifacts or issues with the quality of the scan. Reasoning: Specific sequences for artifact reduction: The report mentions using specialized sequences like diffusion-weighted imaging, T1 MPRAGE, T2 FLAIR, and ASL perfusion. These techniques are designed to minimize artifacts and enhance image quality. Detailed visualization: The radiologist was able to clearly visualize and assess critical structures like the optic nerves, optic nerve sheaths, dural veNous sinuses, basilar cisterns, and brain parenchyma. No mention of limitations: Importantly, the report does Not mention any motion artifacts degrading image quality or hindering the interpretation. Radiologists typically Note such limitations if | Yes. Reasoning: Papilledema: This is a serious finding indicating increased pressure around the optic nerve, potentially from elevated intracranial pressure (ICP). Borderline optic nerve sheath fluid volume: While borderline, this also supports the possibility of increased ICP. Narrowing of the distal transverse sinuses: This could be a contributing factor to impaired cerebrospinal fluid (CSF) flow and increased ICP. Report Recommendation: The report explicitly states, "Correlate with the current clinical | Yes Reasoning: Papilledema: The report explicitly states "Imaging findings of mild papilledema," which is a serious condition indicating increased pressure in or around the brain. Borderline optic nerve sheath fluid volume: While "borderline," this finding also suggests increased intracranial pressure. Narrowing of the distal transverse sinuses: This finding further supports the possibility of obstructed cerebrospinal fluid (CSF) flow, contributing to increased intracranial pressure. Idiopathic intracranial hypertension ("pseudotumor cerebri"): The radiologist explicitly suggests this diagNosis, which is a serious condition requiring management to prevent vision loss and other complications. | **Yes** |

| # | Report | Abnormal? (human referee) | Answer Q1 Direct Inquiry | Answer Q2 | Answer Q3 | Answer Q4 | Answer Q5 | Aggregate Stepwise |
|---|---|---|---|---|---|---|---|---|
| | | | (like a lumbar puncture) is needed to confirm the diagNosis, the MRI findings themselves do point towards brain abnormalities related to pressure and fluid dynamics. | | they impact the diagNostic accuracy. | status of the patient. Lumbar puncture with opening pressure measurement can be considered if clinically indicated." This suggests the radiologist sees a need for further assessment based on the patient's symptoms. Important Considerations: Headache Indication: The patient's reason for the MRI was a headache, which aligns with potential symptoms of idiopathic intracranial hypertension (IIH). Clinical Correlation: The radiologist emphasizes the need to correlate these findings with the patient's clinical presentation. A medical professional needs to evaluate the patient's symptoms, medical history, and these imaging results to determine the best course of action. In conclusion, the findings of papilledema, potential CSF flow obstruction, and the report's recommendation | Lumbar puncture recommendation: The report recommends a lumbar puncture for opening pressure measurement, indicating a need to further investigate and potentially alleviate the suspected high intracranial pressure. While the report uses terms like "mild" and "borderline," the overall findings and recommendations point to a concern for potentially serious underlying conditions requiring further investigation and management. | |

| # | Report | Abnormal? (human referee) | Answer Q1 Direct Inquiry | Answer Q2 | Answer Q3 | Answer Q4 | Answer Q5 | Aggregate Stepwise |
|---|---|---|---|---|---|---|---|---|
| | | | | | | for further assessment based on clinical presentation warrant immediate follow-up with a medical professional. | | |
| 22 | Unremarkable brain MRI prior to and following intraveNouscontrast.Requires Act 112 Notification: FalseI have personally reviewed the images and agree with theinterpretation.Act 112^FalseBRAIN MRI, WITHOUT AND WITH CONTRAST:CLINICAL INDICATION: Neuroblastoma. Follow-up.TECHNIQUE: Sagittal 3D T1 gradient echo with axial reformations,axial and coronal TSE T2, axial and coronal FLAIR, axial spinecho T1, arterial spin labeled perfusion imaging, post-contrastsagittal 3D T1 gradient echo with axial and coronal reformations,post-contrast axial spin echo T1 with fat suppression, axialdiffusion imaging were performed on a 3.0 Tesla system.COMPARISON:  Brain MRI dated ——FINDINGS:Ventricular system Normal.  No abnormal parenchymal signal intensity.No intracranial mass lesions; .  No abnormal fluid collections.  .No abnormal contrast enhancement.Visualized major intracranial vessels patent. No definite abnormality in visualized orbits. Middle ear cavities, mastoid air cells, and paranasal sinusesclear. An exception of partially visualized two small mucousretention cysts in the left maxillary sinus. | No | **No**. The report clearly states "Unremarkable brain MRI prior to and following intraveNous contrast." This means the interpreting radiologist did Not find any abnormalities in the brain imaging. Additionally, the report goes on to describe specific structures within the brain as being Normal, further supporting the lack of abnormalities. | No. The report specifically states "No definite abnormality in visualized orbits", which suggests the area outside the brain, but within the skull, appeared Normal. Additionally, the report focuses solely on the brain and its surrounding structures within the skull. It doesn't provide information about any pathology outside the skull. | No. The report does Not mention any motion artifacts or issues with the scan quality. Phrases like "unremarkable," "Normal," and "No abnormal" suggest that the images were clear eNough for the radiologist to make these assessments. Additionally, the report details the various sequences used, further indicating a comprehensive and likely high-quality scan. | No Reasoning: The report clearly states "Unremarkable brain MRI prior to and following intraveNous contrast." This means No significant abnormalities were detected that would require immediate clinical follow-up. The report indicates this is a follow-up for neuroblastoma, so routine follow-up with the patient's oncologist is likely still necessary, but the MRI itself does Not necessitate urgent action. | No. The report repeatedly uses terms like "Normal," "No abnormal," and "unremarkable," indicating that the findings are within the expected range and do Not raise significant concerns. The presence of small mucous retention cysts is Noted but described as an "exception" and likely Not clinically significant in this context. | **No** |
| 23 | Improved paranasal sinus disease with minimal associated reduceddiffusion. Previous described trace small right frontalextra- axial fluid collection is Not clearly appreciated ontoday's examination. If there is continued clinical concern,follow-up MRI should be obtained.I have personally reviewed the images and agree with theinterpretation.BRAIN MRI, WITHOUT AND WITH CONTRAST:CLINICAL INDICATION: Sinusitis with intracranial extension.Follow-up.TECHNIQUE: Sagittal 3D T1 gradient echo with axial reformations,axial and coronal TSE T2, axial and coronal FLAIR, axial spinecho T1, arterial spin labeled perfusion imaging, post-contrastsagittal 3D T1 gradient echo with axial and coronal reformations,post-contrast axial and sagittal spin echo T1 with fatsuppression, axial diffusion imaging were performed on a 3.0Tesla | No | **No**. The report states that there are No areas of abnormal parenchymal signal intensity, No intracranial mass lesions, and the ventricular system is Normal. While a previously Noted small fluid collection is Not clearly appreciated, | No. The report explicitly states "No intracranial mass lesions," indicating that the pathology is confined within the skull and Not extending outside the brain. | No. The report does Not mention any motion artifacts or low-quality scans. In fact, the radiologist states "I have personally reviewed the images and agree with the interpretation", indicating confidence in the image quality. The report focuses on | No. While the report Notes minimal residual reduced diffusion, it also states that the findings are overall significantly improved compared to the previous exam. Additionally, the | No. The report indicates "improved paranasal sinus disease." The area of concern (extra-axial fluid collection) is No longer clearly seen. While follow-up is recommended if clinical concerns persist, the overall tone suggests improvement and does | **No** |

| # | Report | Abnormal? (human referee) | Answer Q1 Direct Inquiry | Answer Q2 | Answer Q3 | Answer Q4 | Answer Q5 | Aggregate Stepwise |
|---|---|---|---|---|---|---|---|---|
|  | system.COMPARISON: MRI brain ——.FINDINGS:Redemonstrated opacification of the paranasal sinuses, overallimproved compared to the —— exam. There is near completeopacification of the left frontal sinus with mild circumferentialmucosal thickening of the right frontal sinus, partialopacification of the bilateral ethmoid sinuses, partialopacification of the bilateral spheNoid sinuses andcircumferential mucosal thickening of the bilateral maxillarysinuses. There may be a small amount of residual reduceddiffusion in the posterior right spheNoid sinus; however,restricted diffusion within the sinuses is overall significantlyimproved.The previously described trace small right frontal extra-axialfluid collection Not clearly appreciated on today's examination.Ventricular system Normal in size, shape, and configuration. No areas of abnormal parenchymal signal intensity.No intracranial mass lesions; .Visualized major intracranial vessels patent. No definite abnormality in visualized orbits. Right greater than left mastoid effusions. |  | this is Not described as abnormal and a follow-up is only suggested if there are clinical concerns. Overall, the report describes improvement in sinus opacification compared to the prior exam. | While the report mentions "Right greater than left mastoid effusions," mastoid effusions are located within the mastoid air cells, which are part of the temporal bone in the skull and Not external to it. | comparing findings to a previous scan and describes the observed anatomy and changes in detail, suggesting adequate image quality for interpretation. | previously observed fluid collection is No longer clearly visible. The report suggests a follow-up MRI only if there is continued clinical concern. This implies that immediate clinical follow-up is Not mandatory based solely on the imaging findings, but rather contingent on the patient's symptoms. | Not convey high concern. |  |
| 24 | 1. No acute intracranial process is identified.2. Other incidental findings as Noted above.Requires Act 112 Notification: FalseI have personally reviewed the images and agree with theinterpretation.Act 112^FalseBRAIN MRI, WITHOUT AND WITH CONTRAST:CLINICAL INDICATION:——; Numbness and tingling in left arm; Pain in righthand; Facial paralysis on left sideTECHNIQUE:Sagittal 3D T1 gradient echo with axial reformations, axial andcoronal TSE T2, axial FLAIR, axial spin echo T1, axial DTI, post-contrast sagittal 3D T1 gradient echo with axial and coronalreformations, post-contrast axial spin echo T1 with fatsuppression, axial ASL and axial diffusion weighted imaging wereperformed on a 3.0 Tesla system.COMPARISON: NoneFINDINGS:Sagittal images suggest profile with mild brachycephaly. Thecorpus callosum is slightly shortened and thickened inappearance, possibly due to the patient's underlying trisomy.Additionally, the pons and bilateral frontal lobes or somewhatdiminutive in appearance.Note is made of a persistent cavum vergae and tiny cavum septumpellucidum, Normal variants. Mild temporal horns dilation. Theventricular system is Normal in size, shape, and configuration.NoNobstructive oval shaped 3 mm isointense structure within theanterior third ventricle, likely representing a cluster ofchoroid plexus. and there is No masseffect or midline shift.. onthese sequences. There are No areas of abnormal enhancement following contrastadministration.. Note ismade of a slightly diminutive A1 segment of the right anteriorcerebral artery, likely a developmental variant.The globes are somewhat misshapen in appearance, possibly due toa stigmatism/glioma/myopia.Note is made of a tiny pars intermedia cyst of the pituitarygland.The suprasellar cistern is slightly asymmetrically larger on theright, and the sella turcica slopes downwards and to the right,likely developmental in nature.Note is made of a tiny pineal cyst.No definite abnormality is seen in the visualized portions of theorbits.There is mild left ward nasal septum deviation.There is trace bilateral mastoid air cells opacification.Trace bilateral ethmoid air cells and spheNoid/maxillary | No | **No.** Reasoning: The report clearly states "No acute intracranial process is identified." It goes on to list several findings, but characterizes them as incidental, Normal variants, developmental, or likely benign. There is No mention of any finding that would be considered definitively abnormal. | Yes. Reasoning: The report specifically states, "No acute intracranial process is identified," meaning there are No acute problems within the brain. However, the report goes on to list several findings outside the brain, including:<br><br>Misshapen globes (eYes): Possibly due to stigmatism, glioma (which can occur outside the brain), or myopia.<br>Nasal septum deviation: This is a structural issue | No. The report doesn't mention any motion artifacts or low-quality issues. It describes the images and findings in detail, suggesting the scan quality was sufficient for interpretation. Additionally, the radiologist "personally reviewed the images and agree[s] with the interpretation," further indicating No significant quality concerns. | No. Reasoning: The report explicitly states "No acute intracranial process is identified". While the report Notes several incidental findings (most described as developmental variants or miNor), None are described in a way that would suggest an immediate need for clinical follow-up. | No. Reasoning:<br><br>No acute findings: The report explicitly states "No acute intracranial process is identified." This means there are No signs of an immediate, serious problem like a stroke, hemorrhage, or large mass.<br>Incidental findings: The report Notes "incidental findings," meaning miNor variations from the Norm that are likely Not causing the patient's symptoms. These findings (like the slightly diminutive A1 segment of the right anterior cerebral artery) are often described as developmental variants, implying they are benign anatomical differences.<br>Lack of concern in language: The report's language is predominantly descriptive and lacks | **No** |

| # | Report | Abnormal? (human referee) | Answer Q1 Direct Inquiry | Answer Q2 | Answer Q3 | Answer Q4 | Answer Q5 | Aggregate Stepwise |
|---|---|---|---|---|---|---|---|---|
| | sinusesmucosal thickening is identified.Note is made of an oral airway in place. | | | within the Nose. Mastoid air cells opacification: This indicates potential inflammation or fluid in the mastoid air cells, located in the skull behind the ear. Ethmoid/spheNoid/maxillary sinus mucosal thickening: This suggests inflammation or fluid in the sinuses, which are air-filled cavities in the skull. These findings suggest that the cause of the patient's symptoms (numbness, tingling, pain, facial paralysis) might originate from issues outside the brain. | | | terms often associated with serious concerns, such as "mass effect," "significant," "urgent," etc. Focus on developmental factors: The report repeatedly links findings to the patient's Down syndrome ("possibly due to the patient's underlying trisomy"). This suggests the radiologist considers these variations likely linked to the pre-existing condition rather than indicating a new, acute issue. Important Note: While the report doesn't indicate high concern from the radiologist's perspective, the patient's symptoms (numbness, tingling, pain, facial paralysis) still require further investigation by their doctor to determine the underlying cause. | |
| 25 | Decreased cerebellar tonsillar ectopia. Findings do Not meetcriteria for Chiari I deformity.Stable small focus dural enhancement right anterior falx,Nonspecific. Attention on follow-up recommended.BRAIN MRI, WITHOUT AND WITH CONTRAST:CLINICAL INDICATION: Progressive worsening of chronic headaches,kNown lymphatic and veNous malformation with communication withdura and history of increased ICP and papilledemaTECHNIQUE: Sagittal 3D T1 gradient echo with axial reformations,axial and coronal TSE T2, sagittal 3D FLAIR with axial andcoronal reformations, arterial spin labeled perfusion imaging,postcontrast sagittal 3D T1 gradient echo with axial and coronalreformations, postcontrast axial spin echo T1, and axialdiffusion imaging were performed on a 3.0 Tesla PET/MRI system.COMPARISON:  CT brain ——, outside MRI —— FINDINGS:Ventricular system Normal.  No abnormal parenchymal signal intensity.  Stable focus ofT2/FLAIR hyperintense signal right anterior limb of internalcapsule.  Few stable Nonspecific scattered subcortical whitematter foci. No intracranial mass lesions; .  No abnormal fluid collections.  Cerebellar tonsils rounded in configuration and terminate 2 mmbelow level of foramen magnum compared to 4 mm on prior. CSFspace at craniocervical junction mildly reduced. | No | Yes. While the cerebellar tonsillar ectopia has decreased and does Not meet the criteria for Chiari I malformation, the report identifies a few other abnormalities: Stable small focus dural enhancement right anterior falx, Nonspecific: This finding suggests there might be an abnormality in the dura mater (one of the layers of tissue | Yes. The report mentions "Diffuse expansion of calvarial marrow similar to prior". The calvarium is the skull bone, which is outside the brain. Therefore, the pathology (diffuse expansion of calvarial marrow) is outside the brain. | No. The report does Not mention any motion artifacts or low-quality scans. The report describes the images as clear and provides detailed findings, suggesting the images were of sufficient quality for interpretation. | No. While the report recommends "attention on follow-up", it does Not indicate any findings requiring immediate clinical follow-up. Here's why: Decreased cerebellar tonsillar ectopia: This is an improvement from the previous study and does Not meet criteria for | No. The report Notes "decreased cerebellar tonsillar ectopia" and states that the findings "do Not meet criteria for Chiari I deformity." While the report recommends "attention on follow-up," the language used ("stable," "Nonspecific," "mild") suggests a moderate level of concern rather than high. | **No** |

| # | Report | Abnormal? (human referee) | Answer Q1 Direct Inquiry | Answer Q2 | Answer Q3 | Answer Q4 | Answer Q5 | Aggregate Stepwise |
|---|---|---|---|---|---|---|---|---|
| | Stable 4 mm focus of enhancement along right aspect of anteriorfalx. No diffuse dural enhancement.Visualized major intracranial vessels patent. No definite abnormality in visualized orbits. Mild mucosal thickening right spheNoid sinus. Middle earcavities, mastoid air cells, and paranasal sinuses otherwiseclear.Diffuse expansion of calvarial marrow similar to prior. | | surrounding the brain). While deemed "Nonspecific," it warrants attention in follow-up examinations. Stable focus of T2/FLAIR hyperintense signal right anterior limb of internal capsule: This indicates a change in signal intensity within the white matter of the brain. While Noted as stable, it's still an area that differs from Normal brain tissue. Few stable Nonspecific scattered subcortical white matter foci: Similar to the above, this signifies small areas within the white matter that show a difference in signal. These are also Noted as stable but are Not the typical appearance of healthy brain tissue. Even though some findings are described as "stable" and "Nonspecific," the presence of these variations from the expected appearance of brain structures constitutes abnormalities. Further investigation and follow-up are recommended to assess their potential clinical significance. | | | Chiari I malformation, which would require more urgent attention. Stable findings: The report uses "stable" multiple times to describe findings, indicating they haven't worsened and don't pose an acute problem. Nonspecific findings: The dural enhancement and white matter foci are Noted as Nonspecific, meaning they are Not definitively linked to a serious condition and require further monitoring rather than urgent intervention. Important Note: This analysis is for informational purposes only and should Not substitute a healthcare professional's judgment. The patient's physician, considering their complete medical history, should determine the appropriate follow-up. | | |

| # | Report | Abnormal? (human referee) | Answer Q1 Direct Inquiry | Answer Q2 | Answer Q3 | Answer Q4 | Answer Q5 | Aggregate Stepwise |
|---|---|---|---|---|---|---|---|---|
| 26 | 1. Questionable globular left hippocampus with asymmetric increased signal, within the limitations Noted above.END OF IMPRESSION: Requires Act 112 Notification:  I have personally reviewed the images and agree with the interpretation.Act 112^FalseBRAIN MRI, WITHOUT CONTRASTCLINICAL INDICATION: Autism, abnormal examination, seizure.TECHNIQUE: Sagittal 3D T1 gradient echo with axial and coronal reformations,axial and coronal TSE T2, axial and coronal FLAIR, axial SWI,axial ASL, axial DTI, and axial diffusion weighted imaging were performed on a 3.0 Tesla system.Study is somewhat technically limited due to shading on the coronal FLAIR sequences, leading to asymmetric increasedbrightness of the left cerebellar and cerebral hemisphere.COMPARISON:  NoneFINDINGS:Within the limitations of the shading Noted above, there is some question of slight fullness and globular appearance of the left hippocampus, which appears somewhat asymmetrically brighter than its counterpart on the right, possibly due to bona fideasymmetric signal abnormality.The ventricular system is Normal in size, shape, andconfiguration.  , and there is .  No abnormal fluidcollections are identified.  There are No additional areas ofabnormal parenchymal signal intensity on these sequences.  . Note ismade of a mildly dominant right vertebral artery.No definite abnormality is seen in the visualized portions of the orbits.Trace bilateral paranasal sinus mucosal thickening is Noted.Partially visualized is an oral airway in place.There is moderate prominence of the adeNoids and bilateralpalatine tonsils. | Yes | **Yes**.<br><br>The report indicates a "questionable globular left hippocampus with asymmetric increased signal". While the report Notes this could be due to technical limitations, it also ackNowledges the possibility of "bona fide asymmetric signal abnormality". This means further investigation may be needed to rule out a true abnormality. | No.<br><br>The report focuses solely on findings within the brain.<br><br>Reasoning:<br><br>Hippocampus: This structure is located within the temporal lobe of the brain. The report describes potential abnormalities in the left hippocampus. Ventricular system: This system is a network of cavities within the brain that contain cerebrospinal fluid. Parenchymal signal intensity: This refers to the appearance of brain tissue on the MRI. Vertebral artery: While this artery runs through the neck, the report specifically mentions the "visualized portions," indicating the focus is on the part of the artery supplying blood to the brain. Other findings: Mentions of paranasal sinus mucosal thickening, adeNoids, and | Yes.<br><br>The report explicitly states that the study is "somewhat technically limited due to shading on the coronal FLAIR sequences." This shading leads to "asymmetric increased brightness," which makes it difficult to determine if the observed signal abnormalities are genuine or artifacts. | Yes.<br><br>Reasoning:<br><br>The report indicates a "questionable globular left hippocampus with asymmetric increased signal". While the report ackNowledges potential technical limitations due to shading, it also states this finding "possibly due to bona fide asymmetric signal abnormality."<br><br>This suggests a potential abnormality requiring further investigation. A medical professional needs to correlate these findings with the patient's clinical history, symptoms, and potentially order follow-up imaging or tests to determine the cause and significance of the hippocampal finding.<br><br>Additionally, the mention of "Act 112 Notification" implies a potential concern for a medical condition that might necessitate reporting, further emphasizing the need for clinical follow-up. | No.<br><br>Here's why:<br><br>"Questionable" and "Slight": The report uses hesitant language like "questionable" and "slight" to describe the findings in the hippocampus. This suggests a degree of uncertainty and that the changes are Not dramatically abnormal. "Possibly due to...": The radiologist ackNowledges that the brighter signal in the left hippocampus could be due to technical factors ("bona fide asymmetric signal abnormality") rather than a definite pathology.<br>Lack of Alarming Language: The report lacks strong language often associated with serious concerns, such as "mass," "lesion," "significant," or "urgent." Focus on Other Findings: The report focuses on other unremarkable findings, indicating that the hippocampal observation is Not the primary concern.<br>Important Note: While the report doesn't indicate high concern from the radiologist, the Act 112 Notification implies this finding might be relevant to the patient's clinical history (autism, abnormal examination, seizure). A medical doctor will need to correlate this imaging report with the patient's overall clinical picture to | **Yes** |

| # | Report | Abnormal? (human referee) | Answer Q1 Direct Inquiry | Answer Q2 | Answer Q3 | Answer Q4 | Answer Q5 | Aggregate Stepwise |
|---|---|---|---|---|---|---|---|---|
| | | | | palatine tonsils are common incidental findings and Not indicative of pathology outside the brain in this context. The report does Not describe any structures or abnormalities outside of the brain. | | | determine the significance of these findings. | |
| 27 | Interval resolution of bilateral subcortical and deep whitematter, and splenium corpus callosum signal abnormalities andreduced diffusion seen on the prior exam. Nonspecific punctate foci of bilateral subcortical and periatrialwhite matter T2/FLAIR hyperintensities.Requires Act 112 Notification: FalseI have personally reviewed the images and agree with theinterpretation.Act 112^FalseBRAIN MRI, WITHOUT CONTRAST:CLINICAL INDICATION: History altered mental status. Follow-up.TECHNIQUE: Sagittal 3D T1 gradient echo with axial reformations,axial and coronal TSE T2, axial and coronal FLAIR, arterial spinlabeled perfusion imaging, axial susceptibility weighted imagingand axial diffusion imaging were performed on a 3.0 Tesla system.COMPARISON: Brain MRI dated ——and outside CT dated——FINDINGS:The previously described symmetric restricted diffusion involvingthe perirolandic, perisylvian subcortical white matter, deepposterior corona radiata white matter, periatrial white matterand splenium of corpus callosum have resolved. There is No newrestricted diffusion.Subtle punctate foci of T2/FLAIR hyperintensity in bilateralsubcortical and periatrial white matter. Ventricular system Normal in size, shape, and configuration. No intracranial mass lesions; .No abnormal fluid collections. Visualized major intracranial vessels patent. No definite abnormality in visualized orbits. Middle ear cavities and mastoid air cells are clear. An exceptionof minimal paranasal mucosal thickening. | No | **No**. Reasoning: The report explicitly states that the previously seen abnormalities, including signal abnormalities and reduced diffusion in various brain regions, have resolved. While the report mentions "subtle punctate foci of T2/FLAIR hyperintensity," it characterizes them as "Nonspecific." This means that these tiny areas of signal change are Not necessarily indicative of a specific disease or abnormality and could be due to a variety of factors. Furthermore, the report emphasizes the absence of concerning findings: No new restricted diffusion. Ventricular system Normal. No intracranial mass lesions. | No. The report focuses solely on findings within the brain. Here's why: Anatomical References: The report consistently describes structures within the brain (subcortical white matter, splenium of the corpus callosum, ventricles, intracranial vessels, orbits). Focus on Brain Imaging: The report is for a BRAIN MRI, indicating the primary focus is on intracranial structures. Outside CT Mention: While an "outside CT" is mentioned for comparison, there's No indication that the CT revealed any pathology, especially | No. The report does Not mention any motion artifact or low quality scan. Reasoning: Absence of keywords: The report does Not contain any terms like "motion artifact," "blurring," "low signal-to-Noise ratio," or other phrases that typically indicate image quality issues. Clear descriptions: The radiologist provides clear and specific descriptions of the findings, suggesting the images were of sufficient quality for interpretation. For example, the report describes "subtle punctate foci" which indicates the radiologist could discern fine details. Resolution statement: The report explicitly states that the previously observed "restricted diffusion" has "resolved." This implies that the comparison between the current and prior studies was possible and reliable, suggesting adequate image quality in both. While the report doesn't | No. Reasoning: The report clearly states "Interval resolution of bilateral subcortical and deep white matter, and splenium corpus callosum signal abnormalities and reduced diffusion seen on the prior exam." This means the concerning findings from the previous scan have resolved. While there are some "Nonspecific punctate foci" Noted, these are miNor and don't indicate an acute issue requiring immediate follow-up. The overall tone of the report suggests improvement and stability. Important Disclaimer: I am Not a medical | No. Reasoning: Resolution of prior abnormalities: The report clearly states that the "signal abnormalities and reduced diffusion" seen on the previous exam have resolved. This indicates improvement, Not worsening, of the condition that prompted the initial imaging. Nonspecific findings: The "punctate foci of T2/FLAIR hyperintensity" are described as "Nonspecific". This means they could be related to a variety of causes, many of which are benign or insignificant, and don't automatically signify a serious problem. Normal findings: The report repeatedly emphasizes Normal findings: ventricles are Normal, No mass lesions, No fluid collections, vessels are patent, etc. This further suggests an absence of major concerns. Follow-up: While the report doesn't specify the reason for the "altered | **No** |

| # | Report | Abnormal? (human referee) | Answer Q1 Direct Inquiry | Answer Q2 | Answer Q3 | Answer Q4 | Answer Q5 | Aggregate Stepwise |
|---|---|---|---|---|---|---|---|---|
| | | | No abnormal fluid collections. Visualized major intracranial vessels patent. Overall, the report suggests improvement from the previous scan and does Not identify any definitive brain abnormalities. | outside the brain. The report emphasizes resolution of findings seen on the prior brain MRI. | explicitly state "No motion artifact," the overall content strongly suggests the images were of diagNostic quality and No significant quality issues hindered the interpretation. | professional. This information should Not be considered medical advice. A physician should always be consulted for any health concerns or before making any decisions related to your health or treatment. | mental status," the fact that this is a follow-up and the concerning signs have resolved suggests the situation is being managed. Important Note: This analysis is based solely on the radiology report. A comprehensive clinical assessment by a medical professional is essential for proper diagNosis and treatment. | |
| 28 | Few punctate foci of Nonspecific FLAIR hyperintensity within theright frontal subcortical white matter and focus ofhyperintensity in the posterior limb of the right internalcapsule/globus pallidus. Follow-up MRI is recommended to confirmstability and for better characterization with a complete MRI.Prominent CSF space in the region of the right quadrigeminalcistern causing mild surrounding mass effect, likely an arachNoidcyst.SCREENING BRAIN MRI, WITHOUT CONTRAST:CLINICAL INDICATION: HeadacheTECHNIQUE: Axial diffusion-weighted, axial GRE, axial FLAIR,coronal T2, and sagittal T1-weighted images of the brain withoutintraveNous contrast were performed on a 3.0 Tesla system perfast screening protocol at the Emergency Department's request. COMPARISON: NoneFINDINGS: Examination Not typical quality of full diagNostic MRI;pathology may be missed.Ventricular system Normal.Few punctate foci of Nonspecific FLAIR hyperintensity within theright frontal subcortical white matter. Minimal FLAIRhyperintensity in the peritrigonal white matter. There is a focusof hyperintensity in the posterior limb of the right internalcapsule/globus pallidus. No intracranial mass lesions; No masseffect or midline shift. There is prominent CSF space in the region of the rightquadrigeminal cistern causing mild surrounding mass effect.No definite abnormality in visualized orbits. Middle ear cavities, mastoid air cells, and paranasal sinusesclear. | Yes | **Yes**. Reasoning: The report indicates several findings that suggest potential abnormalities: "Few punctate foci of Nonspecific FLAIR hyperintensity" in the right frontal subcortical white matter and the posterior limb of the right internal capsule/globus pallidus. While these might be Nonspecific, they require follow-up to determine their nature and stability. "Prominent CSF space in the region of the right quadrigeminal cistern causing mild surrounding mass effect, likely an arachNoid cyst." While likely benign, an arachNoid cyst can still cause symptoms depending on its size and location and requires further investigation. Importantly: The report explicitly states that the examination is Not of diagNostic quality and | No Reasoning: All findings described in the radiology report are located within the skull and are therefore within the brain or its surrounding structures: Right frontal subcortical white matter: This refers to a region within the right frontal lobe of the brain. Posterior limb of the right internal capsule/globus pallidus: These are structures located deep within the brain. Right quadrigeminal cistern: This is a space filled with cerebrospinal fluid (CSF) located within the brain. ArachNoid cyst: This is a fluid- | Yes Reasoning: The report explicitly states, "Examination Not typical quality of full diagNostic MRI; pathology may be missed." This indicates the scan quality is suboptimal and could be due to motion artifacts or other factors affecting image clarity. | No. Reasoning: The report recommends follow-up MRI to confirm stability and better characterize the findings, but doesn't explicitly call for immediate clinical follow-up. The findings, while potentially significant, are Not described in a manner that suggests an urgent or emergent need for intervention based on this screening MRI alone. | No. Reasoning: Nonspecific findings: The report uses terms like "Nonspecific FLAIR hyperintensity" and "few punctate foci," suggesting the findings are small and Not clearly indicative of a serious problem. Recommendation for follow-up: The recommendation for a follow-up MRI to "confirm stability and for better characterization" indicates a watch-and-see approach rather than urgent concern. Possible arachNoid cyst: While the report Notes a "prominent CSF space" that could be an arachNoid cyst, it also mentions "mild surrounding mass effect," suggesting it's Not causing significant problems. Screening context: This MRI was performed as a screening for headaches, which are common and often benign. | **Yes** |

| # | Report | Abnormal? (human referee) | Answer Q1 Direct Inquiry | Answer Q2 | Answer Q3 | Answer Q4 | Answer Q5 | Aggregate Stepwise |
|---|---|---|---|---|---|---|---|---|
| | | | that pathology may be missed. This underscores the need for a follow-up MRI for a more complete evaluation. | filled cyst that can occur within the brain or spinal cord. The report does Not mention any findings outside of the skull, which would be necessary to conclude that the pathology is outside the brain. | | | Important Note: This analysis is for informational purposes only and should Not be interpreted as medical advice. A medical professional familiar with the patient's history and the complete medical context is best equipped to interpret these results and determine the appropriate course of action. | |
| 29 | Mild bilateral cerebellar tonsillar ectopia, as described above.Otherwise, No evidence for intracranial abnormality explainingpatient's symptomatology. Varying degree of paranasal sinus inflammatory changes, worseinvolving the spheNoid sinuses, as described. Requires Act 112 Notification: FalseI have personally reviewed the images and agree with theinterpretation.Act 112^FalseBRAIN MRI, WITHOUT CONTRAST:CLINICAL INDICATION: Headache, chronic, new features or increasedfrequency.TECHNIQUE: Sagittal 3D T1 gradient echo with axial reformations,axial and coronal TSE T2, axial FLAIR, axial spin echo T1,arterial spin labeled perfusion imaging, post-contrast sagittal3D T1 gradient echo with axial and coronal reformations,post-contrast axial spin echo T1 with fat suppression, axialsusceptibility weighted imaging, axial diffusion imaging wereperformed on a 3.0 Tesla system.COMPARISON:  No previous for comparison.FINDINGS:Suggestion of tiny persistent cavum septum pellucidum, Normalvariant. Overall, ventricular system Normal.  No abnormal parenchymal signal intensity.Small scattered slightly dilated perivascular spaces withinbilateral cerebral white matter.Mild mineralization within the globi pallidi.No large intracranial space-occupying mass lesions; No masseffect or midline shift.  No abnormal fluid collections.  Visualized major intracranial vessels patent.  IncidentallyNoted, the right transverse and sigmoid sinuses, and imaged upperinternal jugular vein are dominant.The posterior fossa appears somewhat small, and there isbilateral cerebellar tonsillar ectopia with up to roughly 3.5 mmdescension below the level of the foramen magnum. Cerebellartonsillar tips do Not appear significantly pointed in morphologicconfiguration. There may be minimal/mild crowding at the level ofthe craniocervical junction.  No significant abnormalities areNoted within the visualized upper cervical spinal cord.The pituitary gland demonstrates superior convex margin, withinNormal limits for the patient's demographics. Tiny pineal glandcystic foci.No evidence for significant intraorbital abnormality.  Mild rightward nasal septum deviation.The spheNoid sinuses are near completely opacified containingperipheral polyploid mucosal thickening as well as centralintrasinus fluid, which may be proteinaceous. Fluid is mostNotable on the left. In addition, there is scattered mucosalthickening of a varying degree within the remainder of theparanasal sinuses, including polyploid mucosal thickening withinthe left maxillary sinus. Trace T2 signal hyperintensity withinthe | No | Yes. Reasoning: The report identifies "mild bilateral cerebellar tonsillar ectopia" with up to 3.5 mm of descent below the foramen magnum. While the report Notes this may be a Normal variant in some individuals, it is still considered a structural abnormality and could be clinically significant depending on the patient's symptoms. | Yes. The report states: "Varying degree of paranasal sinus inflammatory changes, worse involving the spheNoid sinuses, as described." This indicates inflammation in the paranasal sinuses, which are outside the brain. The paranasal sinuses are air-filled spaces within the skull bones, specifically around the Nose. | No. The report does Not mention any motion artifacts or low-quality scans. In fact, the report describes the findings in detail, suggesting the images were of sufficient quality for interpretation. The mention of "positional plagiocephaly" indicates the patient's positioning might Not have been perfectly symmetrical, but this is Not described as impacting the image quality or diagNosis. | No. While the report does mention mild cerebellar tonsillar ectopia, it clearly states that there are "No significant abnormalities" within the visualized upper cervical spinal cord and "No evidence for intracranial abnormality explaining patient's symptomatology". The report also specifies that the degree of tonsillar ectopia is mild and doesn't show signs of significant crowding. Therefore, immediate clinical follow-up is Not indicated based on this report alone. However, this is just based on the provided report and does Not substitute for a medical professional's opinion. The | No. The report indicates "mild" cerebellar tonsillar ectopia, which, while abnormal, is Not described in a way that suggests high concern. The report explicitly states there is No evidence of intracranial abnormality explaining the patient's symptoms. Additionally: The language used throughout, such as "suggestion," "small scattered," "mild mineralization," and "trace," points towards miNor findings. The report Notes Normal variants and incidental findings, further indicating that the primary concern (headaches) is Not explained by any significant abnormalities. While follow-up with the referring physician is necessary to discuss the findings and plan further management, the report itself does Not convey a sense of urgency or high concern. | **No** |

| # | Report | Abnormal? (human referee) | Answer Q1 Direct Inquiry | Answer Q2 | Answer Q3 | Answer Q4 | Answer Q5 | Aggregate Stepwise |
|---|---|---|---|---|---|---|---|---|
| | mastoids may represent trace fluid. Middle ear cavitiesgrossly clear. Mildly prominent adeNoids.There is an element of flattening of the bilateral posteriorcalvarium, more so on the left, and likely representingpositional plagiocephaly. | | | | | patient should continue to follow up with their doctor as recommended to discuss the findings and address their symptoms. | | |
| 30 | No MRI evidence of hypoxic ischemic encephalopathy. No acute intracranial abnormality.Paranasal sinus disease is as described. I have personally reviewed the images and agree with theinterpretation.BRAIN MRI, WITHOUT CONTRAST:CLINICAL INDICATION: History of asthma initially presentingfollowing cardiac arrest.TECHNIQUE: Sagittal 3D T1 gradient echo with axial reformations,axial and coronal TSE T2, axial and coronal FLAIR, arterial spinlabeled perfusion imaging, axial SWI and axial diffusion imagingwere performed on a 3.0 Tesla system.COMPARISON:  CT of the brain dated——.FINDINGS:There are No areas of abnormal restricted diffusion to suggesthypoxic injury.There is mild diffuse prominence of the sulci and ventriclesreflecting volume loss.Faint patchy signal is Noted within the bilateral, left greaterthan right thalami. Nonspecific signal abnormality in the rightfrontal subcortical white matter and in the periventricular whitematter along the frontal horns. No intracranial mass lesions; . Noother areas of abnormal parenchymal signal intensity orrestriction on diffusion weighted imaging. No abnormal fluid collections.  Visualized major intracranial vessels patent. Near complete opacification of the maxillary sinuses, and partialopacification of the ethmoid and left frontal sinuses bycircumferential polypoidal mucosal thickening and frothysecretions.Partial opacification of the mastoid air cells, left greater thanright.Secretions are seen in the nasopharynx and is related to thepartly visualized oral and nasal route tubes. | Yes | No.

Reasoning:

While the report mentions "faint patchy signal" and "Nonspecific signal abnormality" in certain areas, it explicitly states there are No areas of abnormal restricted diffusion to suggest hypoxic injury which was a primary concern given the clinical indication.

Additionally:

The report finds No intracranial mass lesions or other areas of abnormal signal intensity.
Major intracranial vessels are patent.
The observations about the sinuses and mastoid air cells are common findings and Not indicative of brain abnormalities.
Important Note: This analysis is for informational purposes only and should Not be considered a substitute for professional medical advice. Please consult with a healthcare professional for any health concerns. | Yes.

The report states "Paranasal sinus disease is as described". This indicates there are findings outside of the brain itself within the paranasal sinuses. The report further details findings in the maxillary, ethmoid, and frontal sinuses, confirming pathology outside the brain. | No.

The report does Not mention any motion artifacts or issues with the quality of the scan. It describes the images as clear eNough to visualize specific details like:

No areas of abnormal restricted diffusion: This implies the diffusion imaging was of sufficient quality to make this assessment.
Mild diffuse prominence of the sulci and ventricles: This detail requires clear visualization of brain structures.
Faint patchy signal...Nonspecific signal abnormality: The radiologist is able to identify and describe subtle signal variations, suggesting image quality is adequate.
Visualized major intracranial vessels patent: This indicates clear visualization of blood vessels.
The report focuses on the findings themselves, which is typical of a radiology report when image quality is sufficient for diagNosis. | No.

Reasoning: The report explicitly states "No MRI evidence of hypoxic ischemic encephalopathy" and "No acute intracranial abnormality." While there are some findings (mild volume loss, Nonspecific signal abnormalities), these are Not characterized as acute or requiring immediate attention. The paranasal sinus disease is Noted as "as described," implying it's a kNown, pre-existing condition. | No.

Reasoning:

The report explicitly states "No MRI evidence of hypoxic ischemic encephalopathy" and "No acute intracranial abnormality." These statements indicate that the primary concerns related to the patient's initial condition (cardiac arrest and potential brain injury) are Not supported by the MRI findings.

While the report Notes some miNor findings (mild volume loss, faint signal abnormalities), these are often Nonspecific and Not described in a way that would suggest serious concern.

The report focuses significantly on paranasal sinus disease, which appears to be a pre-existing and potentially chronic condition. | No |

| # | Report | Abnormal? (human referee) | Answer Q1 Direct Inquiry | Answer Q2 | Answer Q3 | Answer Q4 | Answer Q5 | Aggregate Stepwise |
|---|---|---|---|---|---|---|---|---|
| 31 | No acute infarct. No areas of encephalomalacia.No focal steNosis or major collateral vessels of moyamoya type.Requires Act 112 Notification: FalseI have personally reviewed the images and agree with theinterpretation.Act 112^FalseAct 112^FalseBRAIN MRI, WITHOUT CONTRAST:CLINICAL INDICATION: SCD -SS with TCD with Nonvisualization onthe left side - r/o silent infarction; Sickle-cell disorderTECHNIQUE: Brain: Sagittal 3D T1 gradient echo with axial reformations,axial and coronal TSE T2, axial FLAIR, arterial spin labeledperfusion imaging, and axial diffusion imaging were performed ona 3.0 Tesla system.MRA: MRA examination of the brain was performed utilizing a 3Tesla system via 3D time-of-flight intracranial magneticresonance angiography of the circle of Willis, with volume ofinterest 3D reconstructions using the maximum intensityprojection method.MRA was deemed medically necessary based on suspicion of silentof the infarct.COMPARISON:  Brain MRI dated ——FINDINGS:Brain:No reduced diffusion to suggest acute infarct.No areas of abnormal parenchymal signal intensity.Hyperperfusion is seen symmetrically in both thalami and medialoccipital lobes, doubtful clinical significance.Ventricular system Normal in size, shape, and configuration.  No intracranial mass lesions; .No abnormal fluid collections.  Visualized major intracranial vessels patent. No definite abnormality in visualized orbits. There is stable diffuse marrow expansion and hypointense signalof the calvarium, skull base, facial bones, and upper cervicalvertebrae, in keeping with the kNown diagNosis of sickle celldisease.Redemonstrated a small subcentimeter well-circumscribed T2hyperintense with subtle T2 hypointense signal lesion in theright occipital bone, similar in signal, but more ovoid inappearance previously more rounded. This shows reduced diffusion,is this lesion is along the lambdoid suture most likelyrepresenting an epidermoid cyst.Middle ear cavities, mastoid air cells, paranasal sinuses clear.An exception of a small left maxillary sinus mucus retentioncyst.MRA:The major intracranial arterial structures demonstrate Normalflow-related enhancement.  There is No appreciable alteration incourse or caliber.  No areas of aneurysmal outpouching areappreciated. There is fenestration of the anterior communicatingartery, an anatomic variant.It should be Noted that MRA may be insensitive for small (<4 mm)aneurysms. | Yes | No.<br><br>The report explicitly states "No reduced diffusion to suggest acute infarct" and "No areas of abnormal parenchymal signal intensity," indicating No signs of a stroke or other brain tissue damage. While the report Notes hyperperfusion in the thalami and occipital lobes, it deems this finding of "doubtful clinical significance." The report also rules out other abnormalities like intracranial masses, fluid collections, and issues with major blood vessels. The identified lesion in the occipital bone is Not in the brain itself. | Yes.<br><br>Reasoning: The report explicitly mentions a "small subcentimeter well-circumscribed T2 hyperintense with subtle T2 hypointense signal lesion in the right occipital bone". The occipital bone is part of the skull, which is outside of the brain. | No.<br><br>The report doesn't mention any motion artifacts or low-quality issues. In fact, it explicitly states that the ventricular system is Normal in size, shape, and configuration, which suggests a clear and undistorted image. Additionally, the radiologist was able to identify subtle findings such as a small subcentimeter lesion in the occipital bone and a fenestration of the anterior communicating artery, further indicating good image quality. | No.<br><br>Reasoning: The report indicates No acute issues requiring immediate follow-up.<br><br>No acute infarct: This rules out a stroke, which would require urgent attention.<br>Stable findings: The report Notes stable findings compared to the previous MRI, including the epidermoid cyst and marrow expansion.<br>Anatomic variant: The fenestration of the anterior communicating artery is Noted as an anatomic variant, Not a cause for concern.<br>While the report mentions "hyperperfusion," it's deemed of "doubtful clinical significance."<br><br>However: This response is for informational purposes only and should Not be considered medical advice. It's essential for the patient to discuss the report's findings and any necessary follow-up with their doctor, who has access to their complete medical | No.<br><br>The report explicitly states "No acute infarct" and "No areas of encephalomalacia," which are serious conditions. While the report Notes a stable lesion (likely an epidermoid cyst) and some hyperperfusion, these findings are Not described in a way that suggests high concern. Additionally, the MRA shows Normal flow in major arteries. | No |

| # | Report | Abnormal? (human referee) | Answer Q1 Direct Inquiry | Answer Q2 | Answer Q3 | Answer Q4 | Answer Q5 | Aggregate Stepwise |
|---|---|---|---|---|---|---|---|---|
| | | | | | | history and can provide personalized guidance. | | |
| 32 | MR brain:1. No evidence of acute intracranial disease process or new brainlesion.2. Unchanged small right cerebral periventricular hyperintenselesion, Nonspecific but may represent sequela of prior injury.MR angiogram brain:No evidence of significant arterial steNosis.Requires Act 112 Notification: FalseI have personally reviewed the images and agree with theinterpretation.Act 112^FalseAct 112^FalseBRAIN MRI, WITHOUT CONTRAST:CLINICAL INDICATION: ——chronic headache newfeatures for increasing frequency.TECHNIQUE: Noncontrast MRI and MRA of the brain.COMPARISON:  MRI brain —— .FINDINGS:Ventricular system remains stable. Mild prominence of the leftlateral ventricle has Not significantly changed. The remainderthe ventricular system is within Normal limits in size.Stable Nonspecific focus of T2 FLAIR hyperintensity on the rightatrial subependymal/periventricular white matter. Stable roundedfocal fluid signal deforming the upper margin of the righthippocampus, likely a small choroidal fissure cyst. However,slightly less mass effect is apparent in the right choroidalfissure, perhaps due to interval brain growth/development betweenstudies.No intracranial soft tissue mass lesions; No significant masseffect or midline shift. The pituitary gland is prominent withconvexity of its upper margin, but this is within Normal limitsfor age and gender.No evidence of abnormal fluid collections or hemorrhage.  Paranasal sinuses and mastoid air cells are relatively clear.Ocular globes and intraorbital contents are unremarkable.Arterial flow voids at the skull base appear intact. Noaggressive osseous lesions identified.MRA:Visualized major intracranial vessels patent on the MRA. No MRAevidence of aneurysm or dissection. Incidentally, the rightophthalmic artery arises more proximally and laterally from theright ICA compared to the left, Normal variant. The rightvertebral artery is dominant. | Yes | No. The report explicitly states "No evidence of acute intracranial disease process or new brain lesion." While there are a few miNor findings: Small right cerebral periventricular hyperintense lesion: This is unchanged from a previous scan and likely insignificant, potentially related to a prior injury. Mild prominence of the left lateral ventricle: This has Not changed and is considered within Normal limits. Stable rounded focal fluid signal near the right hippocampus: This is likely a small choroidal fissure cyst, a benign finding. Prominent pituitary gland: This is considered Normal for the patient's age and gender. Variation in the right ophthalmic artery: This is a Normal anatomical variant. None of these findings suggest any current brain abnormality or acute disease process. | No. The report focuses solely on the evaluation of intracranial (within the brain) structures and findings. Here's why: "MR brain" and "MR angiogram brain" indicate the imaging focused specifically on the brain. The report describes structures within the brain: ventricles, white matter, hippocampus, pituitary gland, etc. It mentions findings related to the brain: small hyperintense lesion, choroidal fissure cyst, No mass lesions or hemorrhage. Even the incidental Note about the right ophthalmic artery is discussing its origin within the | No. The report does Not mention any motion artifacts or low quality issues. In fact, the radiologist specifically states that they have "personally reviewed the images" and agree with the interpretation, indicating that the image quality was sufficient for diagNosis. | No. The report explicitly states "No evidence of acute intracranial disease process or new brain lesion." The findings focus on unchanged or stable pre-existing conditions, None of which are described in a way that suggests urgency. | No. The report explicitly states "No evidence of acute intracranial disease process or new brain lesion". The findings describe stable, Nonspecific, and unchanged observations. While the patient has a history of headaches, the report doesn't suggest any alarming findings that would cause high concern. | No |

| # | Report | Abnormal? (human referee) | Answer Q1 Direct Inquiry | Answer Q2 | Answer Q3 | Answer Q4 | Answer Q5 | Aggregate Stepwise |
|---|---|---|---|---|---|---|---|---|
| | | | | skull in relation to the internal carotid artery (ICA), an intracranial blood vessel. The report does Not provide information about structures outside the skull. | | | | |
| 33 | 1. Pineal cyst measuring up to 1.1 cm. While this is within the range of measurements commonly seen, it exerts at minimal mass effect upon the adjacent superior tectum, slightly altering its contours.. This is favored to represent an incidental finding, although it should be correlated with the patient's current symptoms. A long-term follow-up exam can be obtained to assess stability if of clinical concern. 2. Otherwise unremarkable MRI of the brain with and without contrast. I have personally reviewed the images and agree with the interpretation. BRAIN MRI, WITHOUT AND WITH CONTRAST: CLINICAL INDICATION: —— Headache with new visual disturbances. New eye movement discoordination per emergency room provider. TECHNIQUE: Sagittal 3D T1 gradient echo with axial reformations, axial and coronal TSE T2, axial FLAIR, arterial spin labeled perfusion imaging, and axial diffusion imaging were performed on a 3.0 Tesla system. COMPARISON: None FINDINGS: Ventricular system Normal. No abnormal parenchymal signal intensity.. There is a 1.1 x 1.0 x 0.8 cm cystic structure in the pineal region, compatible with a simple pineal cyst. This exerts a minimal mass effect upon the adjacent superior tectal plate, slightly altering its contours, best visualized on sagittal images (series 5, image 90 for example). Cerebral aqueduct remains patent. Cerebellar tonsils Normal in position. Pituitary gland within the range of Normal. Corpus callosum is intact. No abnormal fluid collections. No abnormal contrast enhancement. Visualized major intracranial vessels patent. No definite abnormality in visualized orbits. Mild mucosal thickening in the right frontal sinus, and trace scattered in ethmoid air cells mostly anteriorly. Additional mild mucosal thickening in the left spheNoid sinus, with trace in the right spheNoid sinus and the maxillary sinuses. Trace T2 hyperintense presumed fluid in the mastoid air cells bilaterally. | Yes | **Yes**. Reasoning: The report identifies a pineal cyst measuring 1.1 cm that exerts minimal mass effect on the adjacent superior tectum. While potentially an incidental finding, the report itself states this is an abnormality ("abnormal parenchymal signal intensity"). It recommends correlation with the patient's symptoms (headache, visual disturbances, eye movement discoordination) and potential follow-up. | No. The report explicitly states that all findings are within the brain. The report describes a pineal cyst, which is located within the brain, and mentions other structures such as the superior tectum, cerebral aqueduct, cerebellar tonsils, pituitary gland, and corpus callosum, all of which are brain structures. | No. The report does Not mention any motion artifacts or low-quality issues. Furthermore, the radiologist explicitly states, "I have personally reviewed the images and agree with the interpretation," indicating confidence in the image quality and findings. | Yes. Reasoning: While the report classifies the pineal cyst as "likely incidental," it explicitly recommends correlating the finding with the patient's symptoms due to the mass effect observed on the adjacent superior tectal plate. The patient is experiencing headaches, visual disturbances, and eye movement discoordination – symptoms that could potentially be caused by pressure in this area. Therefore, immediate clinical follow-up is necessary to: Assess the correlation between the patient's symptoms and the cyst's location/mass effect. | No. The report states that the pineal cyst is "within the range of measurements commonly seen" and is "favored to represent an incidental finding." While they Note a "minimal mass effect," they also state the cerebral aqueduct remains patent. The report recommends correlation with the patient's symptoms and a follow-up only "if of clinical concern." The remaining findings are unremarkable. Overall, the language suggests a clinically insignificant finding rather than a highly concerning one | **Yes** |

| # | Report | Abnormal? (human referee) | Answer Q1 Direct Inquiry | Answer Q2 | Answer Q3 | Answer Q4 | Answer Q5 | Aggregate Stepwise |
|---|---|---|---|---|---|---|---|---|
| | | | | | | Determine if further investigation or management of the pineal cyst is required. | | |
| 34 | Unremarkable brain MRI prior to and following intraveNouscontrast.Normal intracranial magnetic resonance veNography.I have personally reviewed the images and agree with theinterpretation.BRAIN MRI, WITHOUT AND WITH CONTRAST:INTRACRANIAL MR VENOGRAPHY, WITHOUT CONTRAST:CLINICAL INDICATION: Headache.TECHNIQUE: Brain: Sagittal 3D T1 gradient echo with axialreformations, axial and coronal TSE T2, axial and coronal FLAIR,axial spin echo T1, arterial spin labeled perfusion imaging,post-contrast sagittal 3D T1 gradient echo with axial and coronalreformations, post-contrast axial spin echo T1 with fatsuppression, axial diffusion imaging were performed on a 3.0Tesla system. MRV: 2D time-of-flight intracranial magnetic resonance veNographywas performed on a 3 Tesla system, in the coronal and axialplanes, with volume of interest 3D reconstructions performed onan independent workstation.MRV was deemed medically necessary based on concern for duralveNous sinus thrombosis.COMPARISON: CT ——FINDINGS:Ventricular system Normal in size, shape, and configuration.  No areas of abnormal parenchymal signal intensity.No intracranial mass lesions; . Ason prior exam, there is a 1.6 x 2.4 x 2.9 cm (AP by TV by CC)arachNoid cyst along the left parasagittal posterior fossawithout any significant mass on the underlying cerebellum.Visualized major intracranial vessels patent. No definite abnormality in visualized orbits. Trace fluid in the mastoid air cells. Mild scattered mucosalthickening within the paranasal sinuses. Small mucous retentioncyst within the left maxillary sinus. Mild deviation of the nasalseptum to the left. The adeNoids are mildly enlarged with tinyretention cysts within.MRV:Major intracranial veNous structures demonstrate Normalflow-related enhancement.  No veNous sinus thrombosis. No flow-limiting veNous steNosis. | No | **No**. The report explicitly states "No areas of abnormal parenchymal signal intensity", "No intracranial mass lesions", and "No definite abnormality in visualized orbits." While miNor findings like an arachNoid cyst and mild mucosal thickening are Noted, these are common and Not described as abnormal in this context. | No. The report explicitly states "No definite abnormality in visualized orbits" and focuses primarily on intracranial structures (within the skull). While it mentions some findings in the sinuses and nasal passages, these are common incidental findings and are Not described in a way that suggests they are the cause of the patient's headaches or are otherwise clinically significant. | No. The report does Not mention any motion artifacts or low-quality issues. In fact, it explicitly states:  "I have personally reviewed the images and agree with the interpretation." This statement from the radiologist indicates confidence in the image quality and interpretation. Detailed description of the techniques used: The report outlines the specific MRI sequences used, suggesting a comprehensive and high-quality study. No mention of limitations: The report would typically mention if motion artifacts or low image quality hampered the interpretation. | No. Reasoning: The report clearly states the brain MRI and MRV are unremarkable, with No abnormalities detected. The findings specifically rule out concerns like intracranial masses, veNous thrombosis, and steNosis. While the report Notes miNor, common findings like a small arachNoid cyst (unchanged from previous scans) and mild sinus thickening, these are Not presented as acute concerns requiring immediate follow-up. | No. The report repeatedly uses language like "Normal," "unremarkable," and "No definite abnormality." The arachNoid cyst is Noted as unchanged from the prior exam, indicating it is Not an acute concern. While the MRV was ordered due to a concern about thrombosis, the results clearly indicate No such finding. | **No** |
| 35 | Subtle, ill-defined FLAIR hyperintensity in the periventricularwhite matter, more so on the right, new since prior examination.It's Nonspecific and may represent chronic silent ischemia.Otherwise, No new lesions. Specifically, No acute ischemia. Stable MR angiogram.Stable bone marrow changes, as detailed above, related to sicklecell disease.Requires Act 112 Notification: FalseAct 112^FalseAct 112^FalseBRAIN MRI, WITHOUT CONTRAST; MR ANGIOGRAPHY OF THE BRAIN, WITHOUTCONTRAST:CLINICAL INDICATION: Sickle cell disease, imaging surveillanceTECHNIQUE: Sagittal 3D T1 gradient echo with axial reformations,axial and coronal TSE T2, axial and coronal FLAIR, arterial spinlabeled perfusion imaging, and axial diffusion imaging wereperformed on a 3.0 Tesla system. MRA examination of the brain wasperformed utilizing a 3.0 Tesla system via 3D time-of-flightintracranial magnetic resonance angiography of the circle ofWillis, with volume of interest 3D | Yes | **Yes**. Reasoning: The report indicates a "subtle, ill-defined FLAIR hyperintensity in the periventricular white matter," which is a new finding compared to the prior examination. While the report suggests this may represent chronic silent ischemia, it Nonetheless points to | No. The report focuses solely on the brain and surrounding structures. Here's why:  Periventricular white matter: This refers to the white matter surrounding the ventricles, which | No. The report does Not mention any motion artifacts or low-quality scan. In fact, it describes the visualized structures with eNough detail to suggest the images were of sufficient quality for interpretation. The radiologist was able to clearly identify specific findings, including subtle ones like the FLAIR | No. Reasoning: Although the report Notes a new, subtle FLAIR hyperintensity that may represent chronic silent ischemia, it explicitly states there is No acute ischemia. The report indicates stability in other | No. The report describes the new finding as "subtle, ill-defined FLAIR hyperintensity" and states it "may represent chronic silent ischemia". The language suggests a possibility, Not a certainty, and the use of "chronic silent ischemia" implies a long-standing, Not acutely dangerous, condition. Additionally: | **Yes** |

| # | Report | Abnormal? (human referee) | Answer Q1 Direct Inquiry | Answer Q2 | Answer Q3 | Answer Q4 | Answer Q5 | Aggregate Stepwise |
|---|---|---|---|---|---|---|---|---|
| | reconstructions using the maximum intensity projection method. MRA was deemed medically necessary based on kNown vasculopathy. COMPARISON: MRI dated ——FINDINGS: No abnormal restricted diffusion to suggest acute ischemia. Subtle ill-defined FLAIR hyperintensity in the periventricular white matter, more so on the right (series 7 image 25), new since prior examination. There is Nonspecific bowel may represent silent chronic ischemia. Ventricular system is stable in size, shape, and configuration. There is No hydrocephalus. No additional areas of abnormal parenchymal signal intensity. No intracranial mass lesions; .Stable mild right cerebellar tonsillar ectopia. The CSF space at the craniocervical junction is preserved. No abnormal fluid collections. Visualized major intracranial vessels patent. No definite abnormality in visualized orbits. Middle ear cavities, mastoid air cells, paranasal sinuses clear. Persistent signal abnormality and expansion of the calvarial marrow, left pterygoid process. Some of this may reflect chronic bone infarctions. Persistent dark T1/T2 signal within the visualized upper cervical spine. The major intracranial arterial structures demonstrate Normal flow-related enhancement. The A1 segment of the right anterior artery appears mildly hypoplastic, anatomic variant. The left internal carotid artery appears diffusely mildly larger than the right, with tortuous caverNous segment, unchanged. The left superior cerebellar artery is duplicated (anatomic variant). There is No appreciable alteration in caliber., Noting that MRA may be insensitive for small (<4mm) aneurysms. | | an abnormality within the brain. | are spaces within the brain. Cerebellar tonsillar ectopia: This describes a structural variation in the cerebellum, a part of the brain. Craniocervical junction: This is the area where the skull connects to the spine, still relevant to the brain and its protection. Intracranial: This term specifically means "within the skull" further indicating the report's focus on the brain. While the report mentions "bone marrow changes" and "signal abnormality and expansion of the calvarial marrow," these are interpreted in the context of sickle cell disease and its kNown effects on bone, including the skull. There's No indication of pathology outside the skull or vertebral column. | hyperintensity and the hypoplastic A1 segment of the right anterior artery. | findings, and the overall tone suggests No need for immediate clinical follow-up. Routine follow-up with the ordering physician is likely appropriate for further management. | "No acute ischemia" is directly stated, ruling out an immediate concern. "Stable MR angiogram" indicates No worsening of blood vessel condition. "Stable bone marrow changes" further supports a lack of acute issues. While follow-up might be needed, the report doesn't convey high concern from the radiologist. | |

| # | Report | Abnormal? (human referee) | Answer Q1 Direct Inquiry | Answer Q2 | Answer Q3 | Answer Q4 | Answer Q5 | Aggregate Stepwise |
|---|---|---|---|---|---|---|---|---|
| 36 | Focal area of soft tissue prominence with prominent coursingvessels left parietal scalp region measuring up to 1.6 cm,corresponding to areas of increasing vascularity and priorultrasound, favored to reflect a vascular malformation.2 mm posteriorly projecting outpouching of the caverNous ICA, mayreflect small aneurysm. This can be further evaluated with followup MRA, CTA, or cerebral angiogram, as clinically warranted. No acute intracranial abnormality.I have personally reviewed the images and agree with theinterpretation.BRAIN MRI, WITHOUT AND WITH CONTRAST:MR ANGIOGRAPHY OF THE BRAIN, WITHOUT CONTRAST:CLINICAL INDICATION: History of palpable lump in the scalp, possible vascular malformation. Follow up imaging.TECHNIQUE: MRI of the brain with Sagittal 3D T1 gradient echowith axial reformations, axial and coronal TSE T2, axial andcoronal FLAIR, axial spin echo T1, arterial spin labeledperfusion imaging, post-contrast sagittal 3D T1 gradient echowith axial and coronal reformations, post-contrast axial spinecho T1 with fat suppression, axial diffusion imaging wereperformed on a 3.0 Tesla system. MRA examination of the brain wasperformed utilizing a 3 Tesla system via 3D time-of-flightintracranial magnetic resonance angiography of the circle ofWillis, with volume of interest 3D reconstructions using themaximum intensity projection method.MRA was deemed medically necessary based on clinical history.COMPARISON:  Ultrasound soft tissues —— .FINDINGS:BRAIN:Soft tissue prominence left scalp with coursing vessels andfluid-fluid levels, corresponding to area of vascularity Noted onprior US. This area is approximately 1.6 x 0.8 cm (image97/series 22). Some restricted diffusion is Noted along thisarea, which may reflect recent intralesional hemorrhage. There isNo extension in the calvarium or intracranial vault. There is Nosuspicious contrast enhancement.No abnormal parenchymal signal intensity. No restricteddiffusion.No intracranial mass lesions; .  No abnormal fluid collections.  No abnormal contrast enhancement.Visualized major intracranial vessels patent. No definite abnormality in visualized orbits. Middle ear cavities, mastoid air cells, and paranasal sinusesclear.MRA:Otherwise, the intracranial vasculature is patent. No regions offlow-limiting steNosis are identified. Codominant vertebralarteries. Fetal origin of the right posterior communicatingartery.2 mm posterior lateral projecting outpouching of the leftcaverNous ICA (image 198/series 6). | Yes | **Yes**. Reasoning: The report identifies a "2 mm posteriorly projecting outpouching of the left caverNous ICA" which "may reflect a small aneurysm." While this finding requires further investigation with MRA, CTA, or cerebral angiogram to confirm if it is indeed an aneurysm, it is considered an abnormality until proven otherwise. Additionally, the "focal area of soft tissue prominence with prominent coursing vessels" and "some restricted diffusion" in the left parietal scalp region, while likely a vascular malformation, also point to abnormalities present. | Yes. Reasoning: The report describes a "Focal area of soft tissue prominence with prominent coursing vessels left parietal scalp region" and explicitly states "There is No extension in the calvarium or intracranial vault." This indicates the primary finding is located in the scalp (soft tissue), outside the skull and therefore outside the brain. While the report also mentions a possible aneurysm in the caverNous ICA, this is a separate finding and does Not change the fact that the primary pathology is extracranial. | No. The report does Not mention any motion artifacts or low-quality scans. Furthermore, the radiologist "personally reviewed the images and agree[s] with the interpretation," indicating confidence in the image quality and findings. | Yes. Reasoning: Although the report doesn't describe acute findings requiring immediate intervention, it does mention a "2 mm posteriorly projecting outpouching of the left caverNous ICA" that "may reflect [a] small aneurysm." The report recommends follow-up imaging (MRA, CTA, or cerebral angiogram) to further evaluate this finding. Even small aneurysms carry a risk of rupture, which can be life-threatening. Timely follow-up with a clinician is essential to determine the appropriate course of action, which may include further imaging and potential treatment to mitigate the risk of rupture. | No. While the report does Note a couple of findings (vascular malformation, small possible aneurysm), the language emphasizes further evaluation and follow-up rather than urgency. Here's why: "Favored to reflect": This phrase, used to describe the vascular malformation, suggests the radiologist's leading hypothesis but Not a definitive diagNosis requiring immediate action. "Can be further evaluated": This recommendation for the aneurysm implies it's Not an immediate, pressing concern. "As clinically warranted": This phrase empowers the referring physician to decide on further investigation based on the patient's overall clinical picture and Not solely on the radiology findings. "No acute intracranial abnormality": This statement is reassuring and suggests there is No immediate danger. The report indicates areas of interest that require monitoring and potential further investigation. However, the language doesn't convey a sense of high concern or urgency from the radiologist. | **Yes** |

| # | Report | Abnormal? (human referee) | Answer Q1 Direct Inquiry | Answer Q2 | Answer Q3 | Answer Q4 | Answer Q5 | Aggregate Stepwise |
|---|---|---|---|---|---|---|---|---|
| 37 | Small, entirely Nonspecific focus of signal abnormality withoutenhancement, in the right frontal lobe, unchanged since originalpresentation in ——.No new lesions.Requires Act 112 Notification: FalseAct 112^FalseBRAIN MRI, WITHOUT AND WITH CONTRAST:CLINICAL INDICATION: Right frontal lesion, imaging follow-upTECHNIQUE: Sagittal 3D T1 gradient echo with axial reformations,axial and coronal TSE T2, sagittal 3D FLAIR with axial andcoronal reformations, arterial spin labeled perfusion imaging,postcontrast sagittal 3D T1 gradient echo with axial and coronalreformations, postcontrast axial spin echo T1, and axialdiffusion imaging were performed on a 3.0 Tesla PET/MRI system.COMPARISON: Numerous prior examinations, including most recentavailable MRI dated ——FINDINGS:Stable, 3.4 mm, Nonenhancing focus of FLAIR hyperintensity in thedeep white matter of the right frontal lobe. No new lesions have developed.Ventricular system Normal, Noting asymmetric mild prominence ofthe left lateral ventricle, developmental variant. No abnormal parenchymal signal intensity.No intracranial mass lesions; . No abnormal fluid collections. No abnormal contrast enhancement.Visualized major intracranial vessels patent. No definite abnormality in visualized orbits. Middle ear cavities, mastoid air cells, and paranasal sinusesclear, apart from miNor mucosal thickening in the ethmoid aircells and maxillary sinuses.. | Yes | No. Reasoning: The report explicitly states that the small signal abnormality in the right frontal lobe is "Nonspecific" and "unchanged since the original presentation in ——". Additionally, it emphasizes "No new lesions" and describes the finding as stable. While the report Notes the abnormality, it doesn't categorize it as definitively abnormal or indicative of a specific condition. | No. The report explicitly states the abnormality is within the "right frontal lobe" which is a part of the brain. The report describes a "small, entirely Nonspecific focus of signal abnormality" in the right frontal lobe, indicating an area within the brain tissue itself. | No. The report does Not mention any motion artifacts or low image quality. The radiologist describes the findings clearly and does Not indicate any limitations due to technical factors. Additionally, the report mentions using a 3.0 Tesla PET/MRI system, which is a high-field strength scanner capable of producing high-quality images. | No. Reasoning: The report clearly states the small abnormality in the right frontal lobe is: "entirely Nonspecific" meaning it doesn't have characteristics that point to a specific or worrisome diagNosis. "without enhancement" which generally makes a lesion less concerning. "unchanged since original presentation in —" indicating stability and lack of growth. The report also Notes "No new lesions" and finds No other concerning findings. While routine follow-up may be recommended, the report's language does Not suggest an urgent need for immediate clinical follow-up. | No. Reasoning: "Nonspecific" and "unchanged" suggest low concern: The report uses the term "Nonspecific," implying the finding is Not clearly linked to any particular condition. More importantly, it states the abnormality is "unchanged" since ——, indicating stability and lack of worrisome progression. Absence of alarming findings: The report emphasizes "No new lesions," "No abnormal enhancement," and uses reassuring language like "Normal," "stable," and "No definite abnormality" regarding other crucial structures. Focus on routine follow-up: While the report doesn't state the reason for the initial — scan, the overall tone suggests a routine follow-up for a previously identified, Non-urgent finding. Important Note: This analysis is for informational purposes only and should Not substitute professional medical advice. The significance of any radiological finding must be interpreted within the context of the patient's full medical history and clinical presentation. | No |

| # | Report | Abnormal? (human referee) | Answer Q1 Direct Inquiry | Answer Q2 | Answer Q3 | Answer Q4 | Answer Q5 | Aggregate Stepwise |
|---|---|---|---|---|---|---|---|---|
| 38 | No definite abnormality on this screening brain MRI. SCREENING BRAIN MRI, WITHOUT CONTRAST:CLINICAL INDICATION: Headache and vomiting. History ofmicrocephaly and failure to thrive with dysmorphic features andscoliosis, seizures, developmental delay. TECHNIQUE: Axial diffusion-weighted, axial GRE, axial FLAIR,coronal T2, and sagittal T1-weighted images of the brain withoutintravenous contrast were performed on a 3.0 Tesla system perfast screening protocol at the Emergency Department's request.Electronic medical record clinical notes mention a history ofprenatal toxic exposure and a history of nonaccidental traumaCOMPARISON: Brain MRI ——FINDINGS: This examination is not of the typical quality of a fulldiagnostic MRI and, as such, pathology may be missed.The ventricular system is normal in size, shape, andconfiguration. There are no intracranial mass lesions, and thereis no mass effect or midline shift. No abnormal fluid collectionsare identified. No definite areas of abnormal parenchymal signal. No definite abnormality is seen in the visualized portions of theorbits. The middle ear cavities and mastoid air cells are clear.The paranasal sinuses are clear. | Yes | No.

Reasoning: The report explicitly states "No definite abnormality on this screening brain MRI" and "No definite abnormality is seen in the visualized portions of the orbits."

Important Considerations:

Screening vs. DiagNostic: The report emphasizes this was a screening MRI, done quickly in the Emergency Department, and Not a full diagNostic study. This means subtle abnormalities may be missed.
Clinical History: The patient has a complex medical history. While this MRI didn't show clear brain abnormalities, it doesn't rule out other causes for their symptoms. Further investigation and correlation with their history is needed.
Follow-up: It's likely the patient will require follow-up with their doctor to discuss these results and determine if additional testing is necessary. | No.

Reasoning: The report explicitly states it is a "screening brain MRI" and focuses solely on findings within the brain. It describes the ventricles, fluid collections, and even briefly mentions the orbits (eye sockets), all anatomical structures within the head. There is No mention of imaging or findings related to any areas outside of the brain. | Yes.

The report explicitly states, "This examination is Not of the typical quality of a full diagNostic MRI and, as such, pathology may be missed." This indicates the images were likely Not optimal, possibly due to motion artifact, the "fast screening protocol", or other factors impacting image quality. | No.

Reasoning:

"No definite abnormality" - This statement explicitly states that No significant issues were found on the scan.
Screening Exam Limitations: The report emphasizes that this was a screening MRI performed in the Emergency Department with a fast protocol. This means it's Not as comprehensive as a full diagNostic MRI and might Not detect subtle abnormalities.
Clinical Correlation: While the patient has a complex medical history and presented with headache and vomiting, the report doesn't link these symptoms to any specific finding on the MRI.
Important Note: While the MRI itself doesn't necessitate immediate action, the patient's symptoms and history warrant clinical follow-up with their physician. The physician will consider the MRI results in the context of the patient's overall | No.

Reasoning:

"No definite abnormality" is stated multiple times throughout the report, indicating a lack of strong findings suggesting a serious acute issue.
"Screening" MRI: This type of MRI is used for a general overview, often in an emergency setting as in this case, and is Not as detailed as a full diagNostic MRI.
Limitations ackNowledged: The report clearly states that the quality is Not that of a full diagNostic MRI and "pathology may be missed." This suggests the findings should be interpreted cautiously.
History suggests pre-existing conditions: The patient's history indicates pre-existing neurological conditions (microcephaly, seizures, developmental delay), which could be contributing to the current symptoms (headache, vomiting) rather than a new acute issue.
Important Note: While the report doesn't indicate immediate, high-level concern, it doesn't rule out potential underlying issues. Further investigation and follow-up with the patient's doctor are essential for a complete assessment and appropriate management. | **Yes** |

| # | Report | Abnormal? (human referee) | Answer Q1 Direct Inquiry | Answer Q2 | Answer Q3 | Answer Q4 | Answer Q5 | Aggregate Stepwise |
|---|---|---|---|---|---|---|---|---|
| | | | | | | condition and determine the best course of action. | | |
| 39 | MR brain:No evidence of acute intrarenal disease process or focal brainlesion.MR spine:Unremarkable spine MRI. No MR findings to explain the patient'sfoot drop.I have personally reviewed the images and agree with theinterpretation.BRAIN MRI, WITHOUT CONTRAST:CLINICAL INDICATION: ——presenting with foot drop.TECHNIQUE: Noncontrast MRI brain, thoracic and lumbar spine.COMPARISON:  NoneFINDINGS:Brain:Susceptibility artifact from braces and motion is present, mayreduce sensitivity. Examination remains overall diagNostic.Ventricular system Normal in size, shape, and configuration.  No areas of abnormal parenchymal signal intensity.No intracranial mass lesions; .No abnormal fluid collections.  Visualized major intracranial vessels patent. No definite abnormality in visualized orbits. There is scattered mild paranasal sinus mucosal thickening.Mastoid air cells and paranasal sinuses are clear.Spine:Vertebrae and intervertebral discs Normal. Spinal cord Normal.  Cranio-cervical junction unremarkable.  Conus medullaris Normal shape; ends at L1-L2 level.  Cauda equina and filum terminale unremarkable.  No focal intraspinal lesions.Paraspinal soft tissues unremarkable. Kidneys orthotopic; without hydronephrosis. | Yes | No. The report explicitly states "No evidence of acute intrarenal disease process or focal brain lesion" and "No areas of abnormal parenchymal signal intensity." It describes the ventricular system, intracranial vessels, and visualized orbits as Normal. The only findings mentioned are mild sinus thickening and susceptibility artifact from braces, neither of which are abnormal eNough to be clinically significant in this context. | Yes. The report explicitly states "No MR findings to explain the patient's foot drop." Since foot drop is usually caused by problems affecting the nerves or muscles in the leg or spine, and the report finds No spinal abnormalities, this suggests the pathology is outside the brain and spinal cord. | Yes. The report explicitly states "Susceptibility artifact from braces and motion is present, may reduce sensitivity". This confirms the presence of motion artifact in the brain MRI. While the report mentions that the examination remains overall diagNostic, the presence of artifact means the image quality is Not ideal. | No. The report clearly states "No MR findings to explain the patient's foot drop" and finds No abnormalities in the brain, spine, or surrounding structures that would require immediate clinical follow-up. The report suggests that the cause of the foot drop is Not related to a neurological issue visible on the MRI. While the findings don't necessitate immediate action, the patient will need further investigation to determine the cause of the foot drop. | No. The reasoning is as follows: Unremarkable Findings: The report repeatedly uses terms like "Normal," "unremarkable," and "No evidence" to describe the findings for both the brain and spine. This indicates that the radiologist did Not find any significant abnormalities. No Acute Issues: The report specifically states "No evidence of acute intrarenal disease process or focal brain lesion." This suggests that they were looking for something serious and did Not find it. Foot Drop Cause Not Identified: While the report ackNowledges the patient's foot drop, it clearly states that the MRI did Not reveal a cause. This suggests further investigation is needed, but Not necessarily with a sense of urgency conveyed in the report. Important Note: This analysis is based solely on the radiology report. A medical doctor would consider the patient's full medical history, symptoms, and other diagNostic tests to | **Yes** |

| # | Report | Abnormal? (human referee) | Answer Q1 Direct Inquiry | Answer Q2 | Answer Q3 | Answer Q4 | Answer Q5 | Aggregate Stepwise |
|---|---|---|---|---|---|---|---|---|
| | | | | | | | determine the level of concern and next steps in treatment. | |
| 40 | 1. Flattening of the posterior globes with subtle protrusion of the optic nerve heads, may reflect papilledema versus drusen. Opthalmology consultation is suggested. 2. Patency of the dural veNous sinuses with preferential drainage into the left side. END OF IMPRESSION: Requires Act 112 Notification: TrueI have personally reviewed the images and agree with the interpretation. Act 112^TrueBRAIN MRI, WITHOUT CONTRASTCLINICAL INDICATION: Observed seizure-like activity. Per EMR patient with history ofTrisomy 21 who had an event of loss of awareness while at schoolcharacterized by him lifting his arms and then becoming limp. TECHNIQUE: Sagittal 3D T1 gradient echo with axial and coronal reformations, axial and coronal TSE T2, axial and coronal FLAIR, axial SWI, axial ASL, axial DTI, and axial diffusion weighted imaging wereperformed on a 3.0 Tesla system. COMPARISON:  NoneFINDINGS: The ventricular system is Normal in size, shape, andconfiguration, with a prominent cavum septum pellucidum andvergae. Note is made of subtle scattered bifrontal and parietalperivascular spaces. Nonspecific punctate subcortical bifrontalFLAIR hyperintense signal is Noted, most likely related toadjacent perivascular spaces. Question is raised with respect to mild bilateral frontal lobehypoplasia., and there is No masseffect or midline shift.  No abnormal fluid collections areidentified.  There are No additional areas of abnormalparenchymal signal intensity on these sequences. There is mildly thinning of the corpus callosum..There is mild flattening of the posterior globes, at the opticnerve heads, with optic nerve head elevation/subtle protrusion ofthe optic heads at this level (series 7 image 56-57), concerningfor papilledema versus drusen. The dural veNous sinuses are Normal shaped and signal. There ismild dominant drainage of the left jugular vein. Craniocervical junction is patent, with a Normal atlantoaxialdistance. The middle ear cavities, mastoid air cells, and paranasal sinusesare clear. An exception of mild ethmoidal mucosal thickening. Leftward deviation of the nasal septum is Noted. Partially visualized is an oral airway in place. As made of bilateral enlarged palatine tonsils. Tiny adeNoidal s are Noted. | Yes | **Yes**.

Reasoning:

Possible Papilledema: The report Notes flattening of the posterior globes and protrusion of the optic nerve heads, which could indicate papilledema. Papilledema is the swelling of the optic disc due to increased intracranial pressure, a serious condition. Frontal Lobe Hypoplasia: The report mentions "mild bilateral frontal lobe hypoplasia." While the degree of hypoplasia is Noted as mild, it still represents an abnormality in brain development. Important Considerations:

Ophthalmology Referral: The report explicitly recommends an ophthalmology consultation to further assess the optic nerve head appearance and differentiate between papilledema and drusen (benign deposits).
Clinical Context: The MRI was ordered due to seizure-like activity in a patient with Trisomy 21. Brain | Yes.

The report mentions "flattening of the posterior globes" and "subtle protrusion of the optic nerve heads," which suggests a possible increase in intracranial pressure (ICP). This is further supported by the concern for papilledema, which is swelling of the optic disc often caused by increased ICP. Since the optic nerves are extracranial structures, this indicates that the pathology causing these findings could be located outside the brain. | No.

While the report doesn't explicitly mention image quality, it describes subtle findings and intricate anatomical details (like mild perivascular spaces, subtle optic nerve head protrusion, and mild ethmoidal mucosal thickening). This level of detail suggests the images were of sufficient quality for the radiologist to make these observations. | Yes.

The report states "flattening of the posterior globes with subtle protrusion of the optic nerve heads... concerning for papilledema versus drusen. Ophthalmology consultation is suggested."

Papilledema is a serious condition that can lead to vision loss if left untreated. While the report also mentions drusen as a possibility, which is a benign condition, the suggestion for an ophthalmology consultation indicates a need for immediate follow-up to determine the underlying cause of the optic nerve head changes and rule out papilledema. | Yes.

Here's why:

Papilledema Concern: The report highlights "flattening of the posterior globes" and "subtle protrusion of the optic heads," raising concern for papilledema. Papilledema is swelling of the optic nerve head due to increased intracranial pressure and can be a serious medical condition.
Ophthalmology Referral: The radiologist explicitly recommends an immediate Ophthalmology consultation to further evaluate the possibility of papilledema. This urgency emphasizes the potential seriousness of the finding.
Act 112 Notification: While the specific details of Act 112 are Not provided, the fact that this report triggers a Notification indicates a potential need for protective measures or reporting related to suspected child abuse. This further suggests the findings might be related to a serious underlying condition.
In summary, the combination of suspected papilledema, urgent ophthalmology | **Yes** |

| # | Report | Abnormal? (human referee) | Answer Q1 Direct Inquiry | Answer Q2 | Answer Q3 | Answer Q4 | Answer Q5 | Aggregate Stepwise |
|---|---|---|---|---|---|---|---|---|
| | | | abnormalities, including those mentioned, could be associated with these clinical presentations. Disclaimer: I am an AI and canNot provide medical advice. The information here should Not substitute for professional medical evaluation. | | | | referral, and Act 112 Notification strongly suggests the medical professionals are highly concerned about the patient's condition. | |
| 41 | Chiari 1 deformity.I have personally reviewed the images and agree with theinterpretation.SCREENING BRAIN MRI, WITHOUT CONTRAST:CLINICAL INDICATION: headacheTECHNIQUE: Axial diffusion-weighted, axial GRE, axial FLAIR,coronal T2, and sagittal T1-weighted images of the brain withoutintraveNous contrast were performed on a 3.0 Tesla system perfast screening protocol at the Emergency Department's request. COMPARISON: NoneFINDINGS: Examination Not typical quality of full diagNostic MRI;pathology may be missed.Ventricular system Normal.No definite abnormal parenchymal signal. Cerebellar tonsils extend 11 mm below the foramen magnum withpointed appearance. Slightly dorsiflexed dens measuringapproximately 68 degrees.No intracranial mass lesions; . No abnormal fluid collections. No definite abnormality in visualized orbits. Middle ear cavities, mastoid air cells, and paranasal sinusesclear. | Yes | **Yes**<br><br>Reasoning: The report identifies a Chiari 1 deformity, which is a structural abnormality where the cerebellar tonsils protrude downward into the foramen magnum (the opening at the base of the skull). The report states that the cerebellar tonsils extend 11 mm below the foramen magnum, which is significant. Additionally, a slightly dorsiflexed dens is Noted, which can be associated with Chiari malformations. | Yes<br><br>Reasoning: The report describes a Chiari 1 deformity. This condition involves the cerebellar tonsils (part of the cerebellum, which is a part of the brain) extending downward into the foramen magnum (the opening at the base of the skull where the spinal cord connects to the brain). While the cerebellum is part of the brain, its abnormal positioning in a Chiari 1 malformation affects the spinal cord and can obstruct the flow of cerebrospinal fluid. | No.<br><br>While the report mentions "Examination Not typical quality of full diagNostic MRI; pathology may be missed", it does Not specifically mention motion artifact or low image quality as the reason for this limitation.<br><br>The report indicates that the study was performed as a "fast screening protocol" at the Emergency Department's request. This suggests that the study may have been acquired with fewer images or faster sequences compared to a routine, full diagNostic MRI, which could explain the limitation in sensitivity. However, it doesn't automatically translate to motion artifact or poor image quality. | Yes<br><br>Reasoning: The report indicates a Chiari 1 deformity with cerebellar tonsils extending 11 mm below the foramen magnum. This is a significant finding that may require clinical follow-up, especially in the context of the patient's headaches.<br><br>While the report mentions that the examination was Not of full diagNostic quality, the presence of a Chiari 1 malformation warrants further investigation and potential referral to a neurologist or neurosurgeon.<br><br>It is crucial to emphasize that medical advice should always be sought from qualified healthcare professionals. | No.<br><br>While the report does mention a Chiari 1 deformity (cerebellar tonsils extending 11 mm below the foramen magnum), the report also states "Examination Not typical quality of full diagNostic MRI; pathology may be missed."<br><br>This suggests the findings might Not be entirely conclusive and a full diagNostic MRI might be needed for a better assessment. Additionally, there's No mention of acute findings like hydrocephalus or other alarming signs that would typically indicate high concern in the setting of a Chiari malformation. | **Yes** |